\documentclass[preprint]{aastex}
\usepackage{color}
\newcommand\pcc{\;{\rm cm}^{-3}}
\newcommand\Msun{\; {\rm M}_{\odot}}
\newcommand\kms{\; {\rm km}\;{\rm s}^{-1}}
\newcommand\ergs{\; {\rm erg}\;{\rm s}^{-1}}

\newcommand\cm{\;{\rm cm}}
\newcommand\yr{\; {\rm yr}}
\newcommand\Myr{\;{\rm Myr}}
\newcommand\Gyr{\;{\rm Gyr}}
\newcommand\pc{\;{\rm pc}}
\newcommand\kpc{\;{\rm kpc}}
\newcommand\sfrunit{\Msun \kpc^{-2} \yr^{-1}}
\newcommand\Punit{\pcc\,{\rm K}}
\newcommand\Surf{\Msun\;{\rm pc^{-2}}}
\newcommand\rhounit{\Msun\;{\rm pc^{-3}}}
\newcommand\Kel{\;{\rm K}}
\newcommand\eV{\;{\rm eV}}

\newcommand\simgt{\lower.5ex\hbox{$\; \buildrel > \over \sim \;$}}
\newcommand\simlt{\lower.5ex\hbox{$\; \buildrel < \over \sim \;$}}
\newcommand\pderiv[2]{\frac{\partial {#1}}{\partial {#2}}}

\newcommand\advect[2][\vel]{{#1}\cdot \nabla {#2}}
\newcommand\rbrackets[1]{\left({#1}\right)}
\newcommand\sbrackets[1]{\left[{#1}\right]}
\newcommand\cbrackets[1]{\left\{{#1}\right\}}
\newcommand\abrackets[1]{\left\langle{#1}\right\rangle}
\newcommand\divergence[2][\rbrackets]{\nabla \cdot #1{#2}}

\newcommand\vel{\mathbf{v}}
\newcommand\Om{\mathbf{\Omega}}

\newcommand\xhat{\hat{\mathbf{x}} }
\newcommand\yhat{\hat{\mathbf{y}} }
\newcommand\zhat{\hat{\mathbf{z}} }

\newcommand\kbol{k_{\rm B}}

\newcommand\eff{\epsilon_{\rm ff}}
\newcommand\tff{t_{\rm ff}}
\newcommand\tffmid{t_{\rm ff,0}}
\newcommand\ever{\epsilon_{\rm ver}}
\newcommand\tver{t_{\rm ver}}
\newcommand\tdyn{t_{\rm dyn}}
\newcommand\Msn{m_{\rm *}}
\newcommand\rsh{r_{\rm sh}}

\newcommand\pmax{p_{\rm max}}
\newcommand\Psn{p_{\rm *}}

\newcommand\Rmax{R_{\rm max}}
\newcommand\rhoth{\rho_{\rm cr}}
\newcommand\Tth{T_{\rm cr}}
\newcommand\cth{c_{\rm cr}}
\newcommand\nth{n_{\rm cr}}
\newcommand\SigSFR{\Sigma_{\rm SFR}}
\newcommand\SigSFRlow{\Sigma_{\rm SFR,low}}
\newcommand\SigSFRsun{\Sigma_{\rm SFR,0}}
\newcommand\SigGBC{\Sigma_{\rm GBC}}
\newcommand\Sigdiff{\Sigma_{\rm diff}}
\newcommand\torb{t_{\rm orb}}
\newcommand\Pmax{P_{\rm max}}
\newcommand\Pmin{P_{\rm min}}
\newcommand\nmax{n_{\rm 1}}
\newcommand\nmin{n_{\rm 2}}

\newcommand\rhosd{\rho_{\rm sd}}
\newcommand\ngbc{n_{\rm GBC}}
\newcommand\fdiff{f_{\rm diff}}
\newcommand\fgbc{f_{\rm GBC}}
\newcommand\fc{f_{c}}
\newcommand\fu{f_{u}}
\newcommand\fw{f_{w}}
\newcommand\sz{\sigma_{\rm z}}
\newcommand\szdiff{\sigma_{\rm z,diff}}
\newcommand\vzdiff{v_{\rm z,diff}}
\newcommand\vthdiff{v_{\rm th,diff}}
\newcommand\Hdiff{H_{\rm diff}}
\newcommand\Pth{P_{\rm th}}
\newcommand\Pturb{P_{\rm turb}}

\newcommand\Ptot{P_{\rm tot}}
\newcommand\rhomid{n_0}
\newcommand\rhoDE{\rho_{\rm 0,DE}}
\newcommand\nDE{n_{\rm 0,DE}}
\newcommand\HDE{H_{\rm diff, DE}}
\newcommand\PDE{P_{\rm th, DE}}
\newcommand\PthDE{P_{\rm th, DE}}
\newcommand\PtotDE{P_{\rm tot, DE}}
\newcommand\Ptwo{P_{\rm two}}
\newcommand\Qinit{Q_{\rm init}}

\newcommand\tsf{t_{\rm SF,GBC}}
\newcommand\tausf{\tau_{\rm SF, GBC}}

\newcommand\frad{f_{\rm rad}}

\newcommand\etath{\eta_{\rm th}}
\newcommand\etaturb{\eta_{\rm turb}}

\newcommand\Pdriv{P_{\rm driv}}

\shorttitle{Regulation of Star Formation Rates}
\shortauthors{KIM, KIM, \& OSTRIKER}

\begin{document}

\title{Regulation of Star Formation Rates in Multiphase Galactic Disks:
Numerical Tests of the Thermal/Dynamical Equilibrium Model}

\author{Chang-Goo Kim\altaffilmark{1,2}, Woong-Tae Kim\altaffilmark{1,2,3},
and Eve C.\ Ostriker\altaffilmark{4}}

\affil{$^1$Center for the Exploration of the Origin of the Universe
(CEOU), Astronomy Program, Department of Physics \& Astronomy, Seoul
National University, Seoul 151-742, Republic of Korea}
\affil{$^2$Department of Physics \& Astronomy, FPRD, Seoul National
University, Seoul 151-742, Republic of Korea} 
\affil{$^3$Institute for Advanced Study, Einstein Drive, Princeton,
NJ 08540, USA}
\affil{$^4$Department
of Astronomy, University of Maryland, College Park, MD 20742, USA}
\email{kimcg@astro.snu.ac.kr, wkim@astro.snu.ac.kr,
ostriker@astro.umd.edu} 
\slugcomment{Accepted by the \apj}

\begin{abstract}
We use vertically-resolved numerical hydrodynamic simulations to study
star formation and the interstellar medium (ISM) in galactic disks.
We focus on outer disk regions where diffuse \ion{H}{1} dominates,
with gas surface densities $\Sigma =3-20 \Surf$ and star-plus-dark
matter volume densities $\rhosd=0.003-0.5 \rhounit$.  
Star formation occurs in very dense, self-gravitating clouds that 
form by mergers of smaller cold cloudlets.  
Turbulence, driven by momentum feedback from
supernova events, destroys bound clouds and puffs up the disk
vertically.  Time-dependent radiative heating
(FUV from recent star formation) offsets gas cooling. 
We use our simulations to test a new theory for
self-regulated star formation.  
Consistent with this theory, the disks evolve to a
state of vertical dynamical equilibrium and thermal equilibrium with
both warm and cold phases.  The range of star formation surface
densities and midplane thermal pressures is $\SigSFR \sim 10^{-4} -
10^{-2} \sfrunit$ and $\Pth/\kbol \sim 10^2-10^4 \Punit$.  In
agreement with observations, turbulent velocity dispersions are $\sim
7\kms$ and the ratio of the total (effective) to thermal pressure is
$\Ptot/\Pth \sim 4-5$, across this whole range (provided shielding is
similar to the Solar neighborhood).  
We show that $\SigSFR$ is not well correlated with $\Sigma$ alone, but 
rather with $\Sigma \sqrt{\rhosd}$, because the vertical
gravity from stars and dark matter dominates in outer disks.
We also find that $\SigSFR$ has a strong, nearly linear
correlation with $\Ptot$, which itself is within $\sim 13\%$ of the
dynamical-equilibrium estimate $\PtotDE$.
The quantitative relationships we find between $\SigSFR$ and the
turbulent and thermal pressures show that star formation is highly efficient 
for energy and momentum production, in contrast to the low efficiency of 
mass consumption.
Star formation rates adjust until the ISM's energy and momentum losses are
replenished by feedback within a dynamical time.
\end{abstract}

\keywords{galaxies: ISM --- galaxies: kinematics and
dynamics --- galaxies: star formation --- method: numerical ---
turbulence}


\section{Introduction}\label{sec:intro}

Large-scale star formation rates in galaxies are observed to correlate with
both the gaseous and stellar content, and with the galaxy's
gravitational potential well (e.g.
\citealt{ryd94,ken98,wong02,boi03,sal07,leroy08,big08,big10,big11,gen10,dad10,shi11}).
Empirical fits in disks often adopt power-law (``Kennicutt-Schmidt'') forms for
the relationship among the surface density of star formation
$\SigSFR$, the surface density of gas $\Sigma$, the surface density of
the old stellar disk $\Sigma_s$, and the orbital angular velocity
$\Omega$.  

From the ``supply side'' point of view, gas represents the
fuel for star formation, and the stellar disk and dark matter halo
help to define dynamical timescales within the interstellar medium (ISM) that
could affect how rapidly gas collects and collapses: the
galactic orbital time, the vertical oscillation period and flow
crossing time, and the
gravitational free-fall time.  Power laws naturally arise if the star
formation rate is proportional to the ratio of the gas content and one
of these dynamical times.  The observed timescale for gas to be converted
to stars, $t_{\rm SF,gas} \equiv \Sigma/\SigSFR$ is, however,
generally quite long compared to these dynamical times.  Together,
the empirical results present a picture of star formation 
that is sensitive to both fuel supply and ambient
environmental conditions, and that has low apparent efficiency.

In recent work, \citet{oml10} (hereafter OML10) and \citet{ost11}
(hereafter OS11) have argued that star formation rates 
respond to demand, as well as supply.  Maintaining an equilibrium
state in the ISM requires constant inputs of energy and momentum, and
contributions from star formation are critical.  Star formation can be
self-regulated via feedback, in such a way that supply and demand
match within the ISM: heating balances cooling, pressure balances gravity, and
turbulent driving balances dissipation.  The theory of OML10 and OS11
proposes that observed star formation rates can be understood as a 
response to the needs of the ISM.  Because each massive star injects so
much energy, only a relatively modest star formation rate (implying a long 
$t_{\rm SF,gas}$) is necessary. From the point of view of energy and
momentum sources and sinks, star formation is in fact quite efficient.

To see why feedback is vital, it is key to consider the
internal thermal and dynamical state of the ISM, rather than just
integrated properties.  The internal vertical dynamical time $\tdyn
\propto (G \rho_{\rm tot})^{-1/2}$, for $\rho_{\rm tot}$ the total (gas +
stellar) density, depends on the thicknesses of the gaseous and
stellar disks.  In particular, the contribution from gas gravity alone
gives $\Sigma/\tdyn \propto \Sigma^{3/2}/H^{1/2}$.  The gas disk thickness
$H$ depends (linearly or quadratically) on the vertical velocity
dispersion of the gas, which includes both thermal and turbulent
terms.\footnote{
  In this work we neglect the magnetic term, which is likely to be small
  (see below) but would provide a minimum vertical support in the limit of 
  vanishing turbulent terms.}
Because thermal energy is radiated away, and turbulent energy
is dissipated (in shocks and shear layers) on timescales $\simlt \tdyn
\ll t_{\rm SF,gas}$, the internal energy must be continuously
replenished.  Young, high-mass stars restore this energy and preserve
the life of the ISM.  If star formation feedback were entirely absent
and the only heating source were the cosmic background radiation,
$\tdyn$ would drop by nearly two orders of magnitude, with a corresponding
(or greater) increase in $\SigSFR$.

OML10 and OS11, considering respectively mid-to-outer disks and
central starburst regions, showed that observed star formation rates
are quantitatively consistent with analytic predictions that follow
from imposing thermal and dynamical equilibrium in the diffuse ISM.  
OS11 also presented initial results of numerical simulations that 
include turbulent driving associated with star formation, 
confirming the analytic theory for molecule-dominated regions.
Additional results from simulations in the starburst regime will be
presented in Shetty \& Ostriker (2011, in preparation).  

In this paper, we use time-dependent numerical simulations to test the
OML10 theory (and extensions based on OS11), for the outer-disk regime
where the ISM is dominated by diffuse atomic gas.  A crucial aspect of
our simulations is that we vertically resolve the disk (our grid scale
is $1\pc$). We shall show that, as assumed by OML10, thermal and
vertical dynamical equilibrium are both satisfied in our numerical
models.  We shall also show that feedback from star formation is
largely responsible for sustaining both the thermal and turbulent
pressure (and energy) in the atomic ISM.  We numerically calibrate the
yield relation between $\SigSFR$ and the thermal and turbulent pressures in
the diffuse ISM, demonstrating that near-linear relations hold for
both $\Pth$ and $\Pturb$.  By combining these feedback relations with
dynamical equilibrium, we show that $\SigSFR$ depends nearly linearly
on the weight of the diffuse ISM (i.e. the dynamical-equilibrium
pressure $\PtotDE \approx \Pth+\Pturb$). 
The correlation between $\SigSFR$ and $\Ptot$ (or $\PtotDE$)
is stronger
and more general than other star formation relations that are commonly
cited.

In addition to testing the thermal/dynamical equilibrium theory of 
star formation, our numerical models allow us to address a number of
interesting issues related to observations of diffuse atomic gas in
the Milky Way and external galaxies 
\citep{dhh90,bra97,van99,heiles03,you03,pet07,dic09,Kal09}.
These observations show that (1) turbulent velocity dispersions are typically 
$\sim 10 \kms$, relatively independent of location or star formation rate; 
(2) both cold and warm atomic gas are pervasive, in
proportions that appear relatively independent of location; (3) 
the thermal pressure is a small fraction of the total pressure.  Our 
numerical results are consistent with these observations, and 
can be understood based on the thermal/dynamical equilibrium model
with energy and momentum feedback from star formation.

Our numerical models are idealized in that they represent
a local patch of unmagnetized gas in a featureless
disk where star formation is primarily responsible for the injection of
thermal and kinetic energies. Thus, in this paper 
we do not capture the potential consequences of galactic structural
features and certain instabilities that may affect ISM dynamics and
star formation. The ISM surface density averaged over  
$\sim$ kpc scales  can be significantly affected by
large-scale gravitational instability (e.g.,
\citealt{wad99,wad07,kim01,kim02,kim07,li05,tas06,tas09,tas11,bou07,bou09,hop11}), 
spiral arm compression (e.g. \citealt{kim02,kim06,she06,
kko08,kko10,db06,db08,dob08,dob11,wad04,wad08,wad11}), and Parker
instability (e.g. \citealt{bas97,kimj98,kimj01,kos02,mou09}).  Since
the timescales to collect gas over $ \simgt$ kpc scales from 
gravitational instabilities and spiral arms are longer than
local dynamical times, our models may nevertheless provide a good
first approximation to the effects of star formation feedback on local
regions within larger gas accumulations.  In addition, initial tests
we have conducted which include magnetic fields (permitting Parker
instability) show similar behavior to our unmagnetized models.

As well as producing $\sim$ kpc-scale overdensities, 
both gravitational instabilities 
(e.g. \citealt{wad02,kos03,kim07,age09,aum10, bou10})
and spiral shocks (e.g. \citealt{kim06,kko06,kko10,dob06}), together
with magnetorotational instabilities 
(e.g., \citealt{kos03,pio04,pio05,pio07}),
drive turbulence in the ISM.  In particular, turbulence levels $\simgt
10\kms$ can be produced by large-scale gravitational instability, 
and may be important during the highly-transient early evolution of
disk galaxies. Several of the above numerical models have
shown, however, that unless energy (representing feedback) is locally injected
into massive, high-density clumps that form, the result is 
irreversible gravitational collapse and star formation far exceeding
observed rates. Stellar feedback therefore 
appears to be crucial for disrupting bound clouds (thus
limiting star formation) and maintaining -- over many galactic orbits --
turbulent ISM levels similar to those observed in nearby galaxies.

The plan of this paper is as follows.  In Section 2, we begin by
summarizing the theory developed in OML10 and OS11.  Section 3
describes the numerical methods and parameters used for our
time-dependent simulations, and Section 4 presents our model results.
These include time averages of star formation rates, thermal and
turbulent pressures, gas layer scale-heights, thermal and turbulent
velocity dispersions, and mass fractions of gas components.  In
Section 5, we use our numerical results to test the validity of the
physical assumptions and adopted parameters in the OML10 theory.
Here, we also
demonstrate the balance between turbulent driving and dissipation
(as in OS11), 
and quantify the feedback yield relations between 
$\Pth$ and  $\Pturb$, and $\SigSFR$.
We compare our numerical results to several simple prescriptions for
star formation in Section 6.  Section 7 summarizes and discusses our
main results.

\section{Summary of Thermal/Dynamical Equilibrium Model}\label{sec:theory}

In this section, we briefly summarize the OML10 thermal/dynamical
equilibrium model, highlighting the fundamental assumptions and
predictions that we shall test in this work.  We then draw on OS11 to
outline additional predictions related to the dynamical state and star
formation rate in disks dominated by turbulent, diffuse gas, and
describe how these hypotheses will be tested.

OML10 considered a multiphase, 
turbulent galactic ISM disk with thermal properties mediated by
stellar heating.  The gaseous disk, with total surface
density of neutral gas $\Sigma$, is immersed within the stellar disk
and dark matter halo, whose combined midplane density is given by
$\rho_s+\rho_{\rm dm}\equiv\rhosd$.  The neutral gas disk 
is composed of two components: diffuse gas, with 
surface density averaged over large scales 
$\Sigdiff$; and gravitationally bound clouds (GBCs)
with surface density averaged over large scales (i.e. many individual bound 
clouds) $\SigGBC=\Sigma-\Sigdiff$.
The diffuse component includes both warm, rarefied gas and
cold, dense gas in cloudlets that are not massive enough to be 
gravitationally bound.  Star formation
takes place within the gravitationally-bound component.

The first key assumption of OML10 is that the volume-filling diffuse
ISM disk is in force balance in the vertical direction. The combined
inward gravitational force of the stars, dark matter, and gas
(both diffuse and GBC components) must be matched by the outward
pressure forces within the diffuse gas.  Averaging the vertical
component of the momentum equation over time and in the
horizontal direction, OML10 showed that in a state of dynamical
equilibrium, $P_{\rm tot}=\PtotDE$ for 
\begin{equation}\label{eq:PtotDE}
\PtotDE \equiv \frac{\pi G \Sigdiff^2}{4}
\cbrackets{1+2\frac{\SigGBC}{\Sigdiff}+
\sbrackets{\rbrackets{1+2\frac{\SigGBC}{\Sigdiff}}^2+
\frac{32\zeta_d c_w^2 \tilde{f}_w \alpha}{\pi G}
\frac{\rhosd}{\Sigdiff^2}}^{1/2}};
\end{equation}
that is, the total effective midplane pressure\footnote{As discussed
  in OML10, $P_{\rm tot}$ is actually a pressure {\it difference}
  between the midplane and the top of the neutral layer.  Thus, if the
  cosmic-ray and magnetic scale heights far exceed that of the neutral
  gas, there is not a significant contribution to $P_{\rm tot}$ from
  magnetic or cosmic-ray terms (even if their midplane pressures are
  large), and the weight of the diffuse neutral layer must be
  supported primarily by turbulent and thermal pressure.}
$P_{\rm tot}$ must support the weight of the overlying diffuse gas in the
total gravitational field.  Although we use the symbol 
$\PtotDE$ to denote the vertical weight, it is important to note that 
the weight and effective pressure balance only if equilibrium holds, and 
only in an averaged sense. 

In equation (\ref{eq:PtotDE}), $\alpha$ is the ratio of total
(effective) pressure to thermal pressure in the diffuse medium,
$\zeta_d$ is a dimensionless parameter characterizing the gas density
profile ($\zeta_d=1/\pi$ for a Gaussian profile), $c_w=(k
T_w/\mu)^{1/2}$ is the thermal speed of the warm gas, and $\tilde{f}_w
=\vthdiff^2/c_w^2$ for $\vthdiff$ the mass-weighted thermal velocity
dispersion in the diffuse gas.  The quantity $\tilde f_w$ is also
equal to $\rho_w/\rho_0$ for $\rho_w$ the warm medium density and
$\rho_0$ the volume-averaged density of the diffuse medium (including
cold cloudlets, assumed to be in pressure equilibrium with the warm
medium) at the disk midplane.  The mass fraction of the warm medium in
the diffuse gas is comparable to $\tilde f_w$ (see OML10).  In a state
of dynamical equilibrium, the midplane diffuse-gas thermal pressure
$\Pth = \rho_0 \vthdiff^2$ is equal to
\begin{equation}\label{eq:PthDE}
\PthDE=\frac{\PtotDE}{\alpha}
\label{eq:PDE}
\end{equation}
(see equation 11 of OML10).  If the dominant contributions to the 
total effective pressure are thermal and turbulent terms 
with $\Pturb=\rho_0 \vzdiff^2$, then 
$\alpha =(\vthdiff^2 + \vzdiff^2)/\vthdiff^2=\szdiff^2/\vthdiff^2$ 
for $\vzdiff$ the turbulent vertical velocity dispersion and 
$\szdiff$ the total vertical velocity dispersion in the diffuse gas 
($\szdiff$ is a direct observable for a face-on disk).  Note that in equation 
(\ref{eq:PtotDE}), the product $c_w^2\tilde f_w \alpha = \Ptot/\rho_0$, which 
is equal to  $\szdiff^2$ if turbulent and thermal terms dominate the effective
pressure, i.e.
$\Ptot \approx \Pth +\Pturb$.

Next, OML10 assumed that the diffuse ISM is in a state of thermal
equilibrium, in which cold and warm atomic phases coexist at a
midplane thermal pressure $P_{\rm th,TE}$.  In order for the diffuse
gas to be in the two-phase regime, $P_{\rm th,TE}$ must fall between
the minimum pressure $\Pmin$ for the cold phase and the maximum
pressure $\Pmax$ for the warm phase (cf. \citealt{fie69}).  Both
$\Pmax$ and $\Pmin$ depend linearly on the local radiative heating
rate per particle, $\Gamma$, which itself depends approximately
linearly on the locally-averaged star formation rate surface
density, $\SigSFR$, if young massive stars are responsible for
most of the heating.  Motivated by detailed modeling of heating and
cooling in the Solar neighborhood \citep{wol03} and numerical
simulations of turbulent multiphase gas \citep{pio05,pio07}, OML10
assumed that $P_{\rm th,TE}$ is comparable to the geometric-mean
pressure $\Ptwo\equiv(\Pmin\Pmax)^{1/2}$. Based on the results of
\citet{wol03}, OML10 adopted a geometric mean ``two-phase'' pressure given by
\begin{equation}\label{eq:Ptwo}
\Ptwo/\kbol=3\times10^3 
\Punit\frac{4G_0'}{1+3Z_d'(\Sigma/\Sigma_0)^{0.4}},
\end{equation}
where $\kbol$ is the Boltzmann constant, $G_0'=J_{\rm FUV}/J_{\rm
  FUV,0}$ is the mean FUV intensity relative to the
Solar neighborhood value $J_{\rm FUV,0}=2.2\times10^{-4}\ergs \cm^{-2}
{\rm sr}^{-1}$, $\Sigma_0=10\Msun\pc^{-2}$ is the surface density of
neutral gas at the Solar circle \citep{Dic90,Kal09}, and $Z_d'$ is the
dust abundance relative to Solar neighborhood value.  In the Solar 
neighborhood, $\Ptwo/\kbol=3000 \Punit$ for the OML10 prescription.

In a state of simultaneous thermal and dynamical equilibrium, heating
and cooling are in balance so that $\Pth=P_{\rm th,TE}\sim \Ptwo$, and
vertical forces are in balance so that $\Pth=P_{\rm th,DE}$.  With
$\Ptwo \propto G_0' \propto J_{\rm FUV} \propto \SigSFR$, the surface density of
star formation should be proportional to $P_{\rm th}$.
Thus, equating (\ref{eq:PDE}) and (\ref{eq:Ptwo}) yields an expression
for the star formation rate, with $\SigSFR$ proportional to the
right-hand side of equation (\ref{eq:PtotDE}) -- i.e. to the weight of
the diffuse gas layer in the total gravitational field.  In
low-density outer-disk regions where the diffuse gas dominates GBCs
($\Sigdiff\rightarrow\Sigma$ and $\SigGBC\rightarrow0$), an
approximate form for $\SigSFR$ is then given by
\begin{eqnarray}\label{eq:sfrlow}
\SigSFRlow &\approx& 3\times10^{-4}\sfrunit 
\rbrackets{\frac{\Sigma}{10\Surf}}
\sbrackets{1+3\rbrackets{\frac{Z_d'\Sigma}{10\Surf}}^{0.4}}\times\nonumber\\
&&\sbrackets{\frac{2}{\alpha}\rbrackets{\frac{\Sigma}{10\Surf}}+
\rbrackets{\frac{50\tilde{f}_w}{\alpha}}^{1/2}
\rbrackets{\frac{\rhosd}{0.1\rhounit}}^{1/2}}
\end{eqnarray}
(see eqs. 22 and A13 in OML10).  The numerical coefficient 
in equation (\ref{eq:sfrlow}) is calibrated based on 
the local Milky Way value $\Sigma_{\rm SFR,0}=2.5\times10^{-3}\sfrunit$
\citep{fuchs09}.

In the case when $\SigGBC/\Sigma$ is non-negligible, in order to
obtain a closed set of equations, OML10 made the additional assumption
that star formation within GBCs has a gas consumption timescale $\tsf$
so that
\begin{equation}\label{eq:tsf}
\SigSFR=\frac{\SigGBC}{\tsf}=\frac{\Sigma-\Sigdiff}{\tsf}.
\end{equation}
If GBCs have relatively uniform properties, then $\tsf$ will be
relatively constant.  By equating (\ref{eq:PDE}) and
(\ref{eq:Ptwo}), and combining with equation (\ref{eq:tsf}), OML10
obtained a cubic equation that can be solved for $\SigSFR$ as a function of
$\Sigma$ and $\rhosd$ in the general case; an approximate form is
given by
\begin{equation}\label{eq:sfrboth}
\SigSFR\approx\sbrackets{\frac{\tsf}{\Sigma}+\frac{1}{\SigSFRlow}}^{-1}
\end{equation}
(see eqs. 23 and A14 in OML10).  Note that 
for low surface density outer disks, equation (\ref{eq:sfrlow})
is recovered and $\SigSFR$ is independent of $\tsf$ 
-- i.e. the star formation rate becomes independent of the 
rate at which gas in GBCs collapses to make stars.

OML10 took $\tsf=2\Gyr$ based on the empirical linear correlation
\citep{big08} between the molecular mass in CO and the SFR at 750 pc
scale for a set of disk galaxies (at moderate $\Sigma \simlt 100
\Msun\pc^{-2}$), and adopted $\alpha\approx 5$ and $\tilde{f}_w\approx
0.5$ as typical values based on observations of the Milky Way and
other well-studied disk galaxies.  If the same set of parameters is
adopted for all galaxies (note that the dependence on $\tilde f_w/\alpha$
in equation \ref{eq:sfrlow} is weak: $\SigSFR \propto
(\tilde f_w/\alpha)^{0.5}$), $\SigSFR$ is a function of just $\Sigma$ and
$\rhosd$.  OML10 applied this formulation to azimuthally-averaged data
for a sample of spiral galaxies to predict $\SigSFR$ as a function of
galactocentric radius $R$. The resulting predicted profiles of
$\SigSFR$ are overall in remarkably good agreement with the
observations.  For a few galaxies, however, observed values of
$\SigSFR$ are offset from the prediction. The difference may owe to
different values of $\alpha$, $\tilde{f}_w$, and/or $\tsf$ from the
adopted values, or to effects associated with azimuthal averaging when
there is strong spiral structure.  It should also be noted that there
are still significant uncertainties in the observations, which might 
lead to offsets with respect to the theory.  Empirical
determinations of $\Sigma$ and $\tsf$ are uncertain 
since some gas may be undetected
in both 21 cm and CO lines, and since the conversion factor $X_{\rm CO}$ from
 CO to H$_2$ can vary by a factor $\sim 2$ ($X_{\rm CO}$
varies even more at low metallicity, and where $\Sigma \simgt 100
\Msun\pc^{-2}$). The age of the young-star population, as well as the treatment
of extended vs. concentrated tracers of star formation, can also affect the
empirical estimates of $\SigSFR$. In addition, as discussed by OML10, values of
$\rhosd$ are uncertain as stellar disk thickness estimates for face-on
galaxies are obtained via scaling relations rather than being directly
measured.

In this paper, we focus on the low-$\Sigma$ case, corresponding to
outer disks where the gas is primarily diffuse and atomic.  In this
regime, $\SigSFR$ is predicted to depend on $\alpha$ and $\tilde{f}_w$
but not on $\tsf$, according to equation (\ref{eq:sfrlow}).  Using our
numerical simulations, in which $\Sigma$ and $\rhosd$ are independent
variables, we can directly test the primary assumptions of the OML10
theory.  Since we can measure $\alpha$, $\tilde{f}_w$, $\Sigdiff$ (and
$\SigGBC=\Sigma-\Sigdiff$) together with $\Pth$ from the simulation 
outputs for any model, we can
test whether the measured midplane thermal pressure in fact agrees
with the dynamical equilibrium value $P_{\rm th,DE}$ predicted by
equation (\ref{eq:PDE}).  We can also investigate whether the measured
midplane $\Pth$ is close to $\Ptwo$, following the hypothesis of OML10
that the system evolves to a state of thermal equilibrium having both
a warm and cold atomic phase.  Similarly, we can test whether the
sum of the measured thermal and turbulent pressures 
$\Pth + \rho_0\vzdiff^2  =\Ptot$
is consistent with the dynamical equilibrium prediction 
of equation (\ref{eq:PtotDE}) (since the present simulations do not include 
magnetic fields, cosmic rays, or radiation pressure, these terms do not enter 
$P_{\rm tot}$).
Further, we can check whether our numerical 
results for $\alpha$ and $\tilde{f}_w$ agree with
empirically-estimated values, and explore how much variation in $\alpha$ 
and $\tilde f_w$ there is
among models with different $\Sigma$ and $\rhosd$.  Finally, we can
compare the value of $\SigSFR$ from the simulations with the
theoretical prediction based on simultaneous thermal and dynamical
equilibrium (cf. equation \ref{eq:sfrlow}).

In addition to testing the theory of OML10, we can use our numerical
simulations to test more general ideas related to the self-regulation
of star formation, as introduced by OS11.  We consider the situation
in which the ISM is dominated by diffuse gas, so that
$\SigGBC/\Sigdiff \rightarrow 0$ and $\Sigdiff \rightarrow \Sigma$.  We
also assume the effective pressure is dominated by thermal and
turbulent terms\footnote{That is, we assume cosmic ray, magnetic
  field, and radiation effects are unimportant -- see OML10 and OS11
  for an evaluation and discussion of these.}, 
and take $\zeta_d
\approx 1/\pi$ and $c_w^2 \tilde f_w \alpha = \szdiff^2\rightarrow
\sigma_z^2$ so that equation (\ref{eq:PtotDE}) for the weight becomes
\begin{equation}\label{eq:PtotDEdiff}
\PtotDE = \frac{\pi G \Sigma^2}{4}
\cbrackets{1+ \sbrackets{1+
\frac{32 \sz^2}{\pi^2 G}
\frac{\rhosd}{\Sigma^2}}^{1/2}}.
\end{equation}
A simplified expression for $\PtotDE$, within 20\% of 
equation (\ref{eq:PtotDEdiff}), is
\begin{eqnarray}\label{eq:PtotDEap}
\PtotDE &\approx&
\frac{\pi G \Sigma^2}{2} 
+ \Sigma (2 G \rhosd)^{1/2}\sigma_z
\\
&=& 10^4 k_B \Punit
\left(
\frac{\Sigma}{10 \Msun\pc^{-2}}  
\right)
\times
  \nonumber\\
 & &\hskip 0.1cm
\left[0.33\left( \frac{\Sigma}{10 \Msun\pc^{-2}}\right) 
+ 1.4 \left( \frac{\rhosd}{0.1 \Msun\pc^{-3}}\right)^{1/2}
\left(\frac{\sigma_z}{10 \kms}\right)
    \right] \nonumber.
\end{eqnarray}
The vertical dynamical equilibrium equation is
\begin{equation}\label{eq:dyneq}
P_{\rm th} + P_{\rm turb}=\rho_0 \vthdiff^2 + \rho_0 \vzdiff^2
=\rho_0 \sigma_z^2=\PtotDE.
\end{equation}

As noted above, it is expected that 
$P_{\rm th}=\rho_0 \vthdiff^2 \propto \SigSFR$ in a state 
of thermal equilibrium.  In addition,
OS11 argued that if mechanical feedback from star formation 
provides the dominant contribution to the vertical turbulent motions,
then the turbulent pressure $\Pturb$ 
should also scale roughly linearly with $\SigSFR$, as
\begin{equation}\label{eq:PturbSF}
\Pturb=f_p\frac{\Psn}{4\Msn}\SigSFR.
\end{equation}
Here, $\Psn$ is the mean radial momentum injected by each massive
star, $\Msn$ is the total mass in stars formed per massive star, and
the order-unity coefficient $f_p$ parameterizes the details of
turbulent momentum injection and dissipation.  When turbulence
dominates the pressure and self-gravity dominates the vertical weight,
equations (\ref{eq:PtotDEdiff}), 
(\ref{eq:dyneq}) and (\ref{eq:PturbSF}) with $f_p\approx
1$ combine to yield a prediction that $\SigSFR\approx 2 \pi G \Sigma^2
\Msn/\Psn$. OS11 found that this prediction is in good agreement with
both numerical simulations (for a cold-gas dominated ISM) and with
observations of molecule-dominated starburst regions with $\Sigma
\simgt 100 \Msun\pc^{-2}$.

More generally, if star formation is responsible for both heating and driving 
vertical motions in the diffuse ISM, we expect the thermal and turbulent 
pressure contributions to scale roughly linearly with $\SigSFR$.  Normalizing 
relative to convenient dimensional units for observational comparison, 
we can define
\begin{eqnarray}
\frac{P_{\rm th}/k_B}{10^3 \Punit}&\equiv & \eta_{\rm th}  
\frac{\SigSFR}{10^{-3} \sfrunit} 
\label{eq:etath}
\\
\frac{P_{\rm turb}/k_B}{10^3 \Punit}&\equiv& \eta_{\rm turb} 
\frac{\SigSFR}{10^{-3} \sfrunit}. 
\label{eq:etaturb}
\end{eqnarray}
The parameters $\etath$ and $\etaturb$ are yield coefficients that
measure the efficacy of feedback.
For the fiducial parameters adopted in OML10, $\eta_{\rm th}= 1.2
[0.25+0.75 Z_d'(\Sigma/10 \Msun \pc^{-2})^{0.4}]^{-1}$, where the
factor in square brackets is unity in the Solar neighborhood.  For the
fiducial value $\Psn/\Msn=3000 \kms$ adopted in OS11 (assuming
supernovae are the most important sources of momentum), $\eta_{\rm
  turb}= 3.6 f_p$.  Note that with the heating and turbulent driving
yield coefficients as defined in equations (\ref{eq:etath}) and
(\ref{eq:etaturb}), $\alpha = (P_{\rm th} + P_{\rm turb})/P_{\rm th} =
1 + \eta_{\rm turb}/\eta_{\rm th}$ if only thermal and turbulent
stresses contribute to the effective midplane pressure.  We thus
expect $\eta_{\rm th} + \eta_{\rm turb} \sim 1.2 + 3.6\sim 5$ and
$\alpha \sim 1+(3.6/1.2)=4$ under conditions similar to the Solar
neighborhood.  The latter is comparable to the value $\alpha =5$ adopted in
OML10 for comparisons of equation (\ref{eq:sfrboth}) with observations
of $\SigSFR$.  By
exploring the relations between the measured values of $\Pth$,
$\Pturb$, and $\SigSFR$ in our simulations, we can numerically
evaluate $\eta_{\rm th}$ and $\eta_{\rm turb}$, testing whether these
quantities (and therefore $\alpha$) are indeed near-constant.

Combining equations (\ref{eq:dyneq}), 
(\ref{eq:etath}) and (\ref{eq:etaturb}),
the self-regulated star formation rate in a diffuse-gas-dominated 
region where the pressure is 
controlled by energy and momentum feedback from massive stars has the form
\begin{equation}
\SigSFR= 2 \times 10^{-3} \sfrunit 
\left(\frac{\eta_{\rm th} + \eta_{\rm turb}}{5}\right)^{-1}
\frac{\PtotDE/k_B }{10^4 \Punit}.
\label{eq:SFRfb}
\end{equation}
For outer-disk regions, equation (\ref{eq:PtotDEdiff}) or
(\ref{eq:PtotDEap}) may be used for the ISM weight $\PtotDE$.  In
galactic-center regions where the bulge potential exceeds that of the
disk, $\rhosd \rightarrow \rho_b/3$ for $\rho_b$ the bulge stellar
density (see OS11).  

For very dust-poor systems, FUV radiation
escapes more easily from star-forming regions and penetrates further in the
diffuse ISM, which may make the heating yield $\eta_{\rm th}$
comparable to or even larger than $\eta_{\rm turb}$ (see OML10 and
\citealt{bol11}).  Alternatively, in regions where $\Sigma$ is
extremely high and reprocessed IR radiation is trapped, radiation
pressure becomes important and a term $\eta_{\rm rad} \propto \Sigma
\kappa_{\rm IR}$ would be included in equation (\ref{eq:SFRfb}).
Since the cosmic ray and magnetic pressures presumably increase with
higher $\SigSFR$ in analogy with equations (\ref{eq:etath}) and
(\ref{eq:etaturb}), corresponding feedback terms could be included in
equation (\ref{eq:SFRfb}), with the values of $\eta_{\rm CR}$ and $\eta_{\rm
mag}$ appropriately taking account of differing vertical scale heights
compared to the neutral, star-forming gas (see OS11).  

Using our present simulations, we can test whether the generalized
feedback-regulated star formation prediction $\SigSFR \propto \PtotDE$
is satisfied.  We will also compare our results to the power-law form
$\SigSFR \propto \Sigma^{1+p}$ traditionally used in fitting
observations, and to the form $\SigSFR \propto \Sigma \rho_0^{1/2}$
that is frequently adopted in numerical simulations of galaxy
formation/evolution in the cosmological context.

\section{Numerical Methods and Models}\label{sec:method}

\subsection{Basic Equations}\label{sec:eq}

The numerical models of this paper investigate thermal and dynamical
evolution of gas in a vertically stratified, differentially rotating,
self-gravitating galactic disk under the influence of interstellar
cooling, heating, and radiative and mechanical feedback from star
formation. We set up a local Cartesian frame whose center is located
at a galactocentric radius $R_0$ and rotates with an angular
velocity $\Omega=\Omega(R_0)$. In this local frame, $x\equiv R-R_0$,
$y\equiv R_0(\phi-\Omega t)$, and $z$ represent the radial, azimuthal,
and vertical coordinates, respectively. Our simulation domain is a
two-dimensional rectangular region with size $L_x \times L_z$ in the
$\xhat$ -- $\zhat$ plane with $y=0$ (hereafter XZ plane), representing
a radial-vertical slice of the disk, although we implicitly consider
the thickness $L_y (\ll L_x,L_z)$ in the $y$-direction for the
purposes of computing star formation rates and 
momentum feedback (see Section \ref{sec:mec}).
We include nonzero velocity in the $y$-direction in order to treat
epicyclic motions 
self-consistently. The equilibrium background velocity relative to the
center ($x=z=0$) of the simulation domain is given by $\vel_0=-q\Omega
x\yhat$, where $q\equiv -(d\ln\Omega/d\ln R)|_{R_0}$ is the local
dimensionless shear rate. In terms of $q$, the epicycle frequency
$\kappa$ is given by $\kappa^2=(4-2q)\Omega^2$.  We assume a flat rotation 
curve so that $q=1$ and $\kappa= \sqrt{2}\Omega$.

We expand the basic equations of hydrodynamics in the local frame, neglecting
terms arising from the curvilinear geometry. The resulting shearing-sheet
equations (e.g. \citealt{kos02,pio07}) are
\begin{equation}\label{eq:cont}
\pderiv{\rho}{t}+\divergence{\rho\vel}=0,
\end{equation}
\begin{equation}\label{eq:mom}
\pderiv{\vel}{t}+\advect{\vel}=
-\frac{1}{\rho}\nabla{P}-2\Om\times\vel
+2q\Omega^2 x \xhat -\nabla\Phi+\mathbf{g}_{\rm sd},
\end{equation}
\begin{equation}\label{eq:energy}
\pderiv{e}{t}+\divergence{e\vel}=-P\nabla\cdot{\vel}
-\rho\mathcal{L}+\mathcal{K}\nabla^2 T,
\end{equation}
\begin{equation}\label{eq:poisson}
\nabla^2\Phi=4\pi G\rho,
\end{equation}
where $\Phi$ is the self-gravitational potential of the gas, $\mathbf{g}_{\rm
sd}$ is the external gravity from the stellar disk and the dark matter halo,
$\rho\mathcal{L}$ is the net cooling function, and $\mathcal{K}$ is the thermal
conductivity.  Assuming that the gas is predominantly atomic and has cosmic
abundances, $P=1.1n\kbol T$ is the gas pressure where $n=\rho/(1.4 m_p)$ is the
number density of hydrogen nuclei.  We adopt an ideal gas law so that the
internal energy density is given by $e=P/(\gamma-1)$ with index $\gamma=5/3$.
For the external gravity, we take the simple form
\begin{equation}\label{eq:gsd}
\mathbf{g}_{\rm sd}=-4\pi G\rhosd z \zhat,
\end{equation}
where $\rhosd$ is the midplane density of the stellar disk plus that of the
dark matter halo. Since the scale height of the gas is much smaller than those
of the stellar disk and the dark matter halo, $\mathbf{g}_{\rm sd}$ given in
equation (\ref{eq:gsd}), corresponding to vertically-uniform $\rhosd$, is
a reasonable approximation in studying dynamics of the gas.

The net cooling function per volume is given by
$\rho\mathcal{L}\equiv n[n\Lambda(T)-\Gamma]$.
For the cooling rate of the diffuse ISM, we adopt the fitting formula
obtained by \citet{koy02}:
\begin{equation}\label{eq:cool}
 \Lambda(T)=2\times 10^{-19}\exp\left(\frac{-1.184\times10^5}{T+1000}\right)
 +2.8\times10^{-28}\sqrt{T}\exp\left(\frac{-92}{T}\right)
{\rm \;erg} \cm^3 \;{\rm s^{-1}},
\end{equation}
with temperature $T$ in degrees Kelvin. 
Cooling at low $T$ is dominated by the 158$\mu$m fine-structure line 
of \ion{C}{2}, whereas cooling at high $T$ is dominated by Ly$\alpha$
line emission; both lines are collisionally excited.
The heating rate $\Gamma$ is
dominated by the photoelectric effect on small dust grains and
polycyclic aromatic hydrocarbons (PAHs) by FUV photons with energy
$6\eV<h\nu<13.6\eV$ \citep{bak94}.  The diffuse FUV radiation field,
with intensity $J_{\rm FUV}$, is produced by young O and B stars and
therefore should depend on recent star formation.  We thus allow
$\Gamma$ to vary with time, while keeping $\Gamma$ uniform throughout
the simulation box (i.e. $J_{\rm FUV}$ is treated as spatially
constant). We follow \citet{koy02} in adopting a fiducial heating rate
in the Solar neighborhood 
$\Gamma_0=2\times10^{-26}\ergs$.  In
thermal equilibrium $(\rho\mathcal{L}=0)$ for this cooling function,
two stable phases co-exist for a range of densities and pressures: the
maximum pressure for the warm phase is
$\Pmax/\kbol=5.5\times10^3(\Gamma/\Gamma_0)\pcc\Kel$ occurring at
$T_{\rm max}=5000\Kel$ and 
$\nmax =1.0(\Gamma/\Gamma_0)\pcc$, and the minimum pressure for the
cold phase is $\Pmin/\kbol=1.8\times10^3(\Gamma/\Gamma_0)\pcc\Kel$ at
$T_{\rm min}=188\Kel$ and 
$\nmin=8.7 (\Gamma/\Gamma_0)\pcc$. The two-phase pressure is thus
given by $\Ptwo/\kbol\equiv(\Pmin\Pmax)^{1/2}/\kbol
=3.1\times10^3(\Gamma/\Gamma_0)\pcc\Kel$.  For Solar-neighborhood
conditions, $\Ptwo$ is essentially the same as adopted in OML10,
$\Ptwo/\kbol=3000\Punit$ (see equation \ref{eq:Ptwo}).  We describe
our prescription for connecting $\Gamma$ with the (time-dependent)
star formation rate, including metagalactic FUV radiation, in
\S~\ref{sec:rad}.

Thermal conduction plays an important role in the development of
thermal instability (TI). Conduction not only sets the critical
wavelength (the ``Field length'') of TI \citep{fie65}, but also
determines the thickness of interface layers between cold and warm
phases \citep{beg90}. Inclusion of thermal
conductivity is therefore essential to resolve TI in numerical
simulations \citep{koy04,pio04,kko08}.  A realistic value of thermal
conductivity in the diffuse ISM at $T<10^4\Kel$ is $\mathcal{K}\sim
2.5\times10^3 T^{1/2}\ergs\cm^{-1}\Kel^{-1}$ \citep{par53}. The
corresponding Field length is then $\lambda_F\sim 0.2\pc$ for the
typical density $n=1\cm^{-3}$ and temperature $T=10^3\Kel$ of the
thermally unstable gas, which would require an extremely fine
numerical grid $\Delta x \simlt \lambda_F/3$ in order for TI to be
resolved. In addition, hydrodynamic simulations involving supersonic
turbulence inherently suffer from a significant level of numerical
diffusion \citep[e.g.,][]{gaz05,kko08}, which is larger than the
physical conductivity unless $\Delta x$ is extremely small.  Adopting
a realistic value of $\cal K$ is therefore prohibitively expensive for
multi-dimensional simulations in kpc-scale numerical
boxes. Fortunately, however, dynamics on larger scales are not
sensitive to the exact conduction scale, similar to large-scale
dynamics in supersonic flows being insensitive to the exact thickness
of shocks. In this paper, we therefore adopt a numerical conductivity of
$\mathcal{K}=4\times10^7\ergs\cm^{-1}\Kel^{-1}/[1+(0.05\pcc/n)]$ as in
\citet{ko09a}, which enables us to resolve the Field length numerically,
and limits thermal conduction in low-density regions.

We solve the time-dependent partial differential equations
(\ref{eq:cont})-(\ref{eq:poisson}) using a modified version of the
{\it Athena} code \citep{sto08,sto09}. {\it Athena} employs a
single-step, directionally unsplit Godunov method for
(magneto)hydrodynamics in multispatial dimensions, providing several 
schemes for integration in time, spatial reconstruction, and solution
of the Riemann problem. We
use the van Leer algorithm \citep{sto09} for integration, with
piecewise linear reconstruction and the HLLC Riemann solver. We solve
the net cooling function based on implicit time integration using
Simpson's rule \citep[e.g.,][]{ko09a} with a limit for the maximum
temperature change of $50\%$.  We also use an explicit conduction
solver for isotropic thermal conduction, and revert to first order
flux updates if a negative density appears during the higher-order
update \citep{lem09}. The gravitational potential is calculated using
fast Fourier transforms in disk geometry with vacuum boundary
conditions in the $z$-direction \citep{ko09a}. At the $x$-boundaries,
we apply shearing-periodic boundary conditions \citep{hgb95}. In the
$z$-direction, we adopt periodic boundary conditions for the
hydrodynamic variables so as to maintain a constant mass within the
domain. By running comparison models using outflow boundary conditions
in $z$, we have checked that the boundary conditions do not affect the
simulation outcomes significantly.

\subsection{Prescription for Star Formation Feedback}\label{sec:feedback}

In our simulations, self-gravitational collapse and ensuing feedback
from star formation control both thermal and dynamical evolution of
the model ISM.  We consider both mechanical (momentum input) and
radiative (thermal energy input) feedback effects.  Mechanical
feedback drives turbulence that supports the disk in the vertical
direction, while radiative feedback affects the thermal pressure by
changing the heating rate. In this section, we detail our prescription
for star formation feedback.\footnote{Other recent numerical studies 
	of the ISM have used somewhat different prescriptions for radiative and
	mechanical feedback from those we adopt. For example, \citet{jou09} adopted
	$\Gamma\propto \Sigma_{\rm gas}^{0.4}$ together with type-II SN rates scaling
	as $\Sigma_{\rm SN} \propto \Sigma_{\rm gas}^{1.4}$; \citet{age09} included
	feedback from supernovae based on a volumetric star formation rate $\rho_{\rm
	SFR} \propto \rho_{\rm gas}^{1.5}$ but did not include diffuse UV heating;
	and \citet{tas11} adopted a photoelectric heating rate that declines
	exponentially outward, but did not include mechanical feedback from
	supernovae.  }

\subsubsection{Mechanical Feedback}\label{sec:mec}

Star formation in our models occurs only inside
clouds where the gas density is larger than a critical value. The
threshold density $\rhoth$ should be large enough for star formation
to occur only in self-gravitating regions.  In addition, these
self-gravitating regions should be resolved on the grid, i.e. the
Jeans wavelength $\lambda_J (\rhoth) = [\pi \cth^2/(G\rhoth)]^{1/2}$
should exceed the grid spacing $\Delta x$ (taken to be $1$ pc in our
models), where $\cth$ denotes the thermal speed at the
threshold temperature $\Tth$. Since the cooling time is very short,
dense clouds are generally in thermal equilibrium, and 
$\nth=\Gamma/\Lambda(\Tth)$. Equation (\ref{eq:cool}) then yields
\begin{equation}\label{eq:ljeans2}
\lambda_J \approx 1.4\Tth^{3/4}e^{-46/\Tth}
(\Gamma/\Gamma_0)^{-1/2} \pc,
\end{equation}
for $T\lesssim100\Kel$.  For a fixed $\lambda_J$, we obtain $\Tth$
(and hence $\nth$) as a function of $\Gamma/\Gamma_0$. A simple
power-law fit for $\lambda_J=2.7\pc$ gives $\nth \approx 500
(\Gamma/\Gamma_0)^{0.2} \pcc$, which we take as the threshold density
for star formation in our simulations.  Although slightly lower
threshold density would be needed to meet the Truelove criterion
$\lambda_J/\Delta x >4$ \citep{tru97,tru98} and limit
artificial fragmentation in collapsing clouds, our choice is
acceptable in the current context since our aim is not to follow cloud
collapse and fragmentation but instead to disperse self-gravitating 
clouds by turning on
star formation feedback, as explained below.

Not all clouds with $\rho\geq\rhoth$ immediately undergo gravitational collapse
and star formation, since the star formation efficiency and the computational
time step should be considered as well. Let us consider a star-forming region
with density $\rho\geq \rhoth$. Assuming that our simulation domain represents
a two-dimensional slab with thickness $L_y$ in the
y-direction, the mass in the cloud above the threshold 
is $M_{\rm cl}=L_y\int_{\rho\geq \rhoth}
\rho dxdz$. For the thickness of the slab, we take $L_y=2\rsh$, where $\rsh$ is
the initial radius of an SN shell explained below. This choice of $L_y$ is due
to the fact that the most significant feedback in the simulation domain comes
from SN events occurring within $2\rsh$ in the $y$-direction. The SFR expected
from the cloud is
\begin{equation}\label{eq:sfr}
\dot{M}_*=\eff\frac{M_{\rm cl}}{\tff(\rho)}
\end{equation}
where $\eff$ is the star formation efficiency per free-fall time,
$\tff(\rho)\equiv(3\pi/(32G\rho))^{1/2}$. We take $\eff=0.01$ as a fiducial
value consistent with theory and observations \citep{krum05,krum07}. The
probable number of massive stars to form within the cloud in a time interval
$\Delta t$ is then given by
\begin{equation}\label{eq:nsn}
\mathcal{N}_{\rm *}=\frac{\dot{M}_*}{\Msn}\Delta t,
\end{equation}
where $\Msn$ is the total mass of stars in all masses formed per massive star.
We define massive stars as those that undergo supernovae, and 
adopt $\Msn=100\Msun$ for all simulations consistent with the initial mass
function of \citet{kroupa01}.  For a given computational time step $\Delta t$,
$\mathcal{N}_{\rm *}$ calculated from equation (\ref{eq:nsn}) is typically
$\sim 10^{-4}-10^{-3}$ (as small as $\sim 10^{-6}$ immediately after SN
explosions due to small time step), much smaller than unity.
Therefore, 
in zones where $\rho\ge \rhoth$
we
generate a uniform random number $\tilde{\mathcal{N}}\in [0,1)$ at each time
step, and turn on feedback only provided
$\mathcal{N}_{\rm *}>\tilde{\mathcal{N}}$.

We implement mechanical feedback from star formation in a very simple
way, by injecting momentum in the form of an expanding spherical
velocity distribution to represent the radiative stage of a SN 
\citep[cf.,][]{she08}. As the initial radius of the shell in three
dimensions, we take $\rsh=10\pc$, corresponding to the SN shock radius
at the shell formation time \citep{cio88,koo04}.  We assume the center of
the sphere is at a location $y_{\rm off}$  
distributed randomly in the range $|y_{\rm off}| \leq \rsh$, so that 
the initial shell radius in the XZ plane (at $y=0$) is 
$\Rmax \equiv \rbrackets{\rsh^2-y_{\rm off}^2}^{1/2}$, varying 
between 0 and $\rsh$. We use a random number to choose the value of $y_{\rm
  off}$ for each feedback event.  When a feedback event occurs, 
we first redistribute mass, momentum, and  thermal
energy within a circular region of radius $\Rmax$ by taking spatial
averages. We then add to the momentum density in the
$x$- and $z$-directions according to 
\begin{equation}\label{eq:snmom}
\rho \vel_{\rm sh,2D} = \left\{\begin{array}{ll}
     \pmax\rbrackets{\frac{R}{\rsh^2}}\mathbf{R}, & R\leq \Rmax,\\
     0 , & R>\Rmax,
\end{array}\right.
\end{equation}
where $\mathbf{R}$ is the position vector with respect to the center
of the SN sphere in the XZ plane, 
and $\pmax$ is the momentum density at $R=\Rmax$. By
requiring the mean momentum input from equation (\ref{eq:snmom})
(averaged over $y_{\rm off}$) is equal to the outward momentum that a
three-dimensional shell would have, one obtains
$\pmax=15\Psn/(32\rsh^3)$, where $\Psn$ is the total radial momentum
in three dimensions.  In all simulations, we take
$\Psn=3\times10^5\Msun\kms$ corresponding to the late stages of a
single SN with energy $E_{\rm SN}=10^{51}{\rm \,erg}$
\citep[][]{cio88}. 
The velocity profile $v(R)\propto R^2$ is chosen to guarantee an
initially divergence-free velocity field at $R=0$.

We note a few caveats that should be kept in mind regarding our
simplified prescription for star formation feedback.  First, as our
main focus is on the diffuse gas component (which dominates by mass),
our treatment does not attempt to follow the evolution and destruction
of star-forming clouds in detail.  Thus, we do not introduce a time
delay prior to the momentum injection, or separately model effects of
expanding \ion{H}{2} regions or winds (the former was previously considered
in \citealt{ko09a}, which found that only relatively low levels of
turbulence were induced in the diffuse ISM).  In this first study, our
goal is primarily to incorporate turbulent driving in the diffuse ISM
at a realistic level for a given star formation rate, which is
accomplished by simply injecting momentum impulsively.  Future work
should improve this treatment, but experience with numerical models of
turbulent giant molecular clouds has shown that much astronomical
insight can be gained even when idealized treatments of turbulent
driving are adopted \citep{mac04,mo07}.

Second, although our feedback treatment aims to model turbulent
driving in the neutral warm/cold ISM that is induced by SNe, our
approach does not attempt to model the high-temperature interiors of
SN remnants themselves. 
Previous work has shown that it is difficult to model SN
explosions by injecting thermal energy in large scale simulations
because of \emph{overcooling}: radiative energy losses are too
rapid due to lack of spatial resolution \citep{kat92}.  At resolution
levels that are affordable, far too little thermal energy ends up being
converted to kinetic energy; instead it is radiated away. In order to
avoid overcooling, in some simulations radiative cooling is
artificially turned off until blast waves have developed 
\citep[e.g.,][]{tha01,age11},
or the initial sizes of regions where SN energy is injected 
are set such that the gas temperature  $T\sim 10^7\Kel$, where a dip is
present in the cooling function \citep{jou06}.  
For simulations such
as ours which include self-gravity, SN events occur within very dense
regions.  Since the cooling rate is proportional to the square of the
gas density, experiments we conducted with thermal energy injection
and a coronal-gas cooling function showed that 
the cooling time was
still unrealistically short at the resolution of our simulations, 
even if we adjusted the
gas temperature to the dip of cooling function.  Thus, although hot
gas created in SNe may be quite important in many ways (including
driving galactic winds), the present models focus just on the
warm/cold ISM and star formation, and leave the interesting issues of
the hot ISM for future work.

\subsubsection{Radiative Feedback}\label{sec:rad}

Since the photoelectric heating rate is proportional to the intensity
of the FUV radiation field, we simply take $\Gamma \propto J_{\rm
  FUV}$, with a proportionality constant depending on the heating
efficiency of small grains and PAHs (see e.g., \citealt{bak94}).
There are two sources of the FUV radiation field in outer disk:
$J_{\rm FUV,local}$, the FUV radiation emitted by recently-formed OB
stars locally in the disk, and $J_{\rm FUV,meta}$, the metagalactic
FUV radiation field. Radiation originating in the inner regions of the
galaxy could also reach the outer galaxy, but this contribution is
smaller than the local radiation unless the optical depth is very low.  

If FUV escapes into the diffuse ISM from star-forming regions at
the midplane at rate per unit area $\Sigma_{\rm FUV}$, then $J_{\rm
  FUV} = \Sigma_{\rm FUV} [1 -E_2(\tau_\perp/2)]/(4\pi \tau_\perp)$
for $\tau_\perp=\Sigma \kappa_{\rm FUV}$ the optical depth through the
diffuse neutral ISM, and $E_2$ the second exponential integral. As the
radiative transfer factor depends only logarithmically on
$1/\tau_\perp$ at low optical depth, for simplicity OML10 adopted
$J_{\rm FUV}\propto \Sigma_{\rm FUV} \propto \SigSFR$ for application
to galaxies with dust abundance not far from Solar and a moderate range
of diffuse-\ion{H}{1} surface densities.  In galaxies with very low
dust abundance, UV may escape much more easily from star forming regions,
and also travel further through the diffuse ISM.  This would lead to an
increase in both $\Sigma_{\rm FUV}/\SigSFR$ and $J_{\rm FUV} /
\Sigma_{\rm FUV}$ relative to the Milky Way, so that the ratio $J_{\rm
  FUV}/\SigSFR$ could be much higher than in the Solar neighborhood.
	\citet{bol11} found that the warm \ion{H}{1} and star formation content of the
SMC indeed appears to require a higher ratio of 
$J_{\rm FUV}/\SigSFR$ than in normal disks like the Milky Way.

In this work, we assume the heating rate due to local FUV scales
with the local star formation rate as 
$\Gamma/\Gamma_0=\frad\SigSFR/\SigSFRsun$, where
$\SigSFRsun=2.5\times10^{-3}\sfrunit$ is the SFR surface density in
the Solar neighborhood \citep{fuchs09} and  
$\Gamma_0=2\times10^{-26}\ergs$ \citep{koy02}.  The parameter $\frad$
thus implicitly includes the normalized heating 
efficiency of the FUV radiation,
allows for additional forms of heating such as X-rays 
(see \citealt{wol95,wol03}), and would vary depending on 
details of radiative transfer.  Note that
$\frad=4/[1+3(Z_d'\Sigma/10\Surf)^{0.4}]$ is adopted in OML10 based on
the fit in \citet{wol03}; this has $f_{\rm rad}=1$ in the Solar neighborhood.

The total volumetric heating rate is then
written as
\begin{equation}\label{eq:heat}
\Gamma=\Gamma_0\sbrackets{\frad\rbrackets{\frac{\SigSFR}{\SigSFRsun}}
+\rbrackets{\frac{J_{\rm FUV,meta}}{J_{\rm FUV,0}}}}.
\end{equation}
Note that the heating by the metagalactic FUV given by the second term
in equation (\ref{eq:heat}) provides a minimum heating rate when
$\SigSFR$ is extremely small.  We adopt $J_{\rm FUV,meta}=0.0024
J_{\rm FUV,0}$ \citep{ste02}, so that in practice $J_{\rm FUV,meta}$
is negligible in most cases. The cooling and heating rates we adopt
give geometric-mean two-phase pressure equal to 
\begin{equation}\label{eq:Ptwo_ours}
\Ptwo/\kbol=1.2\times10^{3}\Punit\frad\rbrackets{\frac{\SigSFR}{10^{-3}\sfrunit}}.
\end{equation}
Thus, comparing to equation (\ref{eq:etath}), if we were to find $\Pth
= \Ptwo$ for the mean midplane thermal pressure, it would imply
$\etath=1.2 f_{\rm rad}$ for the dimensionless heating-feedback 
yield coefficient.
As we shall show in Section \ref{sec:te}, $\Pth$ at the midplane is 
in fact between 
$+10\%$ and $-40\%$ of $\Ptwo$, so that $\etath$ remains very close 
to $1\times \frad$.

In order to change the heating rate self-consistently, we need to calculate the
recent SFR at each time step.  We do this by counting the 
number of the recent SN
events, so that the SFR surface density is calculated by
\begin{equation}\label{eq:sfrsurf}
\SigSFR=\frac{N_{\rm SN}\Msn}{L_x L_y t_{\rm bin}},
\end{equation}
where $t_{\rm bin}$ is the time bin over which the SFR is averaged,
and $N_{\rm SN}$ denotes the total number of SN events that occurred
during the time span ($t-t_{\rm bin}$, $t$). We note that $\SigSFR$
corresponds to a space and time average of $\dot{M}_*$ divided by the
surface area.  Since only recent star formation contributes to gas
heating via FUV radiation, if the simulations were in three dimensions
and optical depth effects were included, the
averages should be taken at least over $t_{\rm FUV}\times(\pi d^2)$ to
cover the whole domain of influence, where $t_{\rm FUV}\sim 10\Myr$ is the
FUV luminosity-weighted lifetime of OB stars \citep{par03} and $d\sim
200\pc/(\Sigma/10\Surf)$ is the effective in-plane distance for radiation to
travel.\footnote{ The effective in-plane distance for FUV radiation to travel
  is given by $d\sim 2H/(\Sigma\kappa_{\rm FUV})$ where $H$ is the
  scale height of the gas disk and $\kappa_{\rm FUV}\sim
  1-2\times10^{-21}\cm^2 ({\;\textrm{H atom}})^{-1} \sim
  0.1\pc^2\Msun^{-1}$ is the dust opacity in the FUV band.
  By taking $H\sim100\pc$, we have $d\sim200\pc/(\Sigma/10\Surf)$.  }
However, our simulation domain represents a radial-vertical 
slab with effective thickness $L_y=2\rsh=20\pc$ in the $y$-direction, 
with $L_y\ll d$.  Since the size of our domain in the $x$-direction
is large enough ($L_x\simgt d$), it is desirable to take a temporal bin 
at least 
$t_{\rm  bin}\sim t_{\rm FUV} (d/L_y)\sim 10 t_{\rm FUV}/(\Sigma/10\Surf)$
in order to limit stochasticity in the heating rate.
We thus set $t_{\rm bin}$ equal to a half of the orbital period
(see below for definition).  Since our set of model parameters is chosen to
maintain 
$\Omega \propto \Sigma$, this implies $t_{\rm bin}\propto \Sigma^{-1}$. 
With this choice, $t_{\rm bin} (\Sigma/10\Surf)\sim 100 \Myr 
\sim 10 t_{\rm FUV}$.

\subsection{Model Parameters}\label{sec:model}

Since the feedback parameters are all specified, we now turn to the disk
parameters. Our initial conditions for the gaseous disk consist of warm-phase
gas with uniform thermal speed $c_w=7\kms$.  The gravitational
susceptibility of the disk depends on three parameters: gas surface density
$\Sigma$, the angular velocity of galactic rotation $\Omega$, and the stellar
plus dark matter density at the midplane $\rhosd$. Both $\Sigma$ and $\Omega$
enter the Toomre stability parameter
\begin{equation}\label{eq:Qw}
\Qinit \equiv \frac{\kappa c_w}{\pi G \Sigma},
\end{equation}
while $\rhosd$ determines the degree of vertical disk compression induced 
by the stellar disk and dark matter halo.  It is
convenient to define
\begin{equation}\label{eq:s0}
s_0\equiv\frac{\pi G\Sigma^2}{2 c_w^2 \rhosd}
=0.28\rbrackets{\frac{\Sigma}{10\Msun\pc^{-2}}}^2
\rbrackets{\frac{c_w}{7\kms}}^{-2}
\rbrackets{\frac{\rhosd}{0.05\rhounit}}^{-1},
\end{equation}
which measures the relative strengths (in the vertical direction) 
of gas self-gravity and the 
external gravity from stars and dark matter \citep{kos02}.
For Solar-neighborhood conditions, $s_0 \approx 0.3$.
Assuming $s_0 \ll1$, the equilibrium density distribution is a Gaussian profile
\begin{equation}\label{eq:gauss}
\rho (z) = \rho_0 \exp (-z^2/ 2H_w^2),
\end{equation}
where $\rho_0=\Sigma/[(2\pi)^{1/2}H_w]$ and
\begin{equation}\label{eq:hw}
H_w=\frac{c_w}{(4\pi G\rhosd)^{1/2}}
=134\pc\rbrackets{\frac{c_w}{7\kms}}
\rbrackets{\frac{\rhosd}{0.05\rhounit}}^{-1/2},
\end{equation}
is the scale height.

To simulate disk evolution in a range of environments systematically,
we vary $\Sigma$ and $\rhosd$ while keeping $\Qinit=2$ fixed, so that
the angular velocity at the center of the domain varies as
$\Omega=28\kms\kpc^{-1}(\Sigma/10\Msun\pc^{-2})$.  We consider four
main series of models: QA, QB, S, and G. The model parameters are
summarized in Table~\ref{tbl:model}. In Series QA and QB, $\rhosd$
varies as $\rhosd\propto\Sigma^2$ so that the stellar Toomre parameter
$Q_s \propto \Omega /\sqrt{\rho_s} \propto Q \sqrt{s_0}$ 
implicitly has the same value for
all members of each series.  For the QA series, $s_0=0.28$ and for the
QB series $s_0=0.07$.  Thus, models in Series QB have four times
larger $\rhosd$ (i.e. a more confining stellar vertical potential)
than those with the same $\Sigma$ in Series QA. The model Series QA
and QB represent conditions typical in disk galaxies at different
galactocentric radii, from mid-disks (i.e.  slightly inside the Solar
circle) to far outer disks \citep[e.g.,][]{ko09a}.\footnote{
	Very far outer galaxies with negligible stellar disks and only dark matter
	contributing to $\rhosd\propto R^{-2}$ would have 
  $s_0=2(\pi G \Sigma)^2(c_w \Omega_{\rm dm})^{-2}$, which could reach
  unity, but these conditions are not studied in the current work.}
For Series S, we
fix $\rhosd$ and vary $\Sigma$ to explore the effect of the gas
surface density independent of the strength of the external vertical
gravity. In Series G, $\Sigma$ and $\Omega$ are held constant, while
$\rhosd$ varies; this allows us to isolate the effect of the external
vertical gravity. Our fiducial model is Model QA10 with
$\Sigma=10\Surf$, $\Omega=28\kms\kpc^{-1}$, and $\rhosd=0.05\rhounit$;
this model is similar to the Solar neighborhood.  The corresponding
orbital period is $\torb=2\pi/\Omega= 220\Myr
(\Omega/28\kms\kpc^{-1})^{-1}= 220\Myr (\Sigma/10 \Surf)^{-1}$, 
which we use as the time unit in our
presentation.

The model series above all have the same feedback parameters.  In
addition, we consider Series R, in which $\frad$ is varied to explore
the effect of varying heating for a given $\SigSFR$. 
All other parameters in
Series R are the same as Model QA10 (which has $\frad=1$).  We ran
four models labeled R02, R05, R25, and R50 with $\frad=0.25$, 0.5,
2.5, and 5.0, respectively.  Since the ratio of local heating rate to
local SFR surface density $\Gamma/\SigSFR \propto \frad$
(see equation \ref{eq:heat}),
larger $\frad$ implies a higher heating rate for a given $\SigSFR$,
corresponding to lower shielding (e.g. from lower dust abundance)
than in the Solar neighborhood.  Smaller $\frad$ corresponds to higher
shielding.  In reality, $\frad$ should depend on both dust abundance
and the total column of gas, since both of these can affect
shielding. For the present study, we simply treat $\frad$ as an
autonomous variable in order to explore effects of varying shielding
(or heating efficiency, which for present purposes is equivalent).

For the vertical extent of our simulation boxes, we take $L_z=4H_w$ 
(this varies depending on the model; see Table~\ref{tbl:model}).  In
the horizontal direction, we take $L_x=512\pc$ as the standard value.
In order to check the effect of the box size, we have run Model
QA10x2, which has the same parameters as Model QA10 except the
horizontal box size is extended to $L_x=1,024\pc$; this model
confirmed that overall evolution and statistical properties are indeed
similar. We vary the number of zones from model to model to make
the grid spacing $\Delta x =\Delta z= 1\pc$ for all the
models. In order to seed TI, isobaric perturbations consisting of a
Gaussian random field with flat power for $1\leq kL_z/2\pi\leq 8$ and
zero power for $kL_z/2\pi> 8$ are added to the initial density and
temperature distributions.  The amplitude of the initial perturbations is set
to $10\%$ of the midplane density. We evolve each model until $t/\torb=3$,
well beyond the time required for the system to reach a quasi-steady state.

\subsection{Classification of Gas Components}\label{sec:comp}

Before describing the simulation results, we establish terminology for
the various gas components we shall discuss. In the neutral ISM, gas
in GBCs and diffuse gas are distinguished based on whether the
gravitational energy and total pressure significantly exceed that of
the surrounding gas at similar $z$, or not.  In general, the GBC
component consists of the population of giant molecular clouds (GMCs),
including both molecular gas inside GMCs and dense atomic shielding
layers. Observations of the Milky Way \citep{sol87,hey09,rom10} and
Local Group galaxies \citep{bol08} have reported that GMCs have
similar surface densities $\Sigma_{\rm GMC}\sim 100\Msun\pc^{-2}$,
corresponding to $n_{\rm GMC}\sim40\pcc(M_{\rm
  GMC}/10^6\Msun)^{-1/2}$. Since we do not take into account radiative
transfer and formation of hydrogen and CO molecules explicitly, we
cannot directly identify structures in our models that would be
observed as GMCs.  In this work, we simply define gas with $n \geq
\ngbc=50\pcc$ as being within the GBC component, since observed GMCs
have comparable densities. We emphasize that this classification is
essentially a nomenclature shorthand, allowing us to refer to the
densest gas as the ``GBC component''.  The designation of gas as
``GBC'' or ``diffuse'' component is not used in any way within the
simulations themselves.  We note that the density threshold for star
formation 
(see section \ref{sec:mec}),
which is much larger than $\ngbc$, ensures that star
formation in our numerical models takes place only within the GBCs.

The diffuse component, defined as gas with $n<\ngbc$, consists of
thermally-stable cold and warm phases as well as a thermally-unstable
phase.  We classify the phases of the diffuse component based on its
density rather than temperature such that it is warm gas if
$n<\nmax$, cold gas if $n>\nmin$, and unstable gas if $\nmax<n<\nmin$
(see definitions of $\nmax$ and $\nmin$ following equation
\ref{eq:cool}).  Note that $\nmax$ and $\nmin$ depend on $\Gamma$ (and
hence $\SigSFR$) and thus vary with time. 
In what follows, $\fgbc$ and
$\fdiff$ denote the mass fractions of GBC and diffuse components in
the whole gas, respectively. Similarly, the mass fractions of cold,
unstable, and warm phases within the diffuse component are represented by
$\fc$, $\fu$, and $\fw$, respectively.  Note that $\fgbc+\fdiff=1$ and
$\fc+\fu+\fw=1$.

\section{Simulation Results}\label{sec:result}

In this section, we describe results of our numerical simulations.
Our models evolve in a generally similar manner to those of
\citet{ko09a}, which also included self-gravity, radiative heating (at
fixed $\Gamma$) and cooling, and feedback from star formation.  In the
models of \citet{ko09a}, only feedback associated with \ion{H}{2}
regions was considered. \ion{H}{2} regions were modeled by applying
intense heating in dense enough regions that met criteria for star
formation; expansion of the overpressured gas provided turbulent
driving.  Since SN explosions are more energetic than expanding
\ion{H}{2} regions, however, our present models achieve a higher
(more realistic) level of turbulence at saturation than those in
\citet{ko09a}.  Also, the variable radiative heating rate in the
present simulations enables us to explore self-regulation of thermal
pressures.

\subsection{Overall Evolution}\label{sec:evol}

\begin{figure}
\epsscale{0.8} \plotone{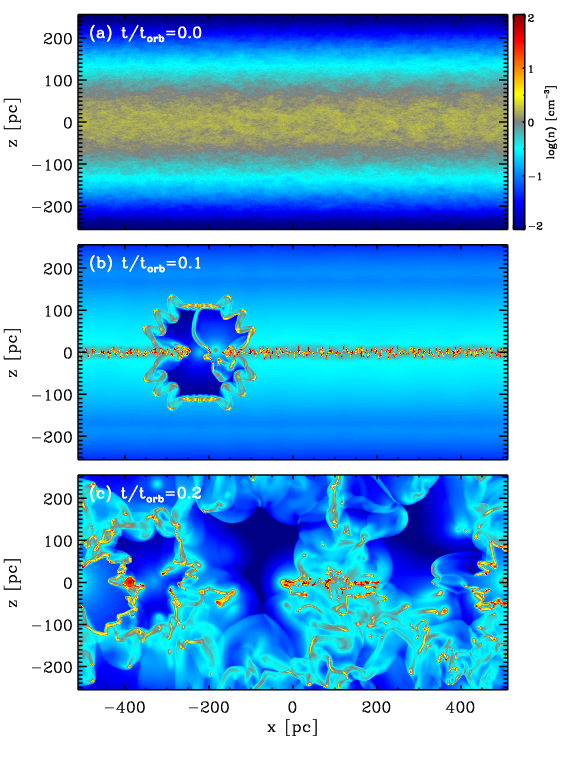} 
\caption{Density snapshots for Model QA10x2 (logarithmic color scale)
  at $t/\torb=0$, $0.1$, and $0.2$.  The initial single-temperature
  gas disk (\emph{a}) evolves rapidly via thermal instability into a
  configuration with midplane cold cloudlets sandwiched by outer
  layers of warm gas.  In (\emph{b}), the first SN explosions occur in
  dense clouds near $x=-200\pc$ produced by mergers and
  self-gravitating contraction of smaller clouds.  Subsequent SN
  explosions disperse the dense clouds and drive the disk into a
  turbulent state (\emph{c}), in which filamentary structures of cold
  gas are found at all heights.
\label{fig:early}}
\end{figure}

We begin by describing evolution of Model QA10x2, which has
$\Sigma=10\Surf$ and $\rhosd=0.05\rhounit$.  Figure~\ref{fig:early}
displays snapshots for Model QA10x2 at $t/\torb=0,$ 0.1, and 0.2 to
show early time evolution.  The initial gas disk has a Gaussian
density profile with scale height $H_w=134\pc$ and constant
temperature, shown in Figure~\ref{fig:early}(\emph{a}).  Since the
initial disk is out of thermal equilibrium, it rapidly evolves and
separates into two phases, with a cold dense layer near the disk
midplane sandwiched by diffuse warm gas at larger $|z|$. At the same
time, TI develops locally, creating numerous cloudlets in the midplane
dense layer. The cold midplane slab has a surface density of $\Sigma_c
= 7\Msun\pc^{-2}$ and a typical sound speed $c_c=1\kms$.  The cold
slab has Toomre stability parameter $Q_c\sim 0.3$ with Jeans length
$\lambda_{{\rm 2D},c} \equiv c_c^2/(G\Sigma_c)= 33\pc$, so that it is
quite gravitationally unstable.  The slab soon fragments
gravitationally to form many dense clouds, which grow in size and mass
by merger with their neighbors.  Massive clouds undergo runaway
collapse as self-gravity dominates the internal pressure, eventually
producing stars and SN explosions when the density
exceeds $\rhoth$. The first SN feedback event occurs at about
$t/\torb=0.1$.  Figure~\ref{fig:early}(\emph{b}) shows formation of
dense clouds and the first SN explosion from Model QA10x2.  Subsequent
SN events drive the gas disk into a turbulent state, as seen in
Figure~\ref{fig:early}(\emph{c}).

The kinetic energy associated with expanding shells disperses dense
clouds in the midplane, and causes the disk to puff up in the vertical
direction.  Successive stages of gravitational contraction and
feedback-induced expansion result in quasi-periodic oscillations of
the disk thickness.  Warm gas located ahead of the expanding shells is
swept up by shocks and collected into the shells. Pre-existing dense
gas becomes even denser from shock compression. Ensuing radiative
cooling in the postshock regions increases the shell density (e.g.,
\citealt{muf74,McCray75}).  High-density expanding shells disintegrate
due to a combination of dynamical processes, forming small dense
cloudlets that subsequently 
merge together to grow into new dense clouds.  These
newly formed dense clouds collapse internally and create additional
stars when their internal density exceeds the threshold value, leading
to further SN feedback events that repeatedly stir up and restructure the
surrounding medium.

Figure~\ref{fig:f_tevol} plots temporal evolution of the mass
fractions of the various gas components, the density-weighted vertical
scale height 
\begin{equation}\label{eq:Hdef}
H\equiv \left(\frac{\int \rho z^2 dx dz}{\int \rho dx dz}\right)^{1/2},
\end{equation}
and the SFR surface density.  The initial changes in the mass
fractions and the disk scale height shown in Figure~\ref{fig:f_tevol}
reflect early-time thermal response of the gas to the net cooling
function.  The formation of new dense clouds is quickest at the
compression phase of the disk oscillation, as evidenced by the
negative correlation between $\fgbc$ and $H$ shown in
Figure~\ref{fig:f_tevol}.  Within a few tenths of an orbit, the system
evolves into a quasi-steady state in the sense that $\SigSFR$, gas
fractions, and other statistical properties fluctuate but do not systematically 
change over time.

Notice that $H$ in Figure~\ref{fig:f_tevol} shows quasi-periodic
oscillations over the entire evolution of Model QA10x2, which also
produces temporal variations in other physical quantities. The
dominant timescale is roughly half of the natural vertical oscillation
period, $\sim 0.5 (\pi/G \rhosd)^{1/2}$. The mean
value and standard deviation of the disk scale height are
$\abrackets{H}=86$ pc and $\Delta H=12$ pc, respectively, where the
angle brackets $\langle\,\rangle$ denote a temporal average over
$2<t/\torb<3$.  
When the disk is compressed vertically, it produces
more dense clouds and hence more active star formation. The enhanced
radiative and mechanical feedback from star formation then increases
the thermal pressure and the velocity dispersion of the gas, causing
the disk to re-expand.  Disk expansion temporarily suppresses star
formation activity, which then reduces the total pressure and leads to
a decrease in the disk scale height.  At saturation, the mass fraction
of the diffuse component in model QA10x2 has a mean value
$\abrackets{\fdiff}=0.77$ and fluctuation amplitude of $\Delta
\fdiff\sim 0.06$. The cold, unstable, and warm phases amount to
fractions $\abrackets{\fc}=0.46$, $\abrackets{\fu}=0.22$, and
$\abrackets{\fw}=0.32$, respectively, of the diffuse gas mass. The SFR
surface density has a mean value $\abrackets{\SigSFR}=1.9\times
10^{-3}\sfrunit$ and standard deviation $\Delta\SigSFR=4.0\times
10^{-4}\sfrunit$. Note that $\Delta\SigSFR$ is small, since in
evaluating $\SigSFR$ we have already time-averaged SN events over
$t_{\rm bin}=0.5\torb$ (cf. Fig.~\ref{fig:f_tevol}).

\begin{figure}
\epsscale{1.0} \plotone{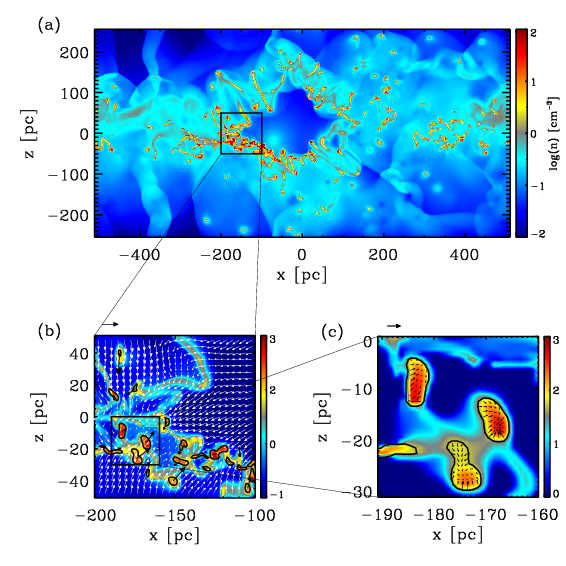} 
\caption{\label{fig:snap} (\emph{a})
  Density structure in the whole simulation domain of Model QA10x2 at
  $t/\torb=2.22$, including a large, fragmented, expanding shell
  produced by a recent SN event. (\emph{b}) The rectangular section in
  (\emph{a}) is enlarged to identify dense clouds ($n> 50\pcc$),
  outlined by black contours, that formed in a region of converging flow
  where the shell collides with surrounding gas. The white arrows
  represent the background velocity field, while the black arrows show
  the mean velocity of each dense cloud. (\emph{c}) The section marked
  in (\emph{b}) is further enlarged to show internal velocity
  structure of three selected dense clouds. The colorbars (whose range
  differs from panel to panel) indicate number density in logarithmic
  scale.  The sizes of the arrows outside the boxes in (\emph{b}) and
  (\emph{c}) correspond to $10\kms$ and $5\kms$, respectively.}
\end{figure}

Figure~\ref{fig:snap} displays the density structure (including newly
formed dense clouds) and velocity field around an expanding shell at
$t/\torb=2.22$, well after Model QA10x2 has reached a quasi-steady
state. The expanding shell, near the center of the simulation box in
Figure~\ref{fig:snap}(\emph{a}), was created by a SN event at
$t/\torb=2.18$. Figure~\ref{fig:snap}(\emph{b}), showing a zoomed-in section
of the shell, illustrates that dense (internal $n\sim 10^2 -10^3
\pcc$) clouds form in regions of converging velocity fields, indicated
as white arrows. The mean velocities of the dense clouds,
represented by black arrows, generally follow the 
background converging velocity
fields (with an additional random component), 
suggesting that cloud collisions will ensue. 
The rectangular section marked in
Figure~\ref{fig:snap}(\emph{b}) is enlarged in Figure~\ref{fig:snap}(\emph{c}) to
show the internal velocity fields of three selected massive dense
clouds. The internal one-dimensional velocity dispersion in each cloud is
$\sim 1\kms$, which is supersonic since the mean sound speed inside
the dense clouds is $\sim 0.5\kms$. The dense cloud near
$(x,z)=(-175,-25)\pc$ will have a star formation event 
at a time $\Delta t= 0.01\torb$ after this snapshot.

Figure~\ref{fig:pdf}(\emph{a}) shows the distribution of the
gas in the $n$--$P$ plane from Model QA10x2, averaged over
$t/\torb=2-3$. The colorbar labels the mass fraction in logarithmic
scale. While a large fraction of the gas remains close to thermal
equilibrium (given by the solid curve), a non-negligible portion
is out of thermal equilibrium ($\sim 18\%$ by mass departs from 
equilibrium by $|\Delta \log P|>0.15$), since the gas is continuously
disturbed by turbulent 
motions.\footnote{The thermal conductivity adopted is somewhat
  larger than the realistic value, and the numerical diffusion caused
  by large flow speeds also contributes, which may increase
  the unstable-mass fraction at the expense of the cold gas in our
  models (e.g., \citealt{kko08}).} 
The thermal equilibrium curve is for 
the time-averaged heating rate; fluctuations 
$\Delta \Gamma =0.16\Gamma_0$ relative to the
mean value $\langle\Gamma\rangle=0.76\Gamma_0$ 
displace the equilibrium 
curve upward and downward.  Variations in heating imply that gas can be 
out-of equilibrium with respect to the mean curve (even if instantaneous 
thermal equilibrium holds). Initially after a SN event, 
some cold gas is converted to the diffuse warm phase, while later 
shock compression and subsequent cooling during later stages
of the shell expansion convert some warm gas to the cold phase.

Figures~\ref{fig:pdf}(\emph{b,c}) plot the probability density functions
(PDFs) of thermal pressure and number density distributions shown in
Figure~\ref{fig:pdf}(\emph{a}), respectively. Thick and thin lines denote
the mass- and volume-weighted PDFs. 
The range of thermal pressure in our models spans more than three orders
of magnitude, although most of the mass is near the mode of the PDF.
The peak value of the pressure PDF corresponds to
the mean thermal pressure at the midplane. The volume-weighted
pressure PDF extends toward very small values mainly due to warm
gas at high altitude, while self-gravitating dense clouds near the
midplane occupy the high end of the mass-weighted pressure PDF. 
The mass-weighted density PDF shows the 
bimodal shape characteristic of the classical two-phase ISM
\citep[e.g.,][]{fie69,wol95,pio04}, although supersonic turbulent
motions  and frequent phase transitions increase the mass fraction in the
unstable phase, making the peaks less prominent
\citep[e.g.,][]{gaz05,gaz09,aud05,aud10,hen07,avi01,avi05,pio05,pio07,jou06,
jou09,ko09a}.

Evolution of other models in Series QA is qualitatively similar to
that of our standard model. One notable trend  
is that physical quantities exhibit larger-amplitude fluctuations 
with decreasing $\Sigma$. In low-$\Sigma$
models where SN events are rare and intermittent, even a single SN
explosion stirs up the whole simulation domain because there is not enough
mass to limit shell expansion.  This gives rise to large
variations in $H$, which in turn increases the dispersions of
$\Pth$ and $\SigSFR$, for lower-$\Sigma$
models. In models with high $\Sigma$, on the other hand, SN
events are frequent and spatially correlated. Shell expansion is 
frequently limited by surrounding dense gas and nearby SN shells.
Consequently, the temporal changes of the disk scale height in these
models are less dramatic than in low-density models. 

Compared to Series QA, models in Series QB have smaller $H$, as a
result of a more-confining vertical gravitational potential (four
times larger $\rhosd$). The resulting SFR is correspondingly larger in
Series QB compared to Series QA. Series S and G also reach
quasi-steady states, and their 
trends with
increasing/decreasing $\Sigma$ or $\rhosd$ follow the same patterns as
in Series QA and QB.  
In particular, independent increases in either $\Sigma$ or $\rhosd$ 
(with the other parameter controlled) produce an increase in $\SigSFR$.
The statistical properties of the models vary
depending on the input ``environmental'' parameters (i.e. $\Sigma$ and
$\rhosd$), as we shall describe and explore in the remainder of this
paper.

\subsection{Statistical Properties of the Gas }\label{sec:stat}

We have seen in Section \ref{sec:evol} that after a brief transient,
our models approach a quasi-steady state, which may be thought of as
an approximate thermal and dynamical equilibrium (with fluctuations
about the mean).  In this subsection, we present the time-averaged
values of the physical quantities that characterize the thermal and
dynamical properties of the gas.  These values will be used in Section
\ref{sec:test} to compare our numerical results with the analytic
predictions summarized in Section \ref{sec:theory}.

\begin{figure}
\epsscale{0.8} \plotone{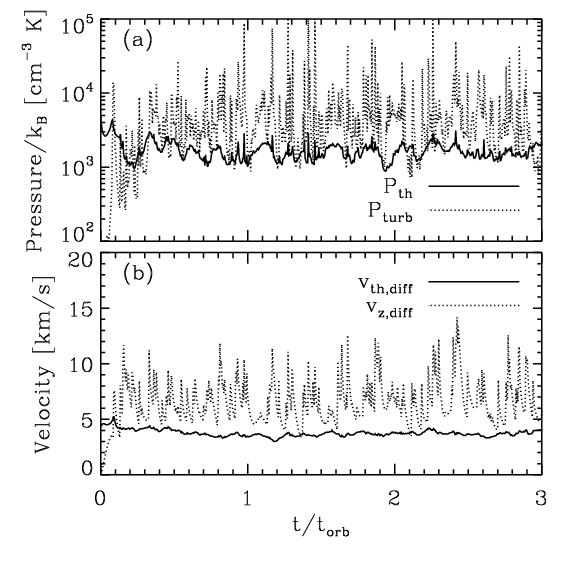} 
\caption{
Time evolution of the midplane thermal and turbulent pressures (\emph{a}) and
the thermal and turbulent velocity dispersions (\emph{b}) of the diffuse
component for model QA10x2. In (\emph{a}), $\Pth$ initially decreases as the
gas cools, while $\Pturb$ increases rapidly after the gas falls toward
midplane and is stirred up by SN explosions. 
After a few cloud formation and feedback cycles (a few 10s of Myr),
$\Pth$ and $\Pturb$ reach saturation values of 
$\abrackets{\Pth/\kbol}\sim 1,680\Punit$ and
$\abrackets{\Pturb/\kbol}\sim 5,440\Punit$, respectively, with a relative
fluctuation amplitudes of $0.21$ and $0.52$.  In (\emph{b}), the velocity
dispersions saturate at 
$\abrackets{\vthdiff}=3.7\kms$ and $\abrackets{\vzdiff}=6.8\kms$, respectively,
with relative fluctuation amplitudes of $5\%$ and $30\%$.
\label{fig:Pv_evol}}
\end{figure}

In our models, the total pressure at the midplane consists only of the
thermal and turbulent components since we do not include a magnetic
field.  We measure these two pressures directly from simulation data as
\begin{equation}\label{eq:Pth}
\Pth = \frac{\int_{z=-\Delta z/2}^{z=+\Delta z/2}\int P\Theta(n\!<\!\ngbc) dx
dz} {\int_{z=-\Delta z/2}^{z=+\Delta z/2}\int\Theta(n\!<\!\ngbc)dx dz}, 
\end{equation}
\begin{equation}\label{eq:Pturb}
\Pturb = \frac{\int_{z=-\Delta z/2}^{z=+\Delta z/2}\int \rho
v_z^2\Theta(n\!<\!\ngbc) dx dz} {\int_{z=-\Delta z/2}^{z=+\Delta
z/2}\int\Theta(n\!<\!\ngbc)dx dz}, 
\end{equation}
where $\Theta(X)$ is 1 if the logical argument `X' is true and 0
otherwise.  These definitions give volume-weighted averages of
pressure for the diffuse component (all gas at $n< \ngbc = 50 \pcc$)
at the midplane (the horizontal planes $z=\pm\Delta z/2$).
Figure~\ref{fig:Pv_evol}(\emph{a}) plots as solid and dotted lines the
midplane thermal and turbulent 
diffuse-gas
pressures, respectively, in Model
QA10x2 as functions of time. After a quasi-steady state is reached
($t/\torb>1$), the mean values are $\abrackets{\Pth/\kbol} =
1,680\Punit$ and 
$\abrackets{\Pturb/\kbol} = 5,440\Punit$, with
fluctuation amplitudes $\Delta{\Pth}/\abrackets{\Pth}=0.21$ and
$\Delta{\Pturb}/\abrackets{\Pturb}=0.52$. Since the
midplane includes high-velocity injection regions associated with SN,
there are large spikes in the midplane value of $\Pturb$.  
The overall fluctuations of $\Pth$ and $\Pturb$ follow the
pattern of variations in $H$ due to vertical oscillations, as shown in
Figure~\ref{fig:f_tevol}.

While we can measure midplane pressures in simulations, the most
direct observables are mass-weighted velocity dispersions.  We
calculate the mass-weighted vertical turbulent and thermal velocity
dispersions of the diffuse component using
\begin{equation}\label{eq:veld}
\vzdiff\equiv\sbrackets{\frac{\int \rho v_z^2\Theta(n\!<\!\ngbc) dxdz}
{\int\rho\Theta(n\!<\!\ngbc) dxdz}}^{1/2}, \quad
\vthdiff\equiv\sbrackets{\frac{\int P\Theta(n\!<\!\ngbc) dxdz}
{\int\rho\Theta(n\!<\!\ngbc) dxdz}}^{1/2}.
\end{equation}
The rms total velocity dispersion of the diffuse component in the vertical
direction is given by $\szdiff\equiv(\vzdiff^2+\vthdiff^2)^{1/2}$.
Figure~\ref{fig:Pv_evol}(\emph{b}) displays the time evolution of $\vthdiff$
and $\vzdiff$ in Model QA10x2 as solid and dotted lines,
respectively.  The vertical turbulent velocity dispersion saturates at
$\abrackets{\vzdiff}=6.8\kms$ with relative fluctuation amplitude
$\Delta\vzdiff/\abrackets{\vzdiff}=0.30$, while the thermal component
has a smaller mean value $\abrackets{\vthdiff}=3.7\kms$ and standard
deviation $\Delta\vthdiff=0.2\kms$.  Many spikes in $\vzdiff$ reflect
energy injection events associated with SN explosions.  
Since the shock-heated gas occupies a very small volume only near the
midplane, the thermal velocity dispersion $\vthdiff$ averaged over the
whole domain varies more smoothly than the volume-weighted mean $\Pth$ 
averaged only near the midplane.

Tables~\ref{tbl:stat1} and \ref{tbl:stat2} list the mean values and standard
deviations of several physical quantities characterizing the gas disk, for all
models. Here and hereafter, we omit angle brackets for convenience; all
the symbols represent time-averages over $t/\torb=2-3$, 
unless stated otherwise.
Column (1) labels each run as in Table~\ref{tbl:model}.  In
Table~\ref{tbl:stat1}, Column (2) gives the logarithm of $\SigSFR$ in
units of $\sfrunit$. Columns (3) and (4) give the logarithm of
$\Pth/\kbol$ and $\Pturb/\kbol$, respectively, in units of $\Punit$.  Column
(5) lists the midplane number density $\rhomid$ of hydrogen 
in units of $\pcc$ defined
in analogy with equation (\ref{eq:Pth}) but for $n$ rather than $P$ in 
the integral. Column (6) gives the
scale height of the diffuse gas $\Hdiff\equiv[\int\rho z^2\Theta(n\!<\!\ngbc)
dxdz/\int\rho\Theta(n\!<\!\ngbc) dxdz]^{1/2}$ in units of $\pc$.  

In Table~\ref{tbl:stat2}, Columns (2) and (3) give the turbulent and thermal
velocity dispersions of the diffuse component in units of $\kms$, while Column
(4) gives the mass-weighted vertical velocity dispersion for all the gas
$\sz\equiv[\int(\rho v_z^2+P)dxdz/\int\rho dxdz]^{1/2}$ in units of $\kms$.
Column (5) lists $\fdiff$, the fraction of mass in the diffuse component
(by definition, all gas at $n<n_{\rm GBC} = 50 \pcc$ is diffuse). 
In Columns
(6) and (7), we list $\alpha\equiv(\vthdiff^2+\vzdiff^2)/\vthdiff^2$ and
$\tilde{f}_w\equiv\vthdiff^2/c_w^2$, respectively; these parameters 
are necessary to test
the OML10 theory. Note that $\tilde{f}_w \approx f_w$ 
(the mass fraction of diffuse gas that is warm) since $\vthdiff^2 = f_w
c_w^2+ f_c c_c^2$ and the thermal speed $c_w$ of the warm medium 
is an order of magnitude larger than that of the cold medium $c_c$.
Also note that $\alpha$ in Table~\ref{tbl:stat2} (based on 
mass-weighted velocities or pressures averaged over the box) is close but
not identical to the ratio $\Ptot/\Pth$ at the midplane.
Finally, Column (8) gives the numerically-measured 
timescale to convert high-density gas 
into stars, $\tausf\equiv(1-\fdiff)\Sigma/\SigSFR$ in Gyr units; 
here $1-\fdiff=f_{\rm GBC}$ is simply defined as the mass fraction at 
$n>n_{\rm GBC}=50\pcc$.

Figure~\ref{fig:v_sfr} plots the mean values of turbulent and total 
velocity dispersions 
(\emph{a}) $\vzdiff$ (\emph{b}) $\szdiff$, 
and (\emph{c}) $\sz$ as functions of $\SigSFR$ for all models except Series R.
The mean values over the whole set of models shown in
Figure~\ref{fig:v_sfr} are $\vzdiff=6.8\pm0.6\kms$, $\szdiff=7.7\pm0.6\kms$,
and $\sz=7.0\pm0.4\kms$. 
It is clear that $\szdiff$ increases slightly as $\SigSFR$ increases, while
$\vzdiff\sim\sz\sim7\kms$ is more-or-less constant in all models 
(excluding Series R). 
The slight increase of $\szdiff$ with $\SigSFR$ is due to an increase of 
$\vthdiff$ with a higher proportion of warm gas at higher 
$\SigSFR$, although thermal speeds (averaged over both 
warm and cold gas) are lower than turbulent speeds for all models except 
R50 (see Table~\ref{tbl:stat2}).  

The nearly constant value of $\vzdiff$, over two orders of magnitude
in $\SigSFR$, owes to a balance between driving and dissipation for
the turbulent momentum (see Section \ref{sec:te} for a detailed
discussion). The total vertical velocity dispersion for the whole
gaseous medium is also nearly constant in all models, $\sz\sim7\kms$
(see Fig.~\ref{fig:v_sfr}\emph{c}). This is because the higher
proportion of warm gas in the diffuse medium (raising $\szdiff$ as
$\SigSFR$ increases) is counterbalanced by a lower proportion of the
gas being in the diffuse component (which has higher velocity
dispersion than the dense, dynamically- and thermally-cold GBC
component) at higher $\SigSFR$.  That is, with $\sz\approx\fdiff^{1/2}\szdiff$,
the larger $\szdiff$ is offset by smaller $\fdiff$, for models with 
higher $\SigSFR$.

We note that the values of velocity dispersions given in
Table~\ref{tbl:stat2} and plotted in Figure~\ref{fig:v_sfr} are
mass-weighted averages over the entire simulation volume rather than just
averages at the midplane (which would be $v_{\rm
  th,mid}=(\Pth/\rho_0)^{1/2}$ and $v_{\rm
  z,mid}=(\Pturb/\rho_0)^{1/2}$, where $\rho_0=1.4m_p\rhomid$).  We report 
volume-averaged values because these are the closest to direct observables.
However, the OML10 theory (and dynamical equilibrium considerations 
more generally) use midplane values of 
the pressure, which depend on midplane velocity dispersions.  We have found 
$\vthdiff/v_{\rm th,mid}\sim1.3$ and $\vzdiff/v_{\rm z,mid}\sim1.3$
for all models.  The reason for this difference is that the gas is 
somewhat differentially stratified, with cold phase preferentially 
concentrated near the midplane, which makes $v_{\rm th, mid}$ slightly 
smaller value than $\vthdiff$.  Also, since the gas density and the
turbulent dissipation rate increase near the midplane, 
$v_{\rm  z,mid}$ is slightly smaller than $\vzdiff$ averaged over the 
whole volume.

Figure~\ref{fig:afw_P} plots the mean values of (\emph{a}) $\alpha$,
(\emph{b}) $\tilde{f}_w$, and (\emph{c}) $f_w\fdiff$ as
functions of $\SigSFR$ for all models except Series R.  There is a
weak decreasing trend of $\alpha$ with $\SigSFR$, but overall $\alpha$
has a small range, $\sim 3-6$.  The small range of
$\alpha=(\vthdiff^2+\vzdiff^2)/\vthdiff^2$ implies that the ratio of
turbulent to thermal pressure 
$\Pturb/\Pth=\vzdiff^2/\vthdiff^2$ 
in the diffuse
gas is close to constant (for a given $\frad$) over a very large range
of $\SigSFR$. The parameter $\tilde{f}_w$ increases as $\SigSFR$
increases since a higher heating rate increases the warm-gas mass
fraction and $\tilde f_w=\vthdiff^2/c_w^2=f_w
+(1-f_w)c_c^2/c_w^2\approx f_w$.  Note that $\alpha = 1+
(\vzdiff^2/c_w^2)/\tilde f_w$, so that with $c_w\sim \vzdiff\sim
7\kms$ (see Fig.~\ref{fig:v_sfr}), the decline in $\alpha \sim 1+
1/\tilde f_w$ from $\sim 6$ to $\sim 3$ is just as expected when
$\tilde f_w$ increases from $\sim 0.2$ to $\sim 0.5$.  The mass
fraction of warm gas in the whole medium $f_w\fdiff$ is nearly
constant, implying the warmer diffuse gas at higher $\SigSFR$ is
offset by a higher fraction of the medium in a very dense component
(here defined as $n>\ngbc=50\pcc$).

For Series R (see Tables~\ref{tbl:stat1} and \ref{tbl:stat2}),
$\vthdiff$ and $\tilde f_w$ increase 
as $\frad$ increases (corresponding to increasing
heating at given $\SigSFR$).  On the
other hand, $\vzdiff$ decreases as $\frad$ increases, for the R
series. Combining these effects, $\alpha= 1+ \vzdiff^2/\vthdiff^2$
decreases by nearly an order of magnitude for increasing $\frad$ in
the R series.  At large $\frad$, $\vthdiff$ exceeds $\vzdiff$.  On the
other hand, $\szdiff$ and $\sz$ decrease only slightly as $\frad$
increases, while $\SigSFR$ decreases by a factor $\sim 2$.  Thus,
$\frad$ appears to affect primarily the energy distribution between
thermal and turbulent components that results from star formation
feedback, together with the proportions of cold and warm gas,
for the parameter regime we have explored.

\section{Test of the Thermal/Dynamical Equilibrium Model}\label{sec:test}

\subsection{Vertical Dynamical Equilibrium}\label{sec:de}

Having obtained the statistical properties of the multiphase,
turbulent gas from our time-dependent numerical simulations, we are
now in a position to examine the validity of the assumptions made in
OML10, and to compare our numerical results with the predictions of
the OML10 analytic theory. 

We first focus on the vertical force balance between (self plus external)
gravity and total (thermal plus turbulent) pressure.  
If dynamical equilibrium holds, the total midplane pressure $\Ptot$ 
should match the vertical weight of diffuse gas, $\PtotDE$.
Taking $\zeta_d=1/\pi$, we rewrite equation (\ref{eq:PtotDE}) in terms of
$\fdiff\equiv\Sigdiff/\Sigma$ and $\szdiff=c_w  (\tilde f_w
\alpha)^{1/2}$ as
\begin{eqnarray}\label{eq:PDE2}
\PtotDE&=&\fdiff\frac{\pi G\Sigma^2}{4}
\cbrackets{(2-\fdiff)+ \sbrackets{\rbrackets{2-\fdiff}^2+
\frac{32\szdiff^2\rhosd}{\pi^2 G\Sigma^2}}^{1/2}}\\
&=&1.7\times10^3\; \kbol \Punit\;\fdiff
\rbrackets{\frac{\Sigma}{10\Surf}}^2 \times\nonumber\\
&& \hskip -1.5cm  
\cbrackets{(2-\fdiff)+\sbrackets{\rbrackets{2-\fdiff}^2+
37\rbrackets{\frac{\szdiff}{7\kms}}^2\rbrackets{\frac{\rhosd}{0.1\rhounit}}
\rbrackets{\frac{\Sigma}{10\Surf}}^{-2}}^{1/2}},
\nonumber
\end{eqnarray}

Since the second term in the square
brackets of equation (\ref{eq:PDE2}) dominates for the range of
parameters we have explored (suitable for outer disks), 
an approximate form for equation (\ref{eq:PDE2}) is:
\begin{eqnarray}\label{eq:PDE3}
\PtotDE &\approx& \fdiff \szdiff\Sigma(2G\rhosd)^{1/2}
\approx \fdiff^{1/2} \sz \Sigma (2G\rhosd)^{1/2}\\
&\approx&1.0\times10^4\kbol\Punit\fdiff^{1/2}
\rbrackets{\frac{\sz}{7\kms}}
\rbrackets{\frac{\Sigma}{10\Surf}}
\rbrackets{\frac{\rhosd}{0.1\rhounit}}^{1/2},
\nonumber
\end{eqnarray}
where we take $\szdiff\approx \fdiff^{-1/2}\sz$ based on the fact
that the velocity dispersions of very dense gas are smaller than those
of the diffuse component. 
This is similar to the formula adopted by \citet{br04,br06}, except
that our expression includes the correction factor $\fdiff$ and 
allows for the dark matter contribution to $\rhosd$ (see also OML10).
Although the factor $\fdiff$ in equation (\ref{eq:PDE3})  
is close to unity in outer disks, this
correction would be quite important in inner-disk regions where gas is
dominated by gravitationally-bound GMCs.
For the current models, 
we note that
$\fdiff \szdiff/c_w= \alpha\fdiff\tilde f_w c_w/\szdiff \sim
\fdiff^{1/2} \sz/c_w \sim 1.0$ insensitive 
to model parameters since 
$\sz\sim \szdiff \sim c_w \sim 7-8 \kms$, $\alpha \sim 4-5$,
and $\fdiff \tilde f_w  \sim 0.2-0.3$ if  $\frad=1$.  
Thus, if dynamical equilibrium is satisfied, we expect the midplane
pressure to correlate well with $\Sigma \sqrt{\rhosd}$.

Figure~\ref{fig:Pcomp}(\emph{a}) plots the midplane total pressure of the
diffuse component $\Ptot\equiv\Pth+\Pturb$ 
measured from the simulations
(as listed in Table~\ref{tbl:stat1}) as a
function of $\Sigma \sqrt{\rhosd}$ for all models. 
The errorbars denote the standard deviations of the pressure fluctuations.
The dynamical-equilibrium prediction of equation (\ref{eq:PDE2}) (or
the approximation in equation \ref{eq:PDE3}) for $\PtotDE$
can be evaluated directly from the 
model inputs $\Sigma$ and $\rhosd$ in Table~\ref{tbl:model} and
simulation results for $\fdiff$, $\vzdiff$, and $\vthdiff$ listed in
Table~\ref{tbl:stat2}.
In the lower panel of
Figure~\ref{fig:Pcomp}(\emph{a}), we plot the relative difference
between the measured $\Ptot$ and $\PtotDE$ computed from equation
(\ref{eq:PDE2}).  These values agree with each other within
$13\%$.  
This close agreement verifies that effective hydrostatic
equilibrium is indeed satisfied. In addition, this suggests that the midplane
total pressure in a star-forming disk is set by environmental
parameters such as the gas surface density, external gravity, the level of the
turbulence, etc.  
Since the parameters appearing in equation (\ref{eq:PDE2}) can be
inferred relatively directly from observables for spatially-resolved
face-on galaxies (modulo uncertainties in the stellar-disk scale
height), the total midplane pressure in diffuse gas is an
empirically-accessible quantity.

Adopting the dependence on ``environmental'' parameters $\Sigma$ 
and $\rhosd$ following equation (\ref{eq:PDE3}),
the numerical results are well fitted by
\begin{equation}\label{eq:Ptotfit}
\Ptot = 9.9\times10^3\kbol\Punit
\rbrackets{\frac{\Sigma}{10\Surf}}
\rbrackets{\frac{\rhosd}{0.1\Msun\pc^{-3}}}^{1/2}.
\end{equation} 
This fit is overplotted as a dotted line in the upper panel of
Figure~\ref{fig:Pcomp}(\emph{a}).  Comparison of the fit 
to the numerical results (equation \ref{eq:Ptotfit}) 
with the analytic prediction (equation \ref{eq:PDE3}) shows
that averaging over our model suite, 
$\fdiff \szdiff \approx \fdiff^{1/2} \sz = 7.0 \kms$.

For accounting purposes, we have arbitrarily adopted the choice
$\ngbc=50\pcc$ as the minimum for the dense-gas GBC component. One
might be concerned that this may significantly affect the value
obtained for $\PtotDE$.  As seen in equation (\ref{eq:PDE3}), however,
$\PtotDE$ for the present models depends on $\ngbc$ just through
$\PtotDE\propto \fdiff^{1/2}$ because $\sz\sim7\kms$ is nearly
constant for all models.  We have checked that if we instead 
chose $\ngbc=100\pcc$, $\fdiff$ increases by about $10\%$, resulting 
in only about $3\%$ change in $\PtotDE$. Thus, for the 
diffuse-dominated regime studied in the present work, $\PtotDE$ does not depend
sensitively on the specific choice for $\ngbc$ as long as it is 
large enough.  In the regime where gravitationally-bound gas is more 
important, or where self-gravity is comparable to the external gravity,
the more exact expression in equation (\ref{eq:PtotDE}) (or equation 
\ref{eq:PDE2}) should be used for $\PtotDE$.

While an empirical measure of total midplane pressure can be obtained from
spatially-resolved observations of $\Sigma$, $\rhosd$, and $\szdiff$,
pressure-sensitive lines can be used to obtain empirical estimates of
$\Pth$ even from unresolved observations.  It is thus useful to consider
how $\Pth$ relates to environmental properties in our models.
Figure~\ref{fig:Pcomp}(\emph{b}) plots the midplane thermal pressure of
the diffuse component $\Pth$ (as listed in
Table~\ref{tbl:stat1}) as a function of $\Sigma \sqrt{\rhosd}$ for all
models except Series R. The lower panel shows the relative difference 
between $\Pth$
and $\PthDE=\PtotDE/\alpha$ as defined in equation (\ref{eq:PthDE}), 
or multiplying equation (\ref{eq:PDE2}) by $1/\alpha$.  
(Note that this differs slightly from the lower panel of 
Figure~\ref{fig:Pcomp}(\emph{a}) because our measured $\alpha$ is
based on volume-averaged rather than midplane pressures.) 
The errorbars denote 
the standard deviations of the pressure fluctuations.
The dynamical-equilibrium prediction $\PthDE$ agrees
with the measured $\Pth$ at the midplane 
within $17\%$, excluding Series R.  The dotted line
in the upper panel of Figure~\ref{fig:Pcomp}(\emph{b}) gives our best fit 
\begin{equation}\label{eq:Pthfit}
	\Pth= 2.2\times10^3 \kbol \Punit \rbrackets{\frac{\Sigma}{10\Surf}}
	\rbrackets{\frac{\rhosd}{0.1\Msun\pc^{-3}}}^{1/2}.
\end{equation}
Multiplying  equation (\ref{eq:PDE3}) by $1/\alpha$, 
the thermal pressure in outer-disk regions is approximately given by 
$\PthDE\approx(\fdiff/\alpha) \szdiff \Sigma(2G\rhosd)^{1/2}$.
Note that the connection 
between thermal pressure and the parameters $\Sigma$ and $\rhosd$ expressed by 
equation (\ref{eq:Pthfit}) 
results from vertical force balance and 
the fact that $\alpha$ and $\fdiff \szdiff \approx \fdiff^{1/2}\sz$
are nearly constant. 

As seen in Section \ref{sec:stat}, since the amount of energy injected
into the thermal component depends on $\frad$, $\Pth$ is proportional
to $\frad$ for Series R, resulting in significant changes of $\Pth$
for the same $\Sigma$ and $\rhosd$ (see Table~\ref{tbl:stat1}).
The relation $\Pth\sim\PthDE$ still approximately holds provided that
the inverse variation of $\alpha$ with $\frad$ is included for varying 
$\frad$ (see equation \ref{eq:alphasol}). 
Although $\sz$ and $\fdiff$ are insensitive to $\frad$, 
the large variation of $\alpha$ with $\frad$ implies that 
the results for Series R significantly depart from equation 
(\ref{eq:Pthfit}).  This is why Series R is omitted from 
Figure~\ref{fig:Pcomp}(\emph{b}).

\begin{figure}
\epsscale{1.1} \plottwo{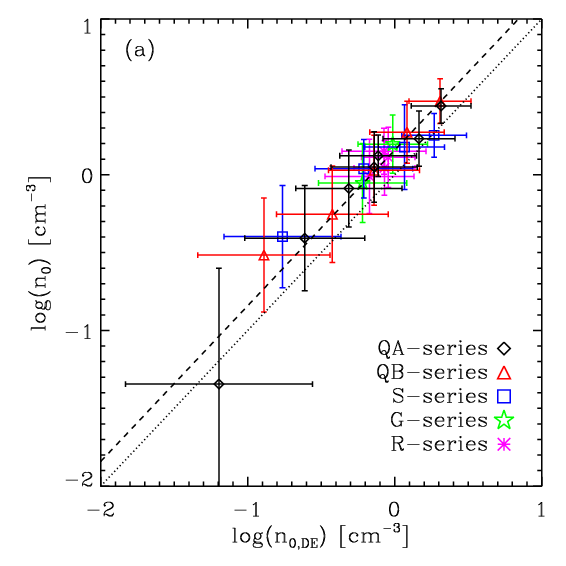}{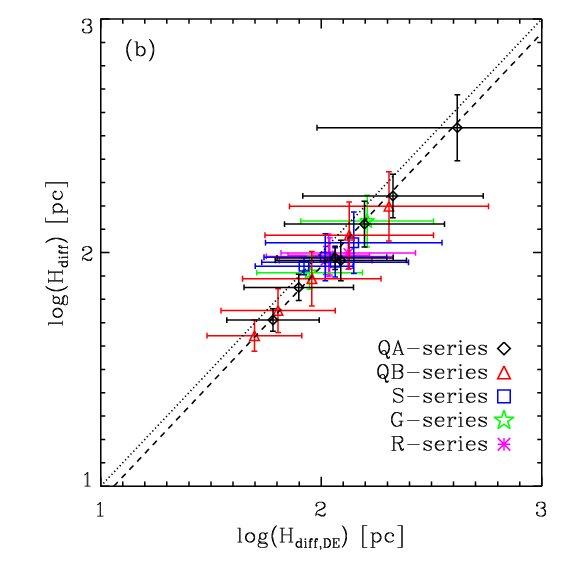} 
\caption{
Measured \emph{versus} estimated values of (\emph{a}) midplane number densities
and (\emph{b}) disk scale heights of the diffuse gas. The
points and errorbars give the mean and standard deviations over
$t/\torb=2-3$.  The estimated midplane number density
$\nDE\equiv\rhoDE/(1.4m_p)$ and scale height $\HDE$ are obtained from dynamical
equilibrium as equations (\ref{eq:rhoDE}) and (\ref{eq:HDE}),
respectively.  The dashed lines show our best fits
$\rhomid=1.4\nDE$ and $\Hdiff=0.87\HDE$ for imposed unity
slopes, while the dotted lines indicate one-to-one
correspondence.  \label{fig:scaleh}}
\end{figure}

It is also possible to estimate the midplane density and the scale height.
If dynamical equilibrium holds, the mean midplane mass density is given by 
$\rhoDE=\PtotDE/\szdiff^2=\PDE/\vthdiff^2$, 
or hydrogen number density
$\nDE=\rhoDE/(1.4 m_p)$ by
\begin{eqnarray}
\nDE &=&
0.20 \pcc \fdiff\rbrackets{\frac{\szdiff}{7\kms}}^{-2}
\rbrackets{\frac{\Sigma}{10\Surf}}^2 \times\nonumber \\
&&\hskip -1.5cm
\cbrackets{(2-\fdiff)+\sbrackets{\rbrackets{2-\fdiff}^2+
37\rbrackets{\frac{\szdiff}{7\kms}}^2\rbrackets{\frac{\rhosd}{0.1\rhounit}}
\rbrackets{\frac{\Sigma}{10\Surf}}^{-2}}^{1/2}}\label{eq:rhoDE}.
\end{eqnarray}
For a Gaussian distribution, the scale height in vertical dynamical 
equilibrium is 
\begin{eqnarray}
\HDE&=&\frac{\fdiff\Sigma}{(2\pi)^{1/2}\rhoDE} \nonumber \\
&=&580\pc\;\rbrackets{\frac{\szdiff}{7\kms}}^{2}
\rbrackets{\frac{\Sigma}{10\Surf}}^{-1} \times\nonumber \\
&&\hskip -1.5cm
\cbrackets{(2-\fdiff)+\sbrackets{\rbrackets{2-\fdiff}^2+
37\rbrackets{\frac{\szdiff}{7\kms}}^2\rbrackets{\frac{\rhosd}{0.1\rhounit}}
\rbrackets{\frac{\Sigma}{10\Surf}}^{-2}}^{1/2}}^{-1}\label{eq:HDE}.
\end{eqnarray}
Figure~\ref{fig:scaleh} plots the measured values of (\emph{a}) the midplane
number density $n_0$ and (\emph{b}) the scale height of
the diffuse gas $\Hdiff$ versus the corresponding 
dynamical-equilibrium estimate given in equation
(\ref{eq:rhoDE}) and (\ref{eq:HDE}), respectively.  Our best fits for imposed
unity slopes give $\rhomid/\nDE=1.4$ and $\Hdiff/\HDE=0.87$. 
These differences owe to small differences between the 
mass-weighted thermal velocity dispersion $\vthdiff$ and the slightly-lower
midplane value $v_{\rm th,mid}$, as discussed in Section ~\ref{sec:stat}.

\subsection{Thermal Equilibrium and Turbulent Balance}\label{sec:te}

As described in Section \ref{sec:theory}, OML10 hypothesized that the gas
disk evolves to a state in which both cold and warm phases can coexist
at the midplane, at the same thermal pressure, with heating balanced
by cooling.  For given heating rate, a range of pressures between
$\Pmin$ and $\Pmax$ permits both a cold and warm phase in thermal
equilibrium. For definiteness, OML10 assumed that the midplane thermal 
pressure $\Pth$ in the diffuse medium is comparable to the geometric-mean
pressure $\Ptwo=(\Pmin\Pmax)^{1/2}$. 

\begin{figure}
\epsscale{1.0} \plotone{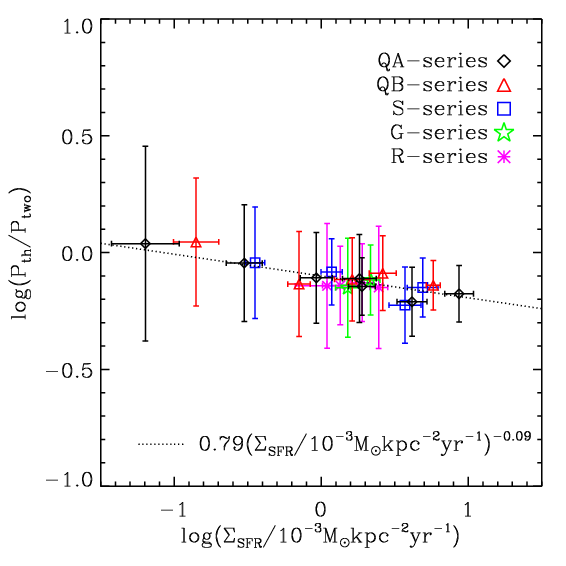} \caption{\label{fig:fth} Measured 
midplane thermal pressure $\Pth$ of the diffuse gas relative to the
two-phase thermal equilibrium pressure $\Ptwo$, as a function of 
$\SigSFR$. The points and errorbars give the mean
and standard deviations over $t/\torb=2-3$ for each model. The dotted
line, with a slope of $-0.09$, gives the best fit.
Heating/cooling and 
mass exchange between warm and cold atomic phases enables the mean
pressure to track the (radiation) energy input from star formation 
$\Pth \propto \Ptwo \propto \SigSFR$ 
over more than two orders of magnitude in $\SigSFR$.
}
\end{figure}

In our numerical models, the heating rate evolves with the SFR
according to equation (\ref{eq:heat}).  Assuming the $J_{\rm FUV,
  meta}$ contribution is negligible, the geometric-mean pressure is
given by equation (\ref{eq:Ptwo_ours}), corresponding to 
$\Ptwo/\kbol=3.1\times10^3\Punit (\frad\SigSFR/\SigSFRsun)$,
where the coefficient is slightly different from that in equation
(\ref{eq:Ptwo}) since the adopted cooling function in our simulations
is slightly different from that in \citet{wol03}.  For each model, the
mean value of $\SigSFR$ measured from the simulation sets the mean of
$\Ptwo$; the mean midplane thermal pressure is also measured (see Section
\ref{sec:de} and Table~\ref{tbl:stat1}). Using these measurements,
Figure~\ref{fig:fth} plots $\Pth/\Ptwo$ as a function of
$\SigSFR$ for all models. The dotted line is our best fit
\begin{equation}
\frac{\Pth}{\Ptwo}=0.79\rbrackets{\frac{\SigSFR}{10^{-3} \sfrunit}}^{-0.09}
\label{eq:fth}.
\end{equation}
The measured thermal pressure of the diffuse gas is thus smaller than the
geometric-mean pressure, but only slightly: $\Pth$ agrees with $\Ptwo$ within
$\sim40\%$ for all models, while 
$\Pth$ varies over more than two orders of magnitude 
for our whole suite of models
(see Table~\ref{tbl:stat1}). This proves that the assumption 
$\Pth\approx\Ptwo$ of the OML10 theory is a reasonable first approximation.

Using the numerical result given in equation (\ref{eq:fth}), we are
now in a position to evaluate the thermal yield from feedback  
$\etath$ defined in equation (\ref{eq:etath}).  We find
\begin{equation}\label{eq:etathnum}
\eta_{\rm th}= 0.99\frad\left(\frac{\SigSFR}{10^{-3}\sfrunit}\right)^{-0.09}.
\end{equation}
Our numerical calibration of 
$\eta_{\rm th}$ gives a value $\sim 30\%$ lower for the Solar neighborhood 
than the value $1.2 \frad$ adopted in OML10, 
and includes a weak decrease of $\eta_{\rm th}$ with increasing 
$\SigSFR$.  

The tendency for $\eta_{\rm th}$ to decrease with increasing $\SigSFR$
can be understood as follows.  Models with higher $\Sigma$ and
$\SigSFR$ have a larger diffuse-gas density, and hence shorter cooling
times, compared to models with lower $\Sigma$ and $\SigSFR$. In the
$n$--$P$ plane, a shorter cooling time implies that $\Pth$ will more
quickly drop towards $\Pmin$, such that $\Pth/\Ptwo$ will be slightly
lower for higher-$\Sigma$, higher-$\SigSFR$ models.  Models with lower
$\Sigma$ have longer cooling times, such that $\Pth$ does not drop as
quickly after heating events, and remains closer to $\Ptwo$.

Under the assumption that the dynamics of the gas disk has reached a
statistical steady state (as Figure~\ref{fig:Pv_evol} indicates), the
rates of turbulent driving and dissipation must balance each
other. For mean momentum $\Psn$ and mass $\Msn$ per supernova, the
rate of injection of vertical momentum per unit area per unit time to
each side of the disk is $\Pdriv\equiv0.25(\Psn/\Msn)\SigSFR$, 
assuming spherical blasts at the midplane (OS11).  If the
injected vertical momentum is preserved until the gas falls back to
the midplane, the vertical momentum flux across the disk $\Pturb$
would be equal to $2\Pdriv$.  If, however, the injected
vertical momentum is dissipated within a vertical crossing time, then
$\Pturb=\Pdriv$.  Finally, if the space-time distribution of star
formation sites is such that expanding shells collide with each other
in the vertical direction, then partial cancellation of injected
momentum would yield $\Pturb<\Pdriv$.

OS11 parameterized the uncertainties in dissipation and
driving by introducing a factor $f_p\equiv\Pturb/\Pdriv$. Here, we use
results of our numerical simulations to directly compare the measured
turbulent pressure with the vertical momentum injected by supernovae
in our models.  We characterize the return on mechanical feedback
from star formation using the turbulent yield parameter $\etaturb$
defined in equation (\ref{eq:etaturb}). The parameter $f_p$ is related to
$\etaturb$ by $\etaturb \equiv 3.6 f_p [(\Psn/\Msn)/3000 \kms]$.

Figure~\ref{fig:fp} plots our measurement of the ratio $\Pturb/\Pdriv$
for all models, as a function of $\SigSFR$.  The dotted line shows our
best fit omitting the R series,
\begin{eqnarray}
\frac{\Pturb}{\Pdriv}=0.97\rbrackets{\frac{\SigSFR}{10^{-3}\sfrunit}}^{-0.17}
\label{eq:fp}.
\end{eqnarray}
Our numerical calibration of the mechanical feedback yield is 
therefore
\begin{equation}
\etaturb=3.5\rbrackets{\frac{\SigSFR}{10^{-3}\sfrunit}}^{-0.17}
\label{eq:etaturbnum},
\end{equation}
where we use $\Psn/\Msn=3000\kms$ for all models.  The numerical
result in equation (\ref{eq:fp}) shows that $f_p\approx 1$ provides a
good overall estimate; this is also consistent with the results of
simulations presented in OS11 (for the molecule-dominated starburst
regime).  The numerical result that $f_p$ (and $\etaturb$) decrease 
weakly with increasing $\SigSFR$ suggests that vertical collisions 
of shells become more important at higher star formation rates, as 
would be expected.  On the other hand, disks with lower $\SigSFR$ suffer
somewhat less momentum dissipation because star formation sites are more 
isolated (in space and time), and shells expand into a more rarefied medium.

As discussed in OS11, the result $\Pturb \sim \Pdriv$ is equivalent to
having the dissipation time of turbulence comparable to the flow
crossing time over the largest energy-containing scale
\citep{sto98,mac98}, which here is the vertical disk thickness
$\Hdiff\sim\HDE$.  Feedback provides an input momentum per unit time per 
unit area of $\sim \Pdriv \sim \SigSFR \Psn/\Msn $.  For a dissipation time 
$\sim \Hdiff/\vzdiff$, the dissipation rate of vertical 
momentum in the diffuse ISM, per unit time per unit area
is $\sim \Sigma \vzdiff^2/\Hdiff \sim \rho \vzdiff^2 \sim \Pturb$.  Thus, driving 
is balanced by dissipation on a crossing time 
provided $\Pturb \sim \Pdriv$, as in equation (\ref{eq:fp}).

Combining equations (\ref{eq:etath}) and (\ref{eq:etaturb}), 
we have 
\begin{equation}\label{eq:PtotSFR}
\frac{\Ptot/\kbol}{10^3\Punit} \equiv \eta\frac{\SigSFR}{10^{-3}\sfrunit},
\end{equation}
where $\eta\equiv\eta_{\rm th} +\eta_{\rm turb}$ is the combined 
yield of thermal and mechanical feedback, with 
the respective contributions given 
in equations (\ref{eq:etathnum}) and (\ref{eq:etaturbnum}) from  
our numerical results.  Other sources of 
vertical support that are associated with star formation 
(e.g. radiation pressure, cosmic rays, and magnetic 
fields driven by turbulence) would contribute additional terms to $\eta$.
Since $\etath$ and $\etaturb$ decrease weakly with $\SigSFR$, the 
increase of $\Ptot$ with $\SigSFR$ is slightly sublinear.

Using equations (\ref{eq:etathnum}) and (\ref{eq:etaturbnum}),
we obtain an expression for the ratio between total and thermal
pressure in the diffuse gas:
\begin{eqnarray}\label{eq:alphasol} 
\alpha &=& 1+\frac{\eta_{\rm turb}}{\eta_{\rm th}}
\nonumber \\
        &=&
1+3.5 \frad^{-1}\left(\frac{\SigSFR}{10^{-3}\sfrunit}\right)^{-0.08}.  
\end{eqnarray}
This explains the very weak
decreasing trend of $\alpha$ with $\SigSFR$ for $\frad=1$ (see
Fig.~\ref{fig:afw_P}\emph{a}).  
In addition, this implies the value $\alpha \approx 5$ 
adopted by OML10 (based on empirical evidence) is in good agreement with 
the results of numerical simulations (for $\frad\sim 1$). Similarly, since 
$\tilde{f}_w=\vthdiff^2/c_w^2=\szdiff^2/(c_w^2\alpha)$,
$\tilde{f}_w \sim 
[1+3.5\frad^{-1}(\SigSFR/10^{-3}\sfrunit)^{-0.08}]^{-1}(\szdiff/c_w)^2$,
where $\szdiff/c_w = 1.1(\SigSFR/10^{-3}\sfrunit)^{0.04}$ 
for all models (see Fig.~\ref{fig:v_sfr}\emph{b}).  
This form is consistent with the
trend for $\tilde f_w$ to increase slightly with increasing $\SigSFR$, 
and to increase significantly with increasing $\frad$ 
(see Table~\ref{tbl:stat2}).

Finally, we note that although turbulent energy dominates over thermal
energy in equilibrium (unless $\frad$ is large), the radiative heating
{\it rate} exceeds the rate of heating from dissipation of turbulent
energy, except in far outer disks.  The energy input rate ratio is
$\etath/\etaturb$ times the ratio of the turbulent dissipation time
($\sim \Hdiff/\vzdiff$; see Section \ref{sec:sflaw}) to the cooling time
(assuming thermal equilibrium).  In the Solar neighborhood, the
cooling time is $\sim 1 \Myr$, whereas the turbulent dissipation time
is $\sim 20 \Myr$, implying a rate ratio $\sim 5$.  Moving outward in
the disk, the 
radiative-to-turbulent heating 
rate ratio decreases $\propto n \Hdiff/\vzdiff$, which is 
$\propto \Sigma$ for $\vzdiff \sim constant.$

\section{Star Formation Laws}\label{sec:sflaw}

In this section, we compare the SFRs obtained in our numerical
simulations to SFR formulae that are widely used in the literature,
both as fitting functions for empirical studies, and as prescriptions
for star formation in numerical models of galaxy formation/evolution.
We also introduce a new formula that relates $\SigSFR$ to the total
pressure in the diffuse ISM.  This relation follows the general form
expected when thermal and dynamical equilibrium are both satisfied,
and when both thermal and turbulent pressure are controlled by
feedback from star formation.

We begin with the orbital time prescription, expressed as $\SigSFR
\propto \Sigma\Omega$ \citep[][]{ken98}.  A relationship
of this kind is expected if the star formation timescale is
proportional to the orbital time, which would be true if star
formation is governed by large-scale gravitational instabilities and
the Toomre $Q$ parameter is near its critical value
(e.g. \citealt{qui72,wys89,sil97,elm97,kim01,kim07,mo07}).
Figure~\ref{fig:sflaw_Om} plots the mean values of $\SigSFR$ from our
numerical models as a function of $\Sigma\Omega$.  The dotted line is
the our best fit $\SigSFR=0.008\Sigma\Omega$ for an imposed unity
slope, while the dashed line denotes the empirical relation obtained
by \citet{ken98}, $\SigSFR=0.017\Sigma\Omega$.  
The RMS fractional deviation of the measurements compared to the fit is $43\%$.
In our simulations,
the sites of star formation are mainly small-scale dense clouds formed
by local thermal and gravitational instabilities, rather than very
massive clouds formed by large-scale instabilities.  Thus, orbital and
epicyclic motions do not directly control star formation in our
models.  Rather, the similarity between the behavior of $\SigSFR$ and
$ \Sigma \Omega$ in Figure~\ref{fig:sflaw_Om} reflects the 
correlation of input parameters chosen for our simulations: we set
$\Omega\propto\Sigma$ for all models, and since the specific 
star formation rate 
increases with $\Sigma$, it also increases with $\Omega$.

We next consider $\SigSFR$ as a function of $\Sigma$, as shown in
Figure~\ref{fig:sflaw}(\emph{a}). 
Also plotted as filled and empty contours are the recent 
pixel-by-pixel measurements of \citet{big08,big10} for 
$\SigSFR$  and $\Sigma$ in the regions inside and outside
the optical radius, respectively, 
of nearby spiral and dwarf galaxies.
Consistent with the observational results for $\Sigma\simlt 10\Surf$,
Figure~\ref{fig:sflaw}(\emph{a}) shows that there can be significant
variation in $\SigSFR$ at a given value of $\Sigma$. 
A single power-law fit 
to the numerical results 
gives
$\SigSFR=2.2\times10^{-3}\sfrunit(\Sigma/10\Surf)^{1.6}$ (not shown in
Figure~\ref{fig:sflaw}\emph{a}), with $33\%$ RMS fractional deviation.
Although the power law we find is similar to empirical results, our simulations
indicate that a single power-law Kennicutt-Schmidt relation $\Sigma
\propto \Sigma^{1+p}$ is not a good fit in outer-galaxy regions where
$\Sigma\simlt 10\Surf$ and diffuse atomic gas dominates.  Close
inspection of Figure~\ref{fig:sflaw}(\emph{a}) shows that
individually, the QA and QB series each follows a relation close to
$\SigSFR \propto \Sigma^2$, but these relations are vertically offset
from each other.  The reason the QA series has lower SFR than the QB
series is that the latter has four times larger $\rhosd$ at a given
value of $\Sigma$, and the reason both series approximately follow
$\SigSFR \propto \Sigma^2$ is that we have set $\rhosd \propto
\Sigma^2$ in both series, as we shall discuss below.

We remark that the current suite of models is not intended to match
the full parameter range of observed galaxies, but instead to explore
the fundamental physical dependence of star formation on environmental
conditions using carefully controlled numerical models.
Nevertheless, Series QA, which includes Solar neighborhood conditions
and extends to higher and lower $\Sigma$ assuming constant $Q$ and
$s_0$, follows the observed distribution of $\SigSFR$ vs. $\Sigma$
quite well. At very low gas surface density $\Sigma=2.5\Surf$, the results
from our models have higher $\SigSFR$ than much of the
observed distribution for far outer disks.  This is largely because
we chose low input values of $s_0$  to show the
effects of stellar gravity clearly in our controlled series of
models (lower $s_0$ corresponds to higher $\rhosd$ for a given
$\Sigma$ -- see equation \ref{eq:s0}). Realistic values of $s_0$ in far
outer disks are likely to be higher (see Section \ref{sec:model}).
Higher $s_0$ 
would reduce the vertical gravity and hence reduce $\SigSFR$ (following
the secular trend of decreasing $\SigSFR$ 
with increasing $s_0=0.02$ to $0.07$ to $0.28$ from Series S to QB to
QA at $\Sigma=2.5\Surf$).
In addition, Series QA, QB, and S fix $\frad=1$,
whereas $\frad$ is likely to increase in far outer disks because of
lower shielding where the dust abundance and $\Sigma$ are lower (see
Section \ref{sec:rad}).  The models of Series R show that $\SigSFR$
systematically decreases with increasing $\frad$ for fixed $\Sigma$
and $\rhosd$.  Thus, the difference between the present model
results and observations at low $\Sigma$ is simply due to differences between
model inputs and ambient conditions of gravity and shielding 
in outer galaxies.  This emphasizes once again 
that $\Sigma$ alone does not determine $\SigSFR$.

For typical parameters in outer disks, the weight associated with the
external (star+dark matter) gravity term $\propto \rhosd^{1/2}$
exceeds the weight associated with gaseous self-gravity in equation
(\ref{eq:PtotDEdiff}) (or \ref{eq:PtotDEap}) for the dynamical-equilibrium
diffuse-ISM pressure $\PtotDE$, which is equal to the 
diffuse-ISM weight.  Since the external-gravity dominates, we have
$\PtotDE\propto\Sigma\rhosd^{1/2}\sz$ 
as in equation (\ref{eq:PDE3})
(see also Figure~\ref{fig:Pcomp}\emph{a}), and 
$\PtotDE \propto \eta \SigSFR$ (equation \ref{eq:PtotSFR}) so that 
$\SigSFR\propto\Sigma\rhosd^{1/2}\sz/\eta$ for $\eta=\etath+\etaturb$.  
Since $\sz$ and the yield parameters  
$\etath$, $\etaturb$ are all close to constant
(see Fig.~\ref{fig:v_sfr} and equations \ref{eq:etathnum} and 
\ref{eq:etaturbnum}), we expect $\SigSFR \propto \Sigma\rhosd^{1/2}$.

Figure~\ref{fig:sflaw}(\emph{b}) plots results from the 
simulations for $\SigSFR$ vs. $\Sigma\rhosd^{1/2}$, showing
a much tighter relationship than $\SigSFR$ vs. $\Sigma$
in Figure~\ref{fig:sflaw}(\emph{a}).  Comparing measured values to the
fit in equation (\ref{eq:PthSFR2}) below, the RMS fractional deviation 
is $24\%$.  This is consistent with recent 
empirical findings that star formation is correlated with the 
stellar, not just the gaseous, content of galactic disks (see Section 
\ref{sec:intro}).

In both panels of Figure~\ref{fig:sflaw}, we overplot the simultaneous
solutions of equations (\ref{eq:tsf}), (\ref{eq:etath}), and 
(\ref{eq:PDE2}), adopting $\sz=7\kms$, $\alpha=5$, 
and $\tsf=1.3\Gyr$, along with the numerical fit for $\etath$ 
(equation~\ref{eq:etathnum} with $\frad=1$).  
If we instead adopt
$\etath=1$, the results are quite similar since $\etath$ is nearly
constant. 
The black dot-dashed curve takes
$s_0=0.28$ as in Series QA, the red dashed curve takes $s_0=0.07$ as
in Series QB, and the blue dotted ($s_0=0.02$) and green long-dashed
($s_0=1.1$) curves bracket the overall range of $s_0$ for our model suite (see
Table \ref{tbl:model}).  
The predicted curve for $s_0=0.28$ (as in Series QA)
follows the observations quite well within the optical radius.
As discussed above, larger values of $s_0$ and $\frad$ are likely 
present in far outer disks, which would produce a steeper 
$\SigSFR$ vs. $\Sigma$ relation moving to very low $\Sigma$ (outside
typical optical radii). 
The agreement between numerical models and the simultaneous solution
of equations (\ref{eq:tsf}), (\ref{eq:etath}), and 
(\ref{eq:PDE2}) confirms the analytic thermal/dynamical equilibrium theory
for star formation developed in OML10. In that work, comparison
to individual galaxies shows excellent agreement when both $\Sigma$
and $\rhosd$ in the theory are set from the observations.

In panel Figure~\ref{fig:sflaw}(\emph{b}), the black solid line denotes 
the power-law solution obtained by combining 
equations (\ref{eq:etath}), (\ref{eq:Pthfit}), 
and (\ref{eq:etathnum}) (with $\frad=1$)
to obtain a prediction for $\SigSFR$:
\begin{equation}\label{eq:PthSFR2}
\SigSFR=2.4\times10^{-3}\sfrunit
\rbrackets{\frac{\Sigma}{10\Surf}}^{1.1}
\rbrackets{\frac{\rhosd}{0.1\rhounit}}^{0.55}.
\end{equation}
We note that for outer disk regions, the focus of the present models,
the approximation $\fdiff \approx 1$ is satisfied, such that the single 
equation (\ref{eq:PDE3}) takes the place of the simultaneous solution of 
equations (\ref{eq:tsf}) and (\ref{eq:PDE2}). That is, the prediction 
for outer-disk star formation is independent of $\tsf$.  
If, rather than using the numerical fit (\ref{eq:etathnum}) for $\etath$, 
we had instead simply adopted a constant value of 
$\etath\approx 1$, then we would obtain a very similar form to 
equation (\ref{eq:PthSFR2}), except the exponent of $\rhosd$ would be 0.5, the 
exponent of $\Sigma$ would be 1, and the coefficient in front would be 
$2.2  \times10^{-3} \etath^{-1}\sfrunit$.
Small differences between the numerical results and the analytic prediction
for outer disk regions
are due to the fact that some of the idealizations of equation (\ref{eq:PDE3})
are not fully satisfied in the numerical models.  For example, 
the S series models
at $\Sigma=15$ and $20\Surf$ have non-negligible gravity from the gas, which 
increases $\PthDE$ above the estimate in equation (\ref{eq:PDE3}), and results
in $\SigSFR$ exceeding the estimate of (\ref{eq:PthSFR2}), which 
neglects the vertical gas gravity.  Also, we note that the R series, because
it has $\frad \ne 1$, is not expected to agree with equation 
(\ref{eq:PthSFR2}). In fact, members of the R series lie both above and 
below the prediction, consistent with expectations.

A prescription for star formation commonly used in numerical 
simulations of galaxy formation and evolution within a cosmological 
context is to make the star formation timescale  proportional to the 
self-gravitation or free-fall time of the gas, $\propto \rho^{-1/2}$.
In the context of disks, it is natural to adopt the
mean midplane density $\rho_0$ as a reference value, so that the SFR surface
density would be given by
\begin{equation}\label{eq:sfrtff}
\SigSFR\equiv\eff(\rho_0)\frac{\Sigma}{\tffmid},
\end{equation}
where $\tffmid=[3\pi/(32G\rho_0)]^{1/2}$ is the free-fall time at the
midplane and $\eff(\rho_0)$ is a star formation efficiency per
free-fall time at the mean midplane density.  
Figure~\ref{fig:sflaw_ff}(\emph{a})
plots $\SigSFR$ from the numerical simulations as a function of
$\Sigma/\tffmid$. The dotted line shows our best fit
$\SigSFR=0.008(\Sigma/\tffmid)$ for an imposed unity slope. Note that
$\eff(\rho_0)=0.008$ is similar to (but slightly smaller than) the value
$\eff(\rhoth)=0.01$ imposed at high density
($\nth\sim500 \pcc$) 
for star formation to occur in the numerical models.  
The free-fall time prescription gives a tighter
relation than $\SigSFR$ vs. $\Sigma\Omega$ or $\SigSFR$ vs. $\Sigma$,
but there is still scatter.

Although the free-fall time is commonly adopted as the controlling
dynamical timescale, in many circumstances self-gravity is less
important in confining and condensing gas than the gravity of the
stars and dark matter.  For a given total velocity dispersion
$\szdiff$, the vertical dynamical time is related to the disk
thickness by $\tdyn \equiv \Hdiff /\szdiff$.  Since $\Hdiff\equiv
\Sigdiff/(\sqrt{2\pi}\rho_0) = \szdiff /(4 \pi G \rhosd)^{1/2}$ if
external gravity dominates, or $\Hdiff = \szdiff/(\pi^2 G
\rho_0)^{1/2}$ if gas self-gravity dominates, $\tdyn \sim 0.3/(G
\rho_{\rm mid})^{1/2}$ with $\rho_{\rm mid}= \rho_0 + \rhosd$ includes
both limits.  If self-gravity dominates, $\tffmid =1.7 \tdyn$, but if
$\rho_0\ll \rhosd$, $\tdyn \ll \tffmid$, and the ``external''
gravity sets $\tdyn$ and $\Hdiff$.

For a disk with significant turbulent contribution to the total
velocity dispersion $\szdiff$, $\tdyn$ is
comparable to the vertical crossing time $\tver\equiv\Hdiff/\vzdiff$.
The vertical crossing time is the timescale for turbulence to be
dissipated, reducing the disk thickness and raising $\rho_0$.
For a low filling-factor cloudy medium,
small, cold, dense clouds can also ``fall'' to the midplane due to the
combined vertical gravitational force of stars, dark matter, and gas.
When they reach the midplane, these small, dense clouds collide
dissipatively, collecting into high-mass clouds that are internally
gravitationally unstable and make stars.  
The vertical crossing time
$\tver$ is thus expected to control how 
rapidly the diffuse cold component collects into self-gravitating clouds 
and initiates star formation.

Figure~\ref{fig:sflaw_ff}(\emph{b}) plots $\SigSFR$
from the numerical simulations as a function of $\Sigma/\tver$.  The
dotted line indicates our best fit $\SigSFR=0.0025(\Sigma/\tver)$
for an imposed unity slope.  The coefficient of this fit
denotes the star formation efficiency per vertical dynamical time
$\ever=0.0025$. The measured SFR surface density is well described by
the vertical dynamical time prescription, although 
there is still scatter (but  slightly less than in 
Figure~\ref{fig:sflaw_ff}\emph{a}). 
The RMS fractional deviations of measured $\SigSFR$ compared to the estimated
$\SigSFR$ are $26\%$ and $21\%$ for the free-fall time and the
vertical dynamical time prescriptions, respectively.

The good correlations shown in Figure~\ref{fig:sflaw_ff} for both the
$\tffmid$ and $\tver$ prescriptions are presumably because both
implicitly have similar scaling to $\SigSFR\propto\Sigma\sqrt{\rhosd}$
(shown in Figure~\ref{fig:sflaw}\emph{b}).  Since
$\rho_0\sim\rhoDE\propto \Sigma/\Hdiff \propto \Sigma\rhosd^{1/2}$
(when external gravity dominates), $\Sigma/\tffmid$ is basically
proportional to $\Sigma^{3/2}\rhosd^{1/4}$. For Series QA and QB,
$\Sigma^{1/2}\rhosd^{1/4}\propto\rhosd^{1/2}$ because we take
$\rhosd\propto\Sigma^2$ for these models.  Thus, $\Sigma/\tffmid
\propto \Sigma\sqrt{\rhosd}$ for Series QA and QB.  Although Series
S and G have somewhat different input parameter dependence, the
parameter coverage of these model series is not extensive enough to
reveal a clear difference between $\SigSFR \propto \Sigma
\sqrt{\rhosd}$ and $\SigSFR \propto \Sigma/\tffmid$.  For regions
dominated by external gravity, we have $\tver\approx(4\pi
G\rhosd)^{-1/2} \szdiff/\vzdiff$, so that
$\Sigma/\tver\propto\Sigma\rhosd^{1/2}$ since $\szdiff \sim \vzdiff$
for our models (and for the real ISM).  

We note that the vertical dynamical time prescription for star
formation is closely connected to the regulation of turbulent pressure
by feedback from star formation, and to the relationship between input
momentum and mean velocity dispersion in the disk (OS11).  As
shown in Section \ref{sec:te}, a balance between turbulent momentum
driving and dissipation is achieved in our models.  If
$\SigSFR=\ever\Sigma/\tver$, the momentum driving rate per unit mass
becomes
$2\Pdriv/\Sigma=0.5\ever\Psn/(\Msn\tver)$.  Equating this with the expected
turbulence dissipation rate $\sim 0.5 \vzdiff^2/\Hdiff =0.5 \vzdiff
/\tver$, we obtain $\vzdiff \sim \ever\Psn/\Msn$.
Using our adopted value $\Psn/\Msn=3,000\kms$ and the efficiency
$\ever=0.0025$ measured from our numerical models, this yields
$\vzdiff = 7.5 \kms$, remarkably similar to the mean value
$\vzdiff=6.8\kms$ obtained from our numerical simulations.

As argued in Section \ref{sec:theory}, energy and momentum feedback
from star formation are often the dominant sources of heating and
turbulence driving, in which case both $\Pth$ and $\Pturb$ (and
therefore $\Ptot$) 
in the diffuse ISM 
are predicted to vary approximately $\propto
\SigSFR$ (see equations \ref{eq:etath} and \ref{eq:etaturb}).  As we
show in Section \ref{sec:te}, our simulations indeed evidence
near-linear relations.  In Figure~\ref{fig:sfr_P}, we plot the
measured $\SigSFR$ as a function of (\emph{a}) the measured
$(\Pth/\kbol)/\frad$ and (\emph{c}) the measured $\Ptot/\kbol$, for
all of our numerical models.  All quantities are time-averaged.  Note
that the thermal pressure is divided by $\frad$ to compensate for the
effect of the varying assumed heating efficiency ($\Gamma/\SigSFR$).  
The dotted lines in
panels (\emph{a}) and (\emph{c}) are obtained from equations
(\ref{eq:etath}) and (\ref{eq:PtotSFR}), respectively, with numerical
calibrations (\ref{eq:etathnum}) and (\ref{eq:etaturbnum}) for the
feedback yields $\etath$ and $\etaturb$.  The dashed line in
panel (\emph{c}) plots our best fit omitting the R series:
\begin{equation}\label{eq:numSFR}
\SigSFR=2.6\times10^{-3}\sfrunit\rbrackets{\frac{\Ptot/\kbol}{10^4\Punit}}^{1.18}.
\end{equation}
The power slightly steeper than unity reflects the weak decline of 
feedback yields $\etath$ and $\etaturb$ with $\SigSFR$, as 
discussed in Section \ref{sec:te} (cf. equation \ref{eq:PtotSFR}) . 
Comparing
equation (\ref{eq:numSFR}) with equation (\ref{eq:SFRfb}), we see that
our numerical results yield 
$\eta = 3.9 [\Ptot/(10^4 \kbol \Punit)]^{-0.18}$ (for $\frad=1$), 
quite close to the estimate $\eta \sim 5$ obtained by
combining the theory of OML10 and OS11 (see Section \ref{sec:theory}).

In addition to heating/cooling and turbulent driving/dissipation
balance, vertical dynamical equilibrium is expected to apply, so that
the total diffuse-gas pressure at the midplane is equal to the
vertical weight $\Ptot=\PtotDE$.  Thus, a hallmark of self-regulated
star formation, when thermal, turbulent, and dynamical equilibrium are
all satisfied, is that a relation close to $\SigSFR \propto \PtotDE$
is expected to apply (see equation \ref{eq:SFRfb}).  To the extent
that $\alpha \sim const.$, we also expect $\SigSFR \propto
\PthDE/\frad=\PtotDE/(\alpha \frad)$. 
In Figure~\ref{fig:sfr_P} we plot the measured
$\SigSFR$ from numerical simulations as a function of (\emph{b})
$(\PDE/\kbol)/\frad$, and (\emph{d}) $\PtotDE/\kbol$, for all models.
The dynamical-equilibrium pressures are computed from input parameters
$\Sigma$ and $\rhosd$ using equation (\ref{eq:PDE2}) and mean measured 
values of $\fdiff$, $\szdiff$, and $\alpha$ for each model.\footnote{
  If we compute $\PtotDE$ from equation (\ref{eq:PDE2}) using 
  constant values $\fdiff=0.78$ and $\szdiff=7.7 \kms$ (the mean values 
  over the model suite), the best fit to $\SigSFR$ vs. $\PtotDE$
  analogous to equation (\ref{eq:numSFR}) would have a coefficient 
	$2.2 \times 10^{-3}$ and a power 1.05.}
Dotted and dashed lines are as for Figures~\ref{fig:sfr_P}(\emph{a,c}).

Figures~\ref{fig:sfr_P}(\emph{c,d}) show that $\SigSFR$ is extremely well
correlated with $\Ptot$ and $\PtotDE$.  
The RMS fractional deviations of the numerical results from the
relation given in equation (\ref{eq:numSFR}) are only 14\% and 16\%
for $\Ptot$ and $\PtotDE$, respectively.
The correlation is worse if the R series is included.  This 
is because $\etath \propto \frad$, so that higher $\frad$ reduces $\SigSFR$
compared to other models with the same midplane pressure.

Based on the results of our numerical simulations, we conclude that
star formation rates should be most closely correlated with the total
midplane pressure of the diffuse gas, as in equation
(\ref{eq:numSFR}).\footnote{
  Although our current numerical models have only explored the diffuse-dominated
  case, we still expect thermal, turbulent, and vertical dynamical
  equilibrium to hold in the volume-filling diffuse gas even if it is not
  the dominant component of the ISM by mass (see OML10, OS11, and
  Section \ref{sec:theory}).  In this case, equations 
  (\ref{eq:PtotSFR}) and (\ref{eq:SFRfb}) 
  are still expected to hold with near-constant yield 
  coefficients $\eta$, so that $\Ptot$ or $\PtotDE$ would still vary
  nearly linearly with $\SigSFR$.  It is important to note, however,
  that in the GBC-dominated case, this is best interpreted as $\SigSFR$
  setting $\fdiff$ (by equating [\ref{eq:PDE2}] and [\ref{eq:PtotSFR}] with 
  $\SigSFR \approx \Sigma/\tsf$)
  rather than the diffuse-ISM weight setting $\SigSFR$ (see OML10).  
  If GBCs dominate the mass, $\SigSFR$ is controlled by the density (and 
  pressure) within the bound clouds.}
The relation between diffuse-gas pressure and star
formation rate has less scatter than the relation between $\SigSFR$
and the gas surface density $\Sigma$ alone, or the combination $\Sigma
\Omega$.  The relation between $\SigSFR$ and $\Ptot$ is also more
general than $\SigSFR \propto \Sigma \sqrt{\rhosd}$ (which applies
when external gravity exceeds gas self-gravity and $\szdiff \sim
const.$), or $\SigSFR \propto \Sigma/\tver$ (which applies for
turbulence-dominated disks with $\ever \sim const.$).  In regions
dominated by diffuse gas, it is fundamentally the {\it weight of the
  ISM} that regulates star formation rates, since star formation rates
must adjust until the pressure driven by feedback matches this weight.
For outer disks that are diffuse-dominated ($\fdiff \sim 1$), the
weight (or $\PtotDE$, given by equation \ref{eq:PtotDEdiff} or by
the approximation in equation \ref{eq:PtotDEap}) depends only on
$\Sigma$, $\rhosd$, and $\sz$.  As noted above, an increase in $\frad$ 
(which would be associated
with low dust abundance) leads to a decrease in $\SigSFR \propto \PtotDE/\eta$,
 because $\etath \propto \frad$.

\section{Summary and Discussion}\label{sec:snd}

In this paper, we have used time-dependent numerical simulations to
investigate the regulation of star formation, as well as the thermal
and turbulent properties of the gas, in the regime where diffuse
atomic gas dominates the multiphase ISM.  For the Milky Way and
similar galaxies, this corresponds to the outer disk -- i.e. roughly
the Solar circle and beyond.  Physical effects included in our
numerical models (see Section \ref{sec:method}) include differential
rotation, Coriolis forces, 
gaseous self-gravity, vertical gravity due to the stellar
disk and dark matter halo, interstellar cooling and heating, thermal
conduction, and feedback from recent star formation in the form of
radiative and mechanical energy.  Although this initial set of models
involves a number of simplifications (e.g. we consider only a local box
representing thin radial-vertical slices 
so that very large-scale gravitational instabilities are absent;
we do not include galactic magnetic fields and spiral arms; 
we omit hot gas and treat feedback from SNe
via localized momentum injection; we do not explicitly treat radiative
transfer), it captures a very important aspect of real ISM disks that
is missing in many numerical studies of galactic star
formation. Namely, the vertical thickness of the disk, and therefore
the mean gas density, is primarily controlled by
(time-variable) turbulence.  The turbulent vertical velocity
dispersion depends on competition between driving by energy inputs 
from star formation, and dissipation through shocks and the
mode-coupling turbulent cascade.

\begin{figure}
\epsscale{0.8} \plotone{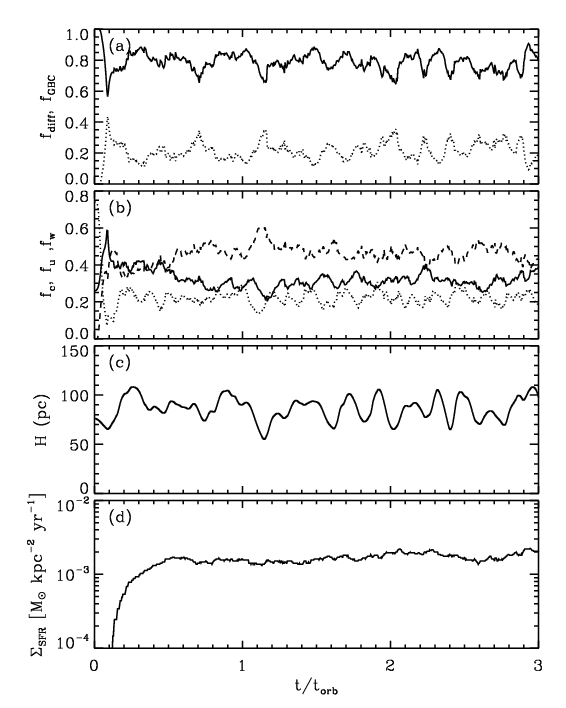} \caption{\label{fig:f_tevol} Time
  evolution in Model QA10x2 of (\emph{a}) the mass fractions of the
  diffuse ($\fdiff$, solid) and GBC ($\fgbc$, dotted) components,
  (\emph{b}) the mass fractions of the cold ($\fc$, dashed), unstable
  ($\fu$, dotted), and warm ($\fw$, solid) phases within the diffuse
  component, (\emph{c}) the density-weighted vertical scale height
  $H$, and (\emph{d}) the SFR surface density $\SigSFR$.  The initial
  increase of $\fgbc$ and $\fc$ stops at 
$t=0.1 \torb = 22 \Myr$ 
when the first
  SN event occurs inside a massive dense cloud. The model reaches a
  quasi-steady state after a few tenths of an orbital time, in the
  sense that the physical quantities fluctuate but do not evolve
  secularly. Note that $\fdiff$ is positively correlated with $H$.}
\end{figure}

To explore the dependence of $\SigSFR$ on environmental parameters, we
run models with varying total gas surface density $\Sigma$ and
midplane density $\rhosd$ of stars plus dark matter. The angular
velocity $\Omega$ is set such the Toomre stability parameter $Q_{\rm init}=2$ for
a velocity dispersion of $7\kms$.  
Our models are highly dynamic, but each reaches a statistical steady state 
within a few tens of Myr.  
In this quasi-steady state,
the star formation rate, disk scale
height, mass fractions of various gas phases, turbulent velocity
dispersion, and other physical properties fluctuate about well-defined
mean values (Fig.~\ref{fig:f_tevol}).  Low-amplitude 
quasi-periodic oscillations of the disk thickness are 
correlated with episodes of bound cloud formation (at maximum
compression) and feedback-driven expansion.  
Small cold clouds repeatedly fall to the midplane and collect (due to 
self-gravity)
into more massive
clouds, which are then dispersed by feedback from star formation.
We use the measured mean properties to test the theory
of star formation and diffuse-ISM regulation developed in OML10 and
OS11, as outlined in Section \ref{sec:theory}.

\begin{figure}
\epsscale{0.8} \plotone{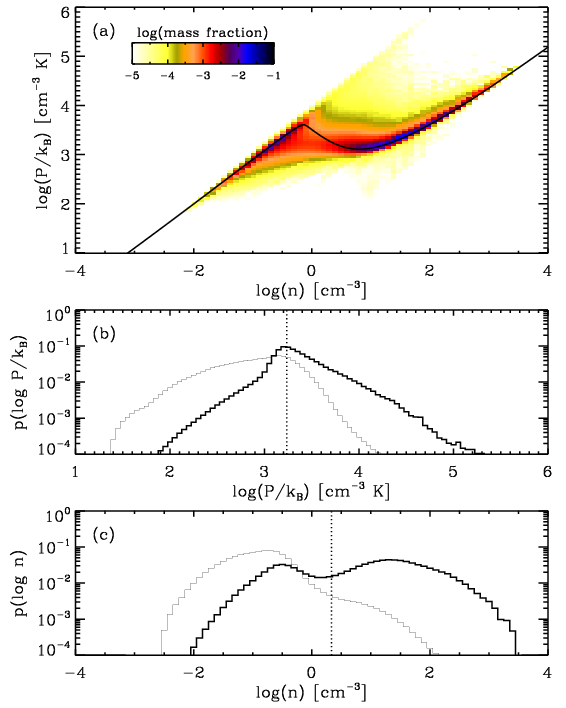} 
\caption{\label{fig:pdf}
(\emph{a}) Distribution of gas in the $n$-$P/\kbol$ plane for Model
QA10x2, averaged over $t/\torb=2-3$. The colorbar gives the mass
fraction in logarithmic scale. The solid curve marks the locus of thermal
equilibrium at the mean heating rate of 
$\abrackets{\Gamma}=0.76\Gamma_0$. Mass-weighted (\emph{thick}) and
volume-weighted (\emph{thin}) probability distribution functions are 
shown for (\emph{b}) thermal pressure and (\emph{c}) number density.
The vertical dotted lines in (\emph{b}) and
(\emph{c}) mark the mean midplane thermal pressure 
and number density, respectively, of the diffuse gas.
These results show that the system evolves to a state in which 
approximate two-phase thermal equilibrium at a common pressure 
holds for the atomic gas.
}
\end{figure}

\begin{figure}
\epsscale{1.0} 
\plottwo{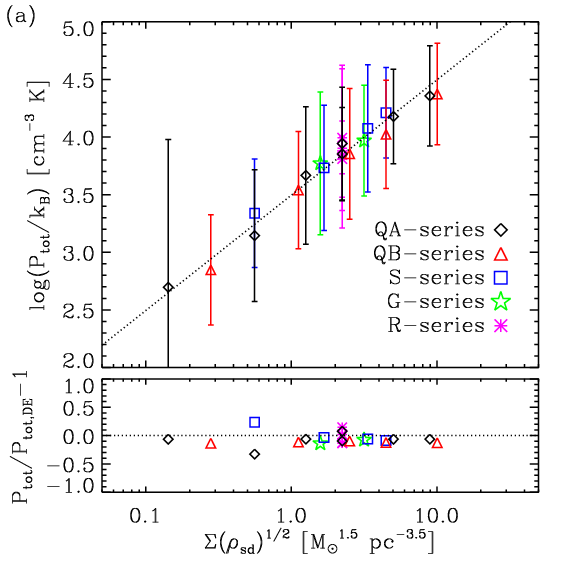}{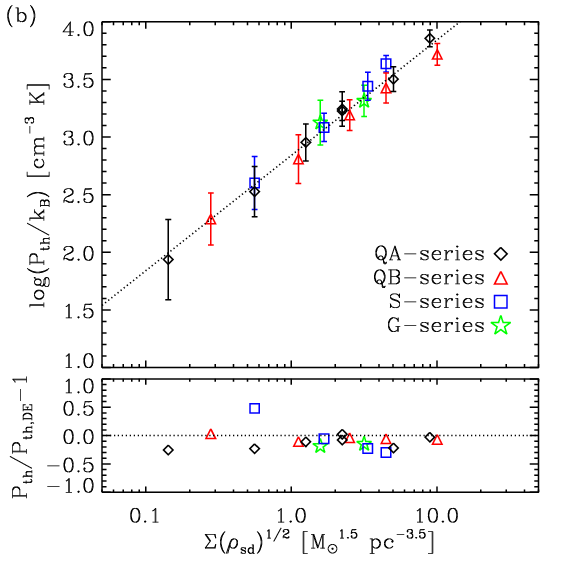}
\caption{\label{fig:Pcomp} 
Top: Midplane (\emph{a}) total and (\emph{b}) thermal pressures of the diffuse
gas as functions of $\Sigma\sqrt{\rhosd}$.  The points 
and errorbars give the mean and standard deviations over $t/\torb=2-3$. 
The dotted lines in upper panels show fits
$\Ptot/\kbol =
9.9\times10^3\Punit (\Sigma/10\Surf)(\rhosd/0.1\Msun\pc^{-3})^{1/2}$
and
$\Pth/\kbol = 2.2\times10^3\Punit
(\Sigma/10\Surf)(\rhosd/0.1\Msun\pc^{-3})^{1/2}$, respectively. 
Bottom: Relative differences between measured midplane pressures
and the dynamical equilibrium estimates using equation (\ref{eq:PDE2}).
The mean midplane pressure $\Ptot$ varies only 13\% relative to
$\PtotDE$, showing that vertical dynamical equilibrium is
an excellent approximation.
}
\end{figure}

The main results from our simulations are as follows:

1.  We find that most of the gas is at pressures, densities, and temperatures
close to thermal equilibrium 
(Fig.~\ref{fig:pdf}). The system evolves to a state in which both
warm and cold stable phases are present, with mean midplane thermal 
pressure $\Pth$ within 
$\sim 40\%$  of the  ``two-phase'' pressure 
$\Ptwo\equiv(\Pmin\Pmax)^{1/2}$, decreasing weakly with increasing
$\SigSFR$ (equation \ref{eq:fth}).  This evolution involves continuous
re-adjustment of the thermal equilibrium curve, as $\Ptwo \propto 
\Gamma \propto \SigSFR$.  Since $\SigSFR$ varies by two orders of
magnitude for our model suite, the thermal equilibrium curve shifts
up and down by the same factor. The midplane
thermal pressure increases from $\Pth/k_B\sim 100 \Punit$ to $\sim 10^4 \Punit$
going from low-$\Sigma$, low-$\rhosd$ to high-$\Sigma$, high-$\rhosd$ models 
(Fig.~\ref{fig:Pcomp}).  This finding is consistent with the
conclusion of \citet{wol03} that \ion{H}{1} should be
found in two phases out to large distances in the Milky Way (based on
an assumed heating rate that declines outward), as well 
as observations indicating both phases are indeed present out to 
$\sim 20-25\kpc$ \citep{dic09}.  Other nearby galaxies also show
evidence for both warm and cold atomic gas \citep{bra97,dic00,you03}.  
The analytic model of OML10 adopted the
assumption that the midplane thermal pressure is equal to $\Ptwo$; our
numerical results show that this is indeed a good first estimate.
The result $\Pth \sim \Ptwo$ implies that radiative heating
approximately balances cooling. From the point of view of thermal 
energy replenishment, this means that star formation is highly efficient.

2. By comparing the prediction of dynamical-equilibrium pressure with
the measured time-averaged midplane pressure in our numerical
simulations, we find that vertical dynamical equilibrium is satisfied
within $13\%$ for the total pressure
(lower panel of Fig.~\ref{fig:Pcomp}\emph{a}).  For the present
models, the total weight of the diffuse ISM ($\PtotDE$, given in
equation \ref{eq:PtotDE} or \ref{eq:PDE2})
is matched by a combination of thermal and turbulent pressure.  In
outer disks, where diffuse gas dominates the total surface density
$\Sigdiff \approx \Sigma$, simplified expressions for the total midplane
pressure in equilibrium are given by equations (\ref{eq:PtotDEdiff})
and (\ref{eq:PtotDEap}).  In many outer-disk regions (including the
Solar neighborhood), the vertical gravity from the stars exceeds that
from the gas, such that in equilibrium $\Ptot \propto \Sigma
\sqrt{\rhosd}$ if the vertical velocity dispersion is constant.  The
results from our simulations fit this form well 
(equation \ref{eq:Ptotfit}), with a
similar result for midplane thermal pressure (equation
\ref{eq:Pthfit}).  The numerical results that $\Pth \approx \PDE$ and
$\Ptot\approx \PtotDE$ demonstrate the validity of the vertical dynamical
equilibrium assumption in the theory of OML10, and confirms prior
findings from simulations by \citet{ko09b}.

\begin{figure}
\epsscale{0.8} \plotone{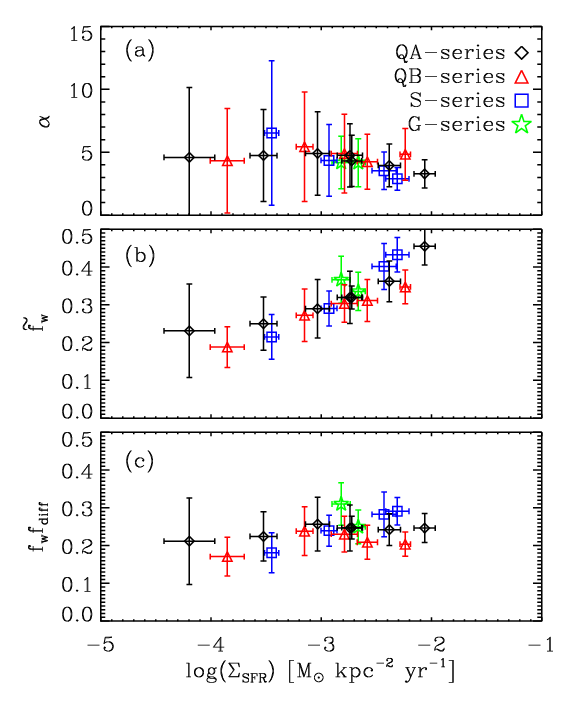} 
\caption{
Computed values of (\emph{a}) the ratio of total-to-thermal 
velocity dispersion for the diffuse gas
$\alpha \equiv 1 + \vzdiff^2/\vthdiff^2$, 
(\emph{b}) the square of mass-weighted 
thermal to warm-medium thermal speed $\vthdiff^2/c_w^2=\tilde{f}_w$, and
(\emph{c}) the product $f_w\fdiff$ (for $f_w\approx\tilde{f}_w$ the warm gas
mass fraction in the diffuse gas and $\fdiff$ the diffuse mass fraction), 
as functions of $\SigSFR$ for all models except Series R.  The points 
and errorbars give the mean and standard deviations over $t/\torb=2-3$.
Over more than two orders of magnitude in $\SigSFR$, 
the balance between energy input (heating, turbulent driving) and
energy output (cooling, turbulent dissipation) maintains nearly 
constant 
$\alpha = \Ptot/\Pth$ and both warm and cold gas phases.
\label{fig:afw_P}}
\end{figure}

3. Based on our numerical measurements of the thermal and turbulent
pressures, we find a ratio 
$\Ptot/\Pth=\alpha \approx 4-5$ 
for essentially all our models (Fig.~\ref{fig:afw_P}\emph{a}) when we
fix $\frad\equiv (\Gamma/\Gamma_0)(\SigSFR/\SigSFRsun)^{-1}=1$ (see
equation \ref{eq:heat}).  This is consistent with the assumption of
OML10 that $\alpha$ is relatively constant for galaxies with shielding
properties (and hence $J_{\rm FUV}/\SigSFR$) similar to the local Milky Way.
The near-constancy of $\alpha$ results from the fact that both thermal
and turbulent pressure are driven by feedback (see below).  When
$\frad$ is varied (for Series R models), corresponding to varying dust
shielding or FUV heating efficiency, $\alpha$ varies because $\Pth
\propto
\frad$ in thermal equilibrium.  Higher $\frad$ (lower shielding)
reduces $\alpha$ following equation (\ref{eq:alphasol});
for large $\frad$, $\Pth$ can exceed $\Pturb$.

4. We find that the fraction of diffuse gas in the warm component 
$\tilde f_w \equiv \vthdiff^2/c_w^2$ 
increases from $\sim 20\%$ to $\sim 50\%$ from low- to
high-$\SigSFR$,  when we hold $\frad=1$, corresponding to 
$\Gamma/\SigSFR=const$.
The upper range, with half of the diffuse 
gas warm (for models similar to the Solar
neighborhood), is comparable to findings of \citet{heiles03} based
on 21 cm emission and absorption observations.  We find 
(for Series R)
that the warm fraction steeply increases as $\frad$ increases (higher 
$\Gamma/\SigSFR$, corresponding to lower
shielding by dust).  This trend is consistent with the finding of
\cite{dic00} that the SMC, with a relative metallicity $\sim 0.2$,
has a much higher warm-to-cold \ion{H}{1} ratio than the Solar neighborhood.
We note that in real galaxies, $\frad$ would be inversely 
correlated with $\Sigma$ (see Section \ref{sec:rad}), which would
increase the warm fraction at low $\Sigma$ compared to the $\frad=1$
models in Series QA, QB, and S presented here.

\begin{figure}
\epsscale{0.8} \plotone{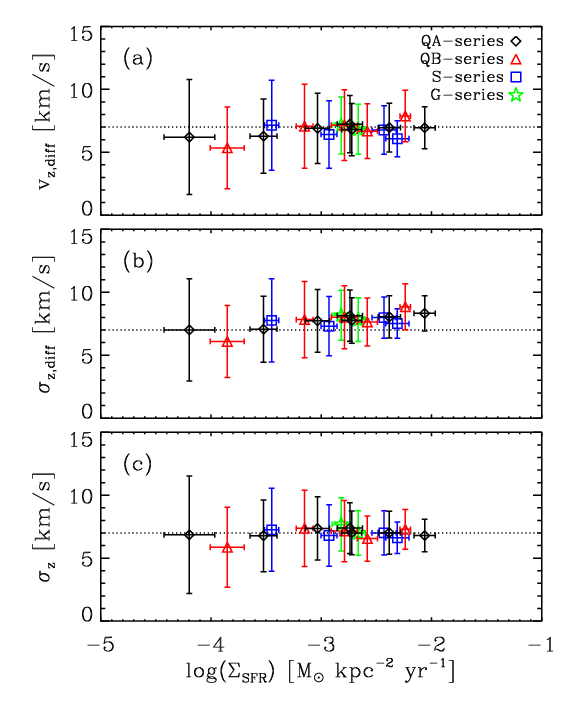} 
\caption{
(\emph{a}) The vertical turbulent velocity dispersion of the diffuse gas
$\vzdiff$, (\emph {b}) the total (turbulent+thermal) 
velocity dispersion of the diffuse gas
$\szdiff$, and (\emph {c}) the total velocity dispersion of all gas $\sz$,
as functions of the SFR surface density $\SigSFR$ for all models except Series
R.  The points and errorbars give the mean and standard deviations
over $t/\torb=2-3$. For the whole set of models shown in this
figure, $\vzdiff=6.8\pm 0.6\kms$, 
$\szdiff =7.7 \pm 0.6\kms$, and
$\sz=7.0\pm 0.4\kms$.  The dotted lines in all panels 
show $7\kms$ for reference.  
\label{fig:v_sfr}
}
\end{figure}

5. The time-averaged turbulent vertical velocity dispersions in all of
our models are $\vzdiff \approx 7\kms$, with no systematic dependence on
$\SigSFR$ (Fig.~\ref{fig:v_sfr}).  Total vertical velocity dispersions 
$\szdiff$ in the diffuse medium are larger by $\sim 1-2 \kms$.  The turbulent
amplitudes we find, and the lack of correlation of $\szdiff$ with
$\SigSFR$, are consistent with observations of \ion{H}{1} velocity
dispersions in the Milky Way and nearby face-on galaxies 
\citep{heiles03,dhh90,van99,pet07,Kal09}.  As discussed in Section 
\ref{sec:sflaw}
(see also OS11), turbulent velocity dispersions 
$\vzdiff \sim \ever\Psn/\Msn$ are expected if the 
star formation efficiency per vertical
crossing time is $\ever \equiv \tver \SigSFR/\Sigma$ 
(for $\tver\equiv\Hdiff/\vzdiff$), and the momentum 
injection per stellar mass from feedback is $\Psn/\Msn$. 
Confirming this expectation, the turbulent amplitudes we find 
are consistent with the mean value 
$\ever =0.0025$ measured from our numerical models, for the momentum 
feedback parameter $\Psn/\Msn=3000\kms$ used in our simulations.

\begin{figure}
\epsscale{1.0} \plotone{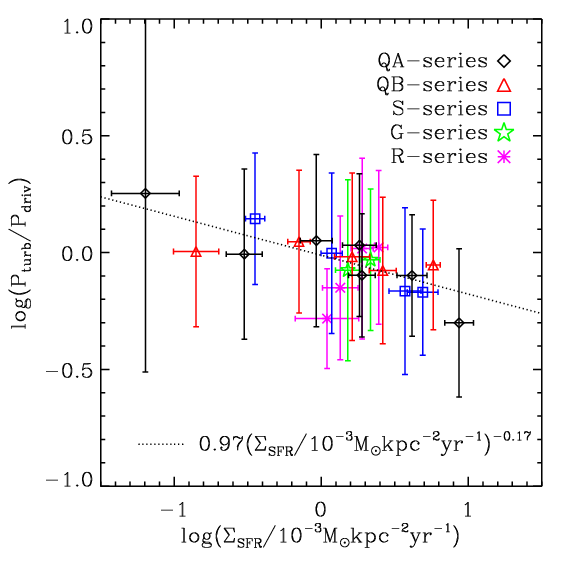} \caption{\label{fig:fp} Measured 
midplane turbulent pressure $\Pturb$ of the diffuse gas relative to the
vertical momentum flux injected by star formation $\Pdriv$, as a function of 
$\SigSFR$. The points and errorbars give the mean
and standard deviations over $t/\torb=2-3$ for each model. The dotted
line with a slope of $-0.17$ gives the best fit.
The result $\Pturb\sim \Pdriv$ indicates that turbulent driving is 
consistently balanced by 
dissipation on approximately a vertical crossing time, even though both
terms vary by more than two orders of magnitude as $\SigSFR$ changes.
}
\end{figure}

6. To assess the balance of turbulent driving and dissipation in our
numerical models, we compare the measured turbulent pressure at the
midplane $\Pturb \equiv \rho_0 \vzdiff^2$ with the fiducial momentum
injection rate per unit area $\Pdriv\equiv0.25(\Psn/\Msn)\SigSFR$ from
star formation feedback.  Fig.~\ref{fig:fp} shows 
that these are approximately equal, decreasing weakly with increasing 
$\SigSFR$ (equation \ref{eq:fp}).  Since $\Pturb$ represents the
characteristic vertical momentum per unit area in the diffuse ISM 
($\Sigdiff \vzdiff$)
divided by $2\Hdiff/\vzdiff$,
this implies the 
momentum dissipation timescale is comparable to the crossing time 
$\tver= \Hdiff/\vzdiff$,
consistent with
previous numerical results on turbulent driving and dissipation 
(e.g. \citealt{sto98,mac98}).  Another way to think of this result is
that the momentum injected in the diffuse ISM by star formation per
unit time is comparable to the existing vertical momentum divided by
the dynamical time.  Thus, from the point of view of momentum
replenishment, star formation is highly efficient.

7. We use our numerical models to calibrate the feedback yield parameters 
$\etath$ and $\etaturb$, respectively 
equal to the ratio $\Pth/\SigSFR$ and $\Pturb/\SigSFR$ in suitable
units (see
equations \ref{eq:etath} and \ref{eq:etaturb}).  Both yield parameters 
decrease only very weakly with increasing $\SigSFR$ (see equations 
\ref{eq:etathnum} and \ref{eq:etaturbnum}), with thermal yield also
depending on the radiation penetration parameter as $\etath \propto \frad$.  
This explains why 
$\alpha = \Ptot/\Pth = 1 + \etaturb/\etath$ is 
nearly constant (for $\frad=1$).  The values $\etath \sim 1 \times
\frad$ and 
$\etaturb \sim 4$ obtained from our numerical models 
are consistent with the analytic predictions of OML10
and OS11, respectively.

\begin{figure}
\epsscale{1.0} \plotone{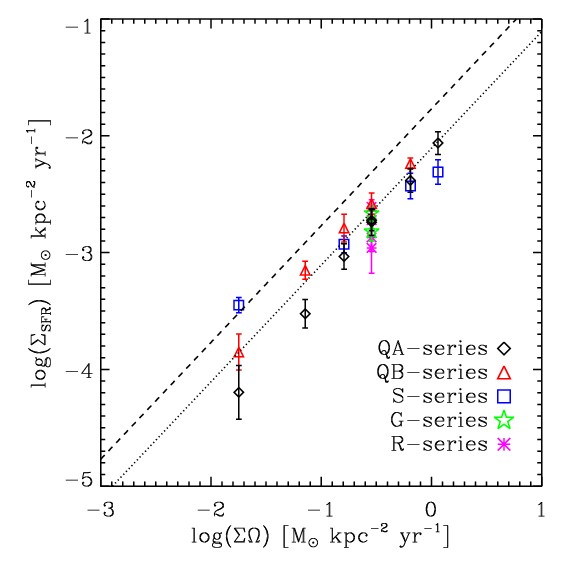}
\caption{
Measured $\SigSFR$ as a function of $\Sigma\Omega$  for all
models. The points and errorbars give mean values and standard
deviations over $t/\torb=2-3$. The dotted line 
shows our best fit $\SigSFR=0.008\Sigma\Omega$ for an imposed unity slope,
and the dashed line shows the empirical result
$\SigSFR=0.017\Sigma\Omega$ of \cite{ken98}.
\label{fig:sflaw_Om}
}
\end{figure}

\begin{figure}
\epsscale{1.1} \plottwo{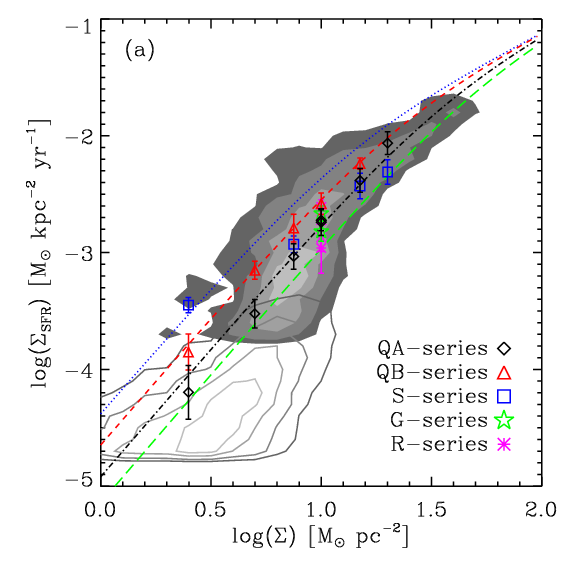}{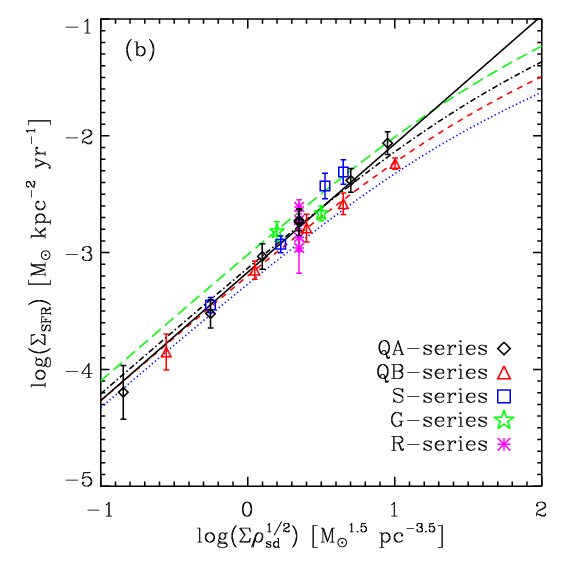}
\caption{\label{fig:sflaw}
SFR surface density $\SigSFR$ as a function of (\emph{a}) $\Sigma$ 
and (\emph{b}) $\Sigma\rhosd^{1/2}$  for all models.  The points and
errorbars give the mean and standard deviations over $t/\torb=2-3$,
respectively. In both panels, blue dotted, red dashed, black dot-dashed, and
green long-dashed lines give the theoretical predictions obtained by solving
equations (\ref{eq:tsf}), (\ref{eq:etath}), and (\ref{eq:PDE2}) 
simultaneously for $s_0=0.02, 0.07, 0.28,$ and $1.10$, respectively. 
The parameters $\sz=7\kms$, $\alpha=5$, 
and $\tsf=1.3\Gyr$ are held fixed for these analytic comparisons,
while $\etath$ varies following 
the numerical fit in equation (\ref{eq:etathnum})
with $\frad=1$.
Filled and empty contours in (\emph{a}) show the observational 
measurements in the regions inside \citep{big08} and outside 
\citep{big10} of the optical radius, respectively, for nearby spirals 
and dwarf galaxies: the contour levels from dark to light correspond 
to 10\%, 25\%, 50\%, and 75\% of the 
data.
With higher $s_0$ and/or $\frad$ 
at low $\Sigma$ (not shown), the models can match  
the observations beyond the optical radius.
The black solid line
in (\emph{b}) denotes the power-law solution for $\SigSFR$ in equation
(\ref{eq:PthSFR2}).  
Note that $\SigSFR$ is much 
better correlated with the combination $\Sigma\rhosd^{1/2}$ than 
with $\Sigma$ alone.
}
\end{figure}

\begin{figure}
\epsscale{1.1} \plottwo{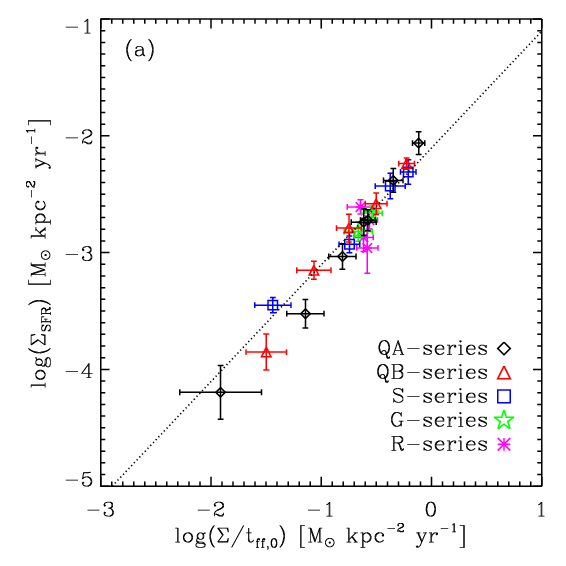}{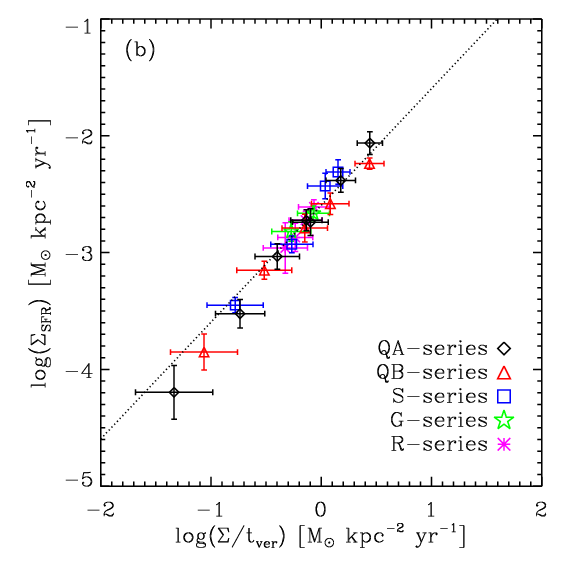}
\caption{
Measured SFR surface density $\SigSFR$ as a function of (\emph{a})
$\Sigma/\tffmid$  and (\emph{b}) $\Sigma/\tver$ for all numerical models, where
$\tffmid=(3\pi/(32G\rho_0))^{1/2}$ and $\tver=\Hdiff/\vzdiff$ 
are computed using time-averaged values of the variables.  The points 
and errorbars give the mean and standard deviations over $t/\torb=2-3$.
The dotted lines in (\emph{a}) and (\emph{b}) show our best fits
for imposed unity slopes,
$\SigSFR=0.008(\Sigma/\tffmid)$ and $\SigSFR=0.0025(\Sigma/\tver)$,
respectively.
\label{fig:sflaw_ff}}
\end{figure}

8. We compare our numerical results for $\SigSFR$ to several
commonly-used formulae,  $\SigSFR \propto \Sigma \Omega$,
$\SigSFR \propto \Sigma^{1+p}$, $\SigSFR = \eff(\rho_0) \Sigma/t_{\rm ff,0}$ 
(see Figs.~\ref{fig:sflaw_Om}, \ref{fig:sflaw}, \ref{fig:sflaw_ff}).  
The first two relations are not well correlated with the numerical
results.  The third relation has improved correlation, but this is in
part because $t_{\rm ff,0} \propto (G \rhosd)^{-1/2}$ for most of our
model suite, and $\SigSFR$ is well correlated with $\Sigma \rhosd^{1/2}$
(Fig.~\ref{fig:sflaw}; see also equation \ref{eq:PthSFR2}).  We also 
compare to the relation $\SigSFR = \ever \Sigma/\tver$ for 
$\tver = \Hdiff/\vzdiff$ the vertical crossing time, which limits how 
rapidly cold clouds can collect at the midplane. The fitted
efficiencies are $\eff(\rho_0)=0.008$ and $\ever=0.0025$, with a
stronger correlation to the vertical crossing-time than free-fall-time 
prescription.

\begin{figure}
\epsscale{1.1} 
\plottwo{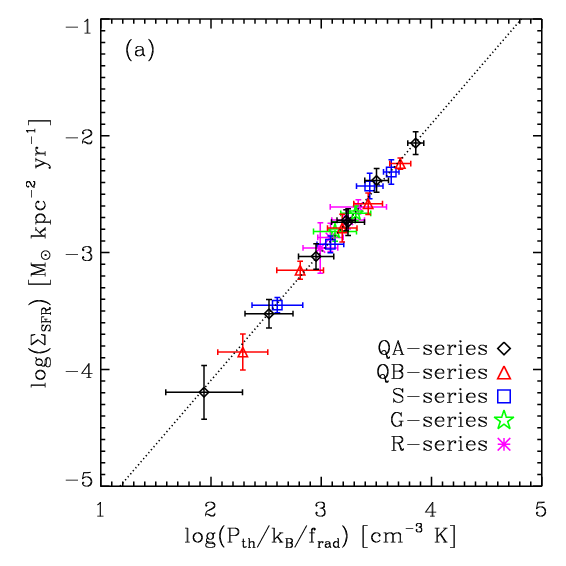}{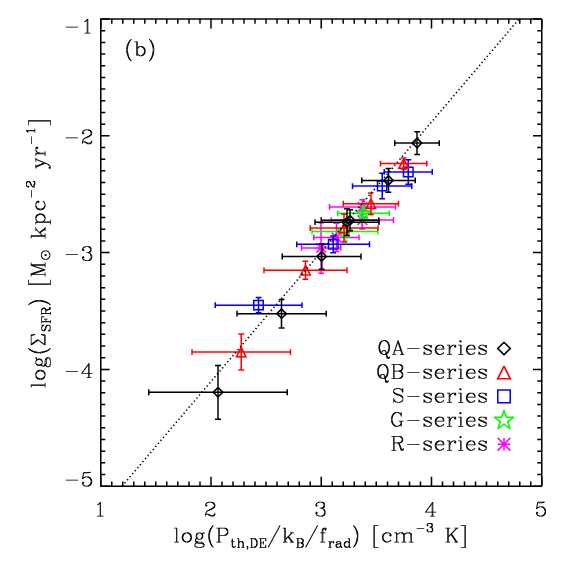}
\plottwo{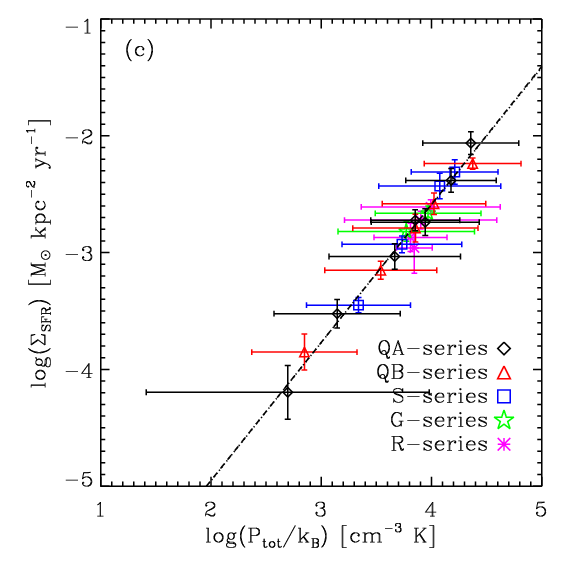}{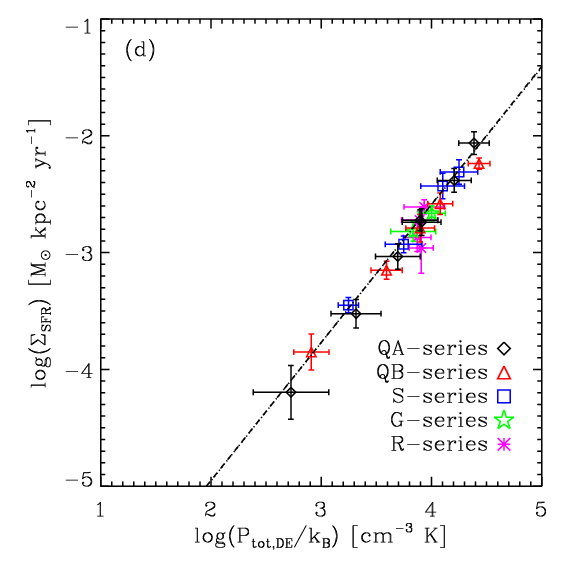}
\caption{\label{fig:sfr_P} Surface density of star formation 
  $\SigSFR$ measured from
  the simulations as functions of (\emph{a}) the measured midplane thermal
  pressure of the diffuse gas $\Pth$,
  (\emph{b}) the predicted midplane thermal pressure $\PDE$, 
(\emph{c}) the measured midplane total
  pressure of the diffuse gas $\Ptot$,
  and (\emph{d}) the predicted midplane total pressure $\PtotDE$.  
  The points and errorbars give the
  mean and standard deviations over $t/\torb=2-3$.  
  Predicted pressures use the dynamical equilibrium equation
  (\ref{eq:PDE2}) 
  and measured values of $\fdiff$, $\alpha$, and
  $\szdiff$ for each model.  In (\emph{a}) and (\emph{b}) $\Pth$ and $\PDE$ are
  divided by $\frad$ to compensate for varying heating efficiency so
  that Series R may be compared with other series.
  In top and bottom
  panels, dotted lines are obtained from equations (\ref{eq:etath})
  and (\ref{eq:PtotSFR}), respectively, using the numerical
  calibrations (\ref{eq:etathnum}) and (\ref{eq:etaturbnum}).  The
  dashed lines in bottom panels show the best fit given by equation
  (\ref{eq:numSFR}).
The pressures and $\SigSFR$ are extremely well correlated,
consistent with the idea that $\SigSFR$ adjusts until the pressures
(driven by feedback) match equilibrium requirements.
}
\end{figure}

9. The best star formation correlation we find is with the total
midplane pressure -- either as measured in the simulations ($\Ptot$),
or as estimated from vertical dynamical equilibrium
($\PtotDE$). Equation (\ref {eq:numSFR}) fits $\SigSFR$ within 16\%
for all models (excluding Series R), as shown in Fig.~\ref{fig:sfr_P}.
Series R shows that $\SigSFR$ drops if the shielding is reduced
(higher $\frad$).
Our numerical result that $\SigSFR$ has a near-linear correlation with
$\PtotDE$ is consistent with the analytic models of OML10 and OS11 for
star formation in diffuse-gas dominated regions -- either outer disks
or starbursts.   The near-linear
relation between $\SigSFR$ and $\PtotDE$ is also consistent with a similar 
empirical result found by \citet{leroy08}, and with the previous 
empirical findings that molecular gas (the immediate precursor of star
formation) increases nearly linearly with the ISM pressure 
\citep{wong02,br04,br06}.  
A relationship of the form $\SigSFR \sim \PtotDE/\eta $ (see
 equation \ref{eq:PtotSFR}) implies that star
formation responds to demand:  the star formation rate increases until
the midplane pressure (controlled by thermal and turbulent feedback) 
balances the vertical weight of the diffuse ISM. 

That energy input from massive stars determines the midplane pressure
and thus self-regulates the star formation rate suggests it is crucial
to include stellar feedback, 
when simulating galactic star formation numerically.
Indeed, work by \citet{hop11} contemporary with the present study
used SPH simulations to show that $\SigSFR$ is consistent with the
observed Kennicutt-Schmidt relations only when feedback is included
(see also \citealt{dob11}). Without feedback,
dense clouds collapse in a runaway fashion,
increasing the star formation rate by $\sim 1-2$ orders of magnitude.
Using grid-based simulations of Milky-Way-type galaxies,
\citet{tas11} similarly found that star formation rates are at least 
an order of magnitude higher than observations if feedback is
not included to drive turbulence and unbind dense clouds that form.
Including feedback is known to strongly affect the star formation
history in long-term simulations of galaxies (e.g. \citealt{gov07}).

Stellar feedback also appears essential for driving and maintaining
turbulence in the direction perpendicular to the disk plane over many
galactic orbits.  Other proposed mechanisms for generating ISM
turbulence include large-scale gravitational instabilities (e.g.,
\citealt{wad02,kos03,kim07,age09,aum10,bou10}),
magnetorotational instabilities (e.g.,
\citealt{kos03,pio04,pio05,pio07}), and non-steady motions generated
in spiral shocks (e.g. \citealt{kim06,kko06,kko10,dob06}). Turbulence
driven by these processes has lower vertical than horizontal velocity
dispersions, because they all tap galactic rotation.  
Rotational-gravitational instabilities are able to produce turbulence
levels comparable to observed values, although vertical dispersions
drop to $\simlt 4\kms$ after several galactic orbits \citep{age09}. In
addition, gravitationally-driven turbulence is dominated by large
scales (i.e., clump-to-clump motions) that do not prevent collapse
within clumps. Without stellar feedback to unbind dense clouds that
form, the resulting star formation rates are too
high. Gravitationally-driven turbulence is likely to be most important
during the transient, gas-rich early stages of galaxy formation at
high redshift (e.g. \citealt{cev10}).  
Characterizing turbulence in galaxies requires
subtraction of a ``background state,'' and this becomes more difficult
to define when there are large secular motions including prominent radial
flows and collapsing clumps.  Even steady spiral shocks create 
a (steady) azimuthal velocity profile that can differ by tens of $\kms$ from 
the background rotation curve. It will be interesting to
analyze in detail how non-stellar processes combine with stellar
feedback to power turbulence over both large and small scales in disk
galaxies, providing a more complete understanding of galactic star
formation over all redshifts.

As noted above, the present numerical models involve radical
simplifications compared to the real star-forming ISM.  Given the
success of these simple models, it is clearly worthwhile to pursue
further computational modeling along similar lines, improving on the
numerical idealizations we have made.  One of the advantages of local
numerical models that resolve $\sim \pc - \kpc$ is that the scales
involved directly correspond to those accessible in high-resolution
observations of nearby galaxies.  Results from successive model
refinements can be compared to observations to identify a ``minimal
physics'' set, incorporating only the most important effects to
minimize computational cost.

By employing high-resolution ISM simulations to identify the
key processes controlling star formation, it will be possible to
enhance subgrid models for computational galaxy formation studies in the
cosmological context.  While feedback to drive turbulent pressure
plays a dominant role in the ISM of the Milky Way and similar
galaxies, feedback to drive thermal pressure is likely to be
increasingly important where there is minimal dust shielding,
potentially leading to 
large $\frad$ and 
$\etath \gg \etaturb$.  Some current
simulations of dust-poor galaxies at high redshift use subgrid
shielding models to estimate the abundance of cold, star-forming gas
(e.g. \citealt{gne09,gne10,kuh11}).  A subgrid model that incorporates
both shielding and turbulence could potentially bridge over a wide
range of redshifts.

\acknowledgements{  
The authors are grateful to the referee for helpful comments on the
manuscript.
The work of C.-G.~K. and W.-T.~K. was supported by the National Research
Foundation of Korea (NRF), funded by the Korean government
(MEST) under grant No.\ 2010-0000712. 
The work of E.~C.~O. was supported by grant AST-0908185 from the
U.S. National Science Foundation.
}

\begin{deluxetable}{lcccccc}
\tabletypesize{\footnotesize}
\tablewidth{0pt}
\tablecaption{Model Parameters\label{tbl:model}}
\tablehead{
\colhead{Model} &
\colhead{$\Sigma$} &
\colhead{$\rho_{\rm sd}$} &
\colhead{$\Omega$} &
\colhead{$H_w$} &
\colhead{$L_z$} &
\colhead{$s_0$} \\
\colhead{}&
\colhead{[$\Msun\,\pc^{-2}$]}&
\colhead{[$\Msun\,\pc^{-3}$]}&
\colhead{[$\kms\,\kpc^{-1}$]}&
\colhead{[$\pc$]} &
\colhead{[$\pc$]} &
\colhead{}
}
\startdata
 QA02 &   2.5 & 0.0031 &   7 & 528 & 2048 & 0.28 \\
 QA05 &   5.0 & 0.0125 &  14 & 269 & 1024 & 0.28 \\
 QA07 &   7.5 & 0.0281 &  21 & 179 &  768 & 0.28 \\
 QA10 &  10.0 & 0.0500 &  28 & 134 &  512 & 0.28 \\
 QA15 &  15.0 & 0.1125 &  42 &  89 &  384 & 0.28 \\
 QA20 &  20.0 & 0.2000 &  56 &  67 &  256 & 0.28 \\
\hline
 QB02 &   2.5 & 0.0125 &   7 & 269 & 1024 & 0.07 \\
 QB05 &   5.0 & 0.0500 &  14 & 134 &  768 & 0.07 \\
 QB07 &   7.5 & 0.1125 &  21 &  89 &  512 & 0.07 \\
 QB10 &  10.0 & 0.2000 &  28 &  67 &  384 & 0.07 \\
 QB15 &  15.0 & 0.4500 &  42 &  44 &  256 & 0.07 \\
\hline
  S02 &   2.5 & 0.0500 &   7 & 134 &  768 & 0.02 \\
  S05 &   5.0 & 0.0500 &  14 & 134 &  768 & 0.07 \\
  S07 &   7.5 & 0.0500 &  21 & 134 &  512 & 0.16 \\
  S10 &  10.0 & 0.0500 &  28 & 134 &  512 & 0.28 \\
  S15 &  15.0 & 0.0500 &  42 & 134 &  512 & 0.62 \\
  S20 &  20.0 & 0.0500 &  56 & 134 &  512 & 1.10 \\
\hline
  G02 &  10.0 & 0.0250 &  28 & 190 &  768 & 0.55 \\
  G05 &  10.0 & 0.0500 &  28 & 134 &  512 & 0.28 \\
  G10 &  10.0 & 0.1000 &  28 &  95 &  512 & 0.14 \\
  G20 &  10.0 & 0.2000 &  28 &  67 &  384 & 0.07
\enddata
\tablecomments{Models 
S05, S10, G05, and G20 are identical to
QB05, QA10, QA10, and QB10 models, respectively.  
All models 
in Series QA, QB, S, and G have $f_{\rm rad}=1$. Models in the R series
(not listed) have the same parameters as model QA10, except $f_{\rm rad}=$ 
0.25, 0.5, 2.5, and 5.0 for R02, R05, R25, and R50, respectively.
All models have $L_x=512\pc$ except 
model QA10x2, which is the same as QA10 but with $L_x=1024\pc$.
}
\end{deluxetable}

\begin{deluxetable}{lccccc}
\tabletypesize{\scriptsize} \tablewidth{0pt}
\tablecaption{Disk Properties 1\label{tbl:stat1}} 
\tablehead{
\colhead{Model} & 
\colhead{$\log\abrackets{\SigSFR}$} &
\colhead{$\log\abrackets{\Pth/\kbol}$} &
\colhead{$\log\abrackets{\Pturb/\kbol}$} &
\colhead{$\abrackets{\rhomid}$} &
\colhead{$\abrackets{\Hdiff}$} \\
\colhead{(1)} & \colhead{(2)} & \colhead{(3)} & \colhead{(4)} &
\colhead{(5)} & \colhead{(6)} }
\startdata
   QA02 & $-4.20\pm 0.23 $ & $ 1.94\pm 0.35 $ & $ 2.61\pm 1.51 $ & $ 0.05\pm 0.08 $ & $ 342\pm 111 $ \\
   QA05 & $-3.52\pm 0.12 $ & $ 2.53\pm 0.22 $ & $ 3.03\pm 0.72 $ & $ 0.39\pm 0.30 $ & $ 174\pm  37 $ \\
   QA07 & $-3.03\pm 0.11 $ & $ 2.95\pm 0.16 $ & $ 3.57\pm 0.73 $ & $ 0.82\pm 0.46 $ & $ 132\pm  29 $ \\
   QA10 & $-2.74\pm 0.11 $ & $ 3.24\pm 0.15 $ & $ 3.85\pm 0.60 $ & $ 1.12\pm 0.58 $ & $  92\pm  18 $ \\
 QA10x2 & $-2.72\pm 0.09 $ & $ 3.23\pm 0.08 $ & $ 3.74\pm 0.52 $ & $ 1.32\pm 0.40 $ & $  94\pm  10 $ \\
   QA15 & $-2.38\pm 0.10 $ & $ 3.50\pm 0.11 $ & $ 4.08\pm 0.51 $ & $ 1.70\pm 0.69 $ & $  70\pm   9 $ \\
   QA20 & $-2.06\pm 0.10 $ & $ 3.86\pm 0.07 $ & $ 4.19\pm 0.63 $ & $ 2.76\pm 0.70 $ & $  51\pm   5 $ \\
\hline
   QB02 & $-3.85\pm 0.15 $ & $ 2.29\pm 0.23 $ & $ 2.71\pm 0.63 $ & $ 0.30\pm 0.26 $ & $ 157\pm  53 $ \\
   QB05 & $-3.15\pm 0.08 $ & $ 2.81\pm 0.21 $ & $ 3.45\pm 0.61 $ & $ 0.56\pm 0.40 $ & $ 118\pm  38 $ \\
   QB07 & $-2.79\pm 0.12 $ & $ 3.19\pm 0.13 $ & $ 3.75\pm 0.71 $ & $ 1.07\pm 0.55 $ & $  77\pm  20 $ \\
   QB10 & $-2.58\pm 0.09 $ & $ 3.43\pm 0.13 $ & $ 3.90\pm 0.62 $ & $ 1.88\pm 0.86 $ & $  56\pm  12 $ \\
   QB15 & $-2.24\pm 0.05 $ & $ 3.72\pm 0.09 $ & $ 4.27\pm 0.55 $ & $ 2.97\pm 0.98 $ & $  44\pm   6 $ \\
\hline
    S02 & $-3.45\pm 0.07 $ & $ 2.60\pm 0.23 $ & $ 3.25\pm 0.56 $ & $ 0.40\pm 0.30 $ & $ 110\pm  33 $ \\
    S07 & $-2.93\pm 0.07 $ & $ 3.08\pm 0.12 $ & $ 3.62\pm 0.68 $ & $ 1.09\pm 0.47 $ & $  90\pm  13 $ \\
    S15 & $-2.43\pm 0.11 $ & $ 3.44\pm 0.12 $ & $ 3.96\pm 0.70 $ & $ 1.50\pm 0.94 $ & $  95\pm  22 $ \\
    S20 & $-2.31\pm 0.10 $ & $ 3.64\pm 0.07 $ & $ 4.08\pm 0.53 $ & $ 1.79\pm 0.58 $ & $  87\pm   7 $ \\
\hline
    G02 & $-2.82\pm 0.08 $ & $ 3.13\pm 0.20 $ & $ 3.66\pm 0.77 $ & $ 0.88\pm 0.52 $ & $ 136\pm  34 $ \\
    G10 & $-2.66\pm 0.06 $ & $ 3.31\pm 0.14 $ & $ 3.86\pm 0.60 $ & $ 1.57\pm 0.68 $ & $  82\pm  13 $ \\
\hline
    R02 & $-2.61\pm 0.06 $ & $ 2.73\pm 0.25 $ & $ 3.97\pm 0.66 $ & $ 0.98\pm 0.54 $ & $  99\pm  15 $ \\
    R05 & $-2.72\pm 0.08 $ & $ 2.94\pm 0.15 $ & $ 3.85\pm 0.77 $ & $ 1.42\pm 0.48 $ & $  96\pm  17 $ \\
    R25 & $-2.87\pm 0.12 $ & $ 3.48\pm 0.12 $ & $ 3.53\pm 0.60 $ & $ 1.11\pm 0.45 $ & $  92\pm  12 $ \\
    R50 & $-2.96\pm 0.22 $ & $ 3.69\pm 0.16 $ & $ 3.31\pm 0.37 $ & $ 1.29\pm 0.58 $ & $  97\pm  20 $ 
\enddata
\tablecomments{
The mean values and 
standard deviations of physical quantities are averaged over $t/\torb=2-3$.
Col. (2): Logarithmic value of the SFR surface density ($\sfrunit$).
Cols. (3)-(4): Logarithmic values of the midplane thermal and 
turbulent pressures over $\kbol$ ($\Punit$).
Col. (5): Midplane number density of hydrogen ($\pcc$).
Col. (6): Scale height of the diffuse component ($\pc$).
See Section \ref{sec:stat} for definitions.
} 
\end{deluxetable}

\begin{deluxetable}{lccccccc}
\tabletypesize{\scriptsize} \tablewidth{0pt}
\tablecaption{Disk Properties 2\label{tbl:stat2}} 
\tablehead{
\colhead{Model} & 
\colhead{$\abrackets{\vzdiff}$} & 
\colhead{$\abrackets{\vthdiff}$} &
\colhead{$\abrackets{\sz}$} &
\colhead{$\abrackets{\fdiff}$} & 
\colhead{$\alpha$} & 
\colhead{$\tilde{f}_w$} & 
\colhead{$\tausf$} \\
\colhead{(1)} & \colhead{(2)} & \colhead{(3)} & \colhead{(4)} &
\colhead{(5)} & \colhead{(6)} & \colhead{(7)} & \colhead{(8)} }
\startdata
   QA02 & $ 6.20\pm 4.57 $ & $ 3.27\pm 0.78 $ & $ 6.86\pm 4.67 $ & $ 0.92\pm 0.07 $ & $ 4.59\pm 5.56 $ & $ 0.23\pm 0.12 $ & $ 3.31\pm 3.35 $ \\
   QA05 & $ 6.28\pm 2.95 $ & $ 3.25\pm 0.43 $ & $ 6.78\pm 2.85 $ & $ 0.90\pm 0.06 $ & $ 4.74\pm 3.65 $ & $ 0.25\pm 0.07 $ & $ 1.73\pm 1.17 $ \\
   QA07 & $ 6.90\pm 2.78 $ & $ 3.49\pm 0.46 $ & $ 7.36\pm 2.51 $ & $ 0.89\pm 0.06 $ & $ 4.90\pm 3.31 $ & $ 0.29\pm 0.08 $ & $ 0.92\pm 0.56 $ \\
   QA10 & $ 7.23\pm 2.28 $ & $ 3.73\pm 0.42 $ & $ 7.39\pm 2.02 $ & $ 0.77\pm 0.09 $ & $ 4.75\pm 2.51 $ & $ 0.32\pm 0.07 $ & $ 1.26\pm 0.60 $ \\
 QA10x2 & $ 6.80\pm 2.07 $ & $ 3.74\pm 0.18 $ & $ 7.01\pm 1.73 $ & $ 0.77\pm 0.06 $ & $ 4.31\pm 2.04 $ & $ 0.32\pm 0.03 $ & $ 1.19\pm 0.40 $ \\
   QA15 & $ 6.95\pm 1.94 $ & $ 4.05\pm 0.30 $ & $ 7.02\pm 1.71 $ & $ 0.67\pm 0.06 $ & $ 3.95\pm 1.70 $ & $ 0.36\pm 0.05 $ & $ 1.20\pm 0.36 $ \\
   QA20 & $ 6.94\pm 1.67 $ & $ 4.59\pm 0.22 $ & $ 6.80\pm 1.30 $ & $ 0.54\pm 0.06 $ & $ 3.29\pm 1.12 $ & $ 0.46\pm 0.05 $ & $ 1.06\pm 0.27 $ \\
\hline
   QB02 & $ 5.35\pm 3.26 $ & $ 2.93\pm 0.39 $ & $ 5.87\pm 3.18 $ & $ 0.91\pm 0.08 $ & $ 4.32\pm 4.15 $ & $ 0.19\pm 0.05 $ & $ 1.64\pm 1.47 $ \\
   QB05 & $ 7.07\pm 3.34 $ & $ 3.36\pm 0.45 $ & $ 7.37\pm 3.02 $ & $ 0.87\pm 0.08 $ & $ 5.44\pm 4.36 $ & $ 0.27\pm 0.07 $ & $ 0.89\pm 0.59 $ \\
   QB07 & $ 7.15\pm 2.81 $ & $ 3.62\pm 0.31 $ & $ 7.16\pm 2.43 $ & $ 0.76\pm 0.09 $ & $ 4.89\pm 3.13 $ & $ 0.30\pm 0.05 $ & $ 1.12\pm 0.53 $ \\
   QB10 & $ 6.68\pm 2.17 $ & $ 3.70\pm 0.32 $ & $ 6.55\pm 1.79 $ & $ 0.67\pm 0.08 $ & $ 4.25\pm 2.19 $ & $ 0.31\pm 0.06 $ & $ 1.26\pm 0.41 $ \\
   QB15 & $ 7.88\pm 2.05 $ & $ 4.02\pm 0.26 $ & $ 7.29\pm 1.58 $ & $ 0.59\pm 0.05 $ & $ 4.83\pm 2.06 $ & $ 0.35\pm 0.04 $ & $ 1.07\pm 0.18 $ \\
\hline
    S02 & $ 7.14\pm 3.58 $ & $ 3.04\pm 0.41 $ & $ 7.25\pm 3.30 $ & $ 0.84\pm 0.08 $ & $ 6.53\pm 5.75 $ & $ 0.22\pm 0.06 $ & $ 1.12\pm 0.62 $ \\
    S07 & $ 6.41\pm 2.68 $ & $ 3.50\pm 0.29 $ & $ 6.80\pm 2.44 $ & $ 0.82\pm 0.05 $ & $ 4.35\pm 2.85 $ & $ 0.29\pm 0.05 $ & $ 1.12\pm 0.38 $ \\
    S15 & $ 6.76\pm 1.92 $ & $ 4.25\pm 0.31 $ & $ 7.01\pm 1.76 $ & $ 0.70\pm 0.10 $ & $ 3.54\pm 1.49 $ & $ 0.40\pm 0.06 $ & $ 1.20\pm 0.51 $ \\
    S20 & $ 6.08\pm 1.43 $ & $ 4.42\pm 0.22 $ & $ 6.62\pm 1.25 $ & $ 0.67\pm 0.05 $ & $ 2.89\pm 0.91 $ & $ 0.43\pm 0.05 $ & $ 1.34\pm 0.37 $ \\
\hline
    G02 & $ 7.13\pm 2.26 $ & $ 4.00\pm 0.34 $ & $ 7.68\pm 2.11 $ & $ 0.85\pm 0.05 $ & $ 4.18\pm 2.09 $ & $ 0.37\pm 0.06 $ & $ 1.01\pm 0.39 $ \\
    G10 & $ 6.82\pm 1.99 $ & $ 3.83\pm 0.30 $ & $ 6.99\pm 1.77 $ & $ 0.74\pm 0.07 $ & $ 4.17\pm 1.91 $ & $ 0.34\pm 0.05 $ & $ 1.18\pm 0.37 $ \\
\hline
    R02 & $ 8.36\pm 2.35 $ & $ 2.28\pm 0.19 $ & $ 7.85\pm 1.92 $ & $ 0.78\pm 0.07 $ & $14.51\pm 7.93 $ & $ 0.11\pm 0.02 $ & $ 0.91\pm 0.32 $ \\
    R05 & $ 6.80\pm 2.01 $ & $ 2.87\pm 0.28 $ & $ 6.84\pm 1.92 $ & $ 0.80\pm 0.06 $ & $ 6.62\pm 3.50 $ & $ 0.19\pm 0.04 $ & $ 1.05\pm 0.37 $ \\
    R25 & $ 5.25\pm 1.77 $ & $ 4.87\pm 0.40 $ & $ 6.72\pm 1.91 $ & $ 0.77\pm 0.06 $ & $ 2.16\pm 0.80 $ & $ 0.54\pm 0.09 $ & $ 1.69\pm 0.65 $ \\
    R50 & $ 4.49\pm 1.85 $ & $ 5.70\pm 0.49 $ & $ 6.78\pm 1.84 $ & $ 0.85\pm 0.07 $ & $ 1.62\pm 0.52 $ & $ 0.73\pm 0.12 $ & $ 1.39\pm 0.92 $ 
\enddata
\tablecomments{
The mean values and standard deviations of physical quantities are
averaged over $t/\torb=2-3$.
Cols. (2)-(3): Vertical turbulent and thermal velocity dispersions of the 
diffuse gas ($\kms$).
Col. (4): Total vertical velocity dispersion for all gas ($\kms$).
Cols. (5)-(7): Mass fraction of the diffuse gas ($\fdiff$), the ratio of total
pressure to turbulent pressure ($\alpha$), and the square of mass-weighted 
thermal to warm-medium thermal speed 
($\vthdiff^2/c_w^2=\tilde{f}_w$) in the diffuse gas.
Col. (8): Timescale to convert dense gas into stars ($\Gyr$).
See Section \ref{sec:stat} for definitions.
}
\end{deluxetable}

\clearpage


\end{document}